\documentclass[11pt,a4paper]{article}
\pdfoutput=1

\usepackage{jheppub}

\usepackage{amsmath}
\usepackage{bbm}
\usepackage{float}
\usepackage{graphicx}	
\usepackage{hyperref}
\usepackage{mathtools}
\usepackage{cancel}

\usepackage[normalem]{ulem}

\usepackage{multirow}
\usepackage{rotating}
\usepackage{subcaption}

\def\lsim{\raise0.3ex\hbox{$\;<$\kern-0.75em\raise-1.1ex
\hbox{$\sim\;$}}}
\def\gsim{\raise0.3ex\hbox{$\;>$\kern-0.75em\raise-1.1ex
\hbox{$\sim\;$}}}

\def\thetitle{ 
Neutrino amplitude decomposition in matter \\
 \vspace{- 8mm}
}
\title{\thetitle}
\hypersetup{pdftitle={\thetitle}}
\author{Hisakazu Minakata}
\affiliation{
Center for Neutrino Physics, Department of Physics, Virginia Tech, Blacksburg, Virginia 24061, USA \\
}

\emailAdd{minakata71@vt.edu}
\date{\today}

\abstract{ 
Observation of the interference between the atmospheric-scale and solar-scale oscillations is one of the challenging and tantalizing goals of the ongoing and upcoming neutrino experiments. An inevitable first step required for such analyses is to establish the way of how the oscillation $S$ matrix can be decomposed into the atmospheric and solar waves, the procedure dubbed as the {\em amplitude decomposition}. In this paper, with use of the perturbative framework proposed by Denton {\it et al.}~(DMP), we establish the prescription for amplitude decomposition which covers the whole kinematical region of the terrestrial neutrino experiments. We analyze the limits to the atmospheric- and solar-resonance regions to argue that the dynamical two modes of the DMP decomposition can be interpreted as the matter-dressed atmospheric and solar oscillations. The expressions of the oscillation probability, which are decomposed into the non-interference and interference terms, are derived for all the relevant flavor oscillation channels. Through construction of the DMP decomposition, we reveal the nature of $\psi$ ($\theta_{12}$ in matter) symmetry as due to the $S$ matrix rephasing invariance. A new picture of the DMP perturbation theory emerged, a unified perturbative framework for neutrino oscillation in earth matter. 

}

\begin{document} % JHEP 

\maketitle

\section{Introduction} 
\label{sec:introduction}

The three-generation structure of the fundamental fermions with complex mass matrices, the masses and the flavor mixing \cite{GellMann:1960np,Cabibbo:1963yz,Maki:1962mu}, has important consequences. When the neutrinos oscillate the two independent modes of oscillations are generated, the $\Delta m^2_{31}$-driven ``atmospheric'' \cite{Fukuda:1998mi} and the $\Delta m^2_{21}$-driven ``solar'' \cite{Eguchi:2002dm} oscillations, whose latter also takes the form of matter-affected \cite{Wolfenstein:1977ue,Mikheev:1986gs} adiabatic flavor transformation \cite{Ahmad:2002jz}. The atmospheric and solar waves interfere with each other as a consequence of the three-generation structure, the viewpoint emphasized in our previous papers~\cite{Huber:2019frh,Minakata:2020ijz}. 
A less obvious but another, probably the most important, consequence is that with $N$-generation fermions, lack of sufficient degrees of freedom of fields that can absorb CP violating phase, starts to manifest at $N=3$, the Kobayashi-Maskawa (KM) mechanism \cite{Kobayashi:1973fv} for CP violation \cite{Christenson:1964fg}. In the quark sector the KM mechanism is beautifully demonstrated experimentally~\cite{Aubert:2001nu,Abe:2001xe}. In the lepton sector CP violation is under active search \cite{Abe:2019vii,Acero:2019ksn,Jiang:2019xwn}. 

It is a tantalizing possibility to experimentally observe the quantum interference between the atmospheric- and the solar-scale oscillation waves. In fact, physics of the interference of these two modes of oscillation has been discussed in various contexts in neutrino physics~\cite{Smirnov:2006sm,Akhmedov:2008qt,Nunokawa:2007qh,Petcov:2001sy,Choubey:2003qx,Learned:2006wy,Klop:2014ima}. Here, we must note that shortage of the list may reflect our ignorance. Nonetheless, if we ask the question of precisely how much is due to the interference effect in a given experimental data, to the best of our knowledge, we do not have the machinery to answer the question. 

In previous papers~\cite{Huber:2019frh,Minakata:2020ijz}, we have started a new approach to the problem of quantifying the interference effect between the atmospheric and the solar waves. We first note that such machinery aiming at detecting the interference has to have the ability to define the atmospheric and the solar oscillation amplitudes. Only after establishing these definitions we can talk about what is the interference between them. Therefore, we need a prescription of how the oscillation $S$ matrix can be decomposed into the atmospheric and the solar oscillation amplitudes. This procedure was named as the {\em ``amplitude decomposition''} in the previous papers, whose establishment in matter is the ultimate goal of this paper. 

It may be worthwhile to pay attention to an interesting contrast between the above two consequences of the three-family structure. Let us focus on the lepton sector assuming the presence of the lepton KM phase $\delta$. A general expectation is that CP violation effect is small, and oscillation-channel dependent. For example, it is likely that CP phase effect is more prominent in the $\nu_{\mu} \rightarrow \nu_{e}$ channel than the $\nu_{\mu} \rightarrow \nu_{\mu}$. In an extreme case, both the CP conserving $\cos \delta$ and violating $\sin \delta$ effects are absent in the $\nu_{e}$ and $\bar{\nu}_{e}$ disappearance channels in vacuum and in matter~\cite{Kuo:1987km,Minakata:1999ze}. On the other hand, the atmospheric and solar wave interference exists universally, and its magnitude is usually not small. Roughly speaking, the interference term in the probability is comparable to the non-interference term~\cite{Huber:2019frh,Minakata:2020ijz}. 

\section{The amplitude decomposition: A brief overview and the paper plan} 
\label{sec:overview}

Let us introduce the problem of amplitude decomposition. We briefly review its current status, and describe a design plan of this paper. We try to make our discussion here very pedagogical.

\subsection{The amplitude decomposition in vacuum}

What is good in vacuum is that we can clearly define what are the atmospheric ($\Delta m^2_{31}$-driven) and the solar ($\Delta m^2_{21}$-driven) amplitudes~\cite{Huber:2019frh}. Given the neutrino oscillation $S$ matrix element 
\begin{eqnarray} 
S_{\alpha \beta} = 
U_{\alpha 1} U^{*}_{\beta 1} e^{ - i \frac{ m^2_{1} }{ 2E } x }
+ U_{\alpha 2} U^{*}_{\beta 2} e^{ - i \frac{ m^2_{2} }{ 2E } x }
+ U_{\alpha 3} U^{*}_{\beta 3} e^{ - i \frac{ m^2_{3} }{ 2E } x }, 
\label{S-matrix-def-original}
\end{eqnarray}
which describes the neutrino oscillation $\nu_\beta \rightarrow \nu_\alpha$ ($\alpha \neq \beta$, or $\alpha = \beta$) from $x=0$ to $x$ in vacuum, $S_{\alpha \beta}$ can be rewritten, after a phase redefinition, as 
\begin{eqnarray} 
S_{\alpha \beta} = 
\delta_{\alpha \beta} 
+ U_{\alpha 2} U^{*}_{\beta 2} 
\left( e^{ - i \Delta_{21} x} - 1 \right) 
+ U_{\alpha 3} U^{*}_{\beta 3} 
\left( e^{ - i \Delta_{31} x} - 1 \right) 
\label{S-matrix-decompose}
\end{eqnarray}
by using unitarity \cite{Nunokawa:2007qh,Bilenky:2012zp,Huber:2019frh}. 
In eqs.~\eqref{S-matrix-def-original} and \eqref{S-matrix-decompose}, $U \equiv U_{\text{\tiny MNS}}$ denotes the lepton flavor mixing matrix~\cite{Maki:1962mu}, $\Delta_{ji} \equiv \frac{ m^2_{j} -  m^2_{i} }{2E}$ and $\delta_{\alpha \beta}$ denotes the Kronecker delta function. Equation~\eqref{S-matrix-decompose} naturally defines the atmospheric and the solar amplitudes 
\begin{eqnarray} 
&& 
S_{\alpha \beta}^{ \text{atm} } 
\equiv  
U_{\alpha 3} U^{*}_{\beta 3} \left( e^{ - i \Delta_{31} x} - 1 \right), 
\nonumber \\
&& 
S_{\alpha \beta}^{ \text{sol} } \equiv  
U_{\alpha 2} U^{*}_{\beta 2}  \left( e^{ - i \Delta_{21} x} -1 \right).
\label{atm-sol-amplitude-vac}
\end{eqnarray}
Each one of $\left( e^{ - i \Delta_{31} x} - 1 \right)$ and $\left( e^{ - i \Delta_{21} x} - 1 \right)$ describe the single $\Delta m^2$ wave with desirable properties of vanishing at $x=0$ and having the conventional oscillation phase dependence. Hence, we call hereafter $\left( e^{ - i \Delta_{31} x} - 1 \right)$ and $\left( e^{ - i \Delta_{21} x} - 1 \right)$, and their extension in matter, as the atmospheric and the solar wave factors, respectively. Thus, the amplitude decomposition in vacuum is nothing but the wave factor decomposition, which is naturally implemented by unitarity. 

\subsection{The amplitude decomposition in matter: The state of the art}
\label{sec:problem-solution}

In our trial of extending the amplitude decomposition into a matter environment in ref.~\cite{Minakata:2020ijz}, we have set up the problem, analyzed it, and tried to solve it by examination of the several perturbative frameworks. In this attempt, we have encountered mainly the following two problems: 

\begin{itemize}
\item 
A simple extension of the vacuum amplitude decomposition~\cite{Huber:2019frh} fails even with an infinitesimal matter potential.\footnote{
%%%%%%%%%%%%%% footnote %%%%%%%%%%%%%%
There is a case in which the vacuum prescription works in matter, the first-order AKS perturbation theory~\cite{Arafune:1997hd}, but it appears to be the unique exception. See ref.~\cite{Minakata:2020ijz}. }

\item 
In matter the similar form of $S$ matrix and the wave factor decomposition as in eqs.~\eqref{S-matrix-def-original} and~\eqref{S-matrix-decompose}, respectively, are known to exist, the Zaglauer-Schwarzer (ZS) construction~\cite{Zaglauer:1988gz}, see below. But, we fail in physical interpretation of the ZS decomposition. 

\end{itemize}
\noindent
Having been faced with these problems, we took a detour in ref.~\cite{Minakata:2020ijz}. Namely, we identified a few kinematical regions in which nature of the matter-effect modified atmospheric and solar waves are understood by suitable appropriate perturbative frameworks that have been developed e.g., in refs.~\cite{Arafune:1997hd,Martinez-Soler:2019nhb,Arafune:1996bt,Cervera:2000kp,Freund:2001pn,Akhmedov:2004ny,Minakata:2015gra}. We then utilized the perturbatively expanded oscillation $S$ matrix to develop the amplitude decomposition in matter under a guidance of the wave factor decomposition. 

The present paper has an overlap in nature with the previous paper~\cite{Minakata:2020ijz}, both of which are devoted to the same subject, the amplitude decomposition in matter. Yet, there is a sharp contrast between them in approaching the problem. In this paper, we squarely tackle the problem starting from the principle of amplitude decomposition in matter.  

To be more specific, the present paper has the following well-defined purposes:

\begin{itemize}
\item
To establish the prescription for amplitude decomposition that covers the whole kinematical region of the terrestrial neutrino experiments. 

\item
To derive the oscillation probability formulas which are decomposed into the non-interference and interference terms in all the relevant flavor oscillation channels with a sufficient accuracy amenable for experimental analyses.

\end{itemize}
\noindent
The region referred in the first item above implies the one of energy-baseline that covers the atmospheric neutrino observation of $E > 100$ MeV by e.g., Super-Kamiokande, which of course includes all the long-baseline neutrino experiments. For this purpose we use the 
perturbative framework proposed by Denton {\it et al.}~\cite{Denton:2016wmg}. 
For brevity, we call this prescription for amplitude decomposition as the ``DMP decomposition''~\cite{Minakata:2020ijz}. 
All these will be explained in due course, and we will make this paper self-contained as much as possible. We start by addressing the principle, which eventually reveals our path to the DMP decomposition. 

In this paper, we re-address the question of what is the correct principle by which the amplitude decomposition can be formulated in a generic matter environment. In section~\ref{sec:principle}, following the discussion of principle of decomposition, we present our solution, the DMP decomposition for the amplitude decomposition in generic matter environment. After introductory description of the DMP perturbation theory in section~\ref{sec:DMP-formulation}, it will be fully developed in section~\ref{sec:DMP-amplitude-decomposition}. The physical interpretation of the independent two dynamical modes is attempted in section~\ref{sec:interpretation} by analyzing the limit toward the atmospheric- and the solar-resonance perturbation theories. Finally in section~\ref{sec:decomposition-matter-P}, we analyze the near vacuum limit of the DMP decomposition to understand how the problem with infinitesimal matter potential is dealt with. Throughout this paper we try to develop a practical framework of amplitude decomposition which can be used in the data analyses.

\section{Principle of amplitude decomposition in matter} 
\label{sec:principle} 

What should be done first to construct the suitable amplitude decomposition scheme in matter is to identify the two independent modes of the three-flavor neutrino oscillation in generic matter environments. Let us call these two dynamical modes of oscillation as the ``A'' and ``S'' modes, the terminologies introduced in ref.~\cite{Minakata:2020ijz}. Let us ask: How can we identify the two dynamical ``A'' and ``S'' modes? The answer to this question is: The Hamiltonian of the system must know it. Namely, the ``A'' and ``S'' modes must show up as the result of diagonalization of the Hamiltonian. 

\subsection{ZS decomposition}
\label{sec:ZS-decomposition}

In fact, the answer to this question is known since long time ago. If the exact $S$ matrix is known the eigenvalues of the Hamiltonian are known, and vice versa, $S = e^{- i H x}$ in the uniform density matter. But, the treatment can be generalized into the varying density matter. For the three-flavor neutrino oscillation in uniform-density matter, the exact $S$ matrix is known as the Zaglauer-Schwarzer solution \cite{Zaglauer:1988gz}. 
\begin{eqnarray} 
S_{\alpha \beta} = 
V_{\alpha 1} V^{*}_{\beta 1} e^{ - i \frac{ \lambda_{1} }{2E} x} 
+ V_{\alpha 2} V^{*}_{\beta 2} e^{ - i \frac{ \lambda_{2} }{2E} x}
+ V_{\alpha 3} V^{*}_{\beta 3} e^{ - i \frac{ \lambda_{3} }{2E} x}. 
\label{S-matrix-ZS} 
\end{eqnarray}
The exact expressions of the eigenvalues $\lambda_{i}$ \cite{Barger:1980tf}, which are the eigenvalues of $2E H$, and the evolution matrix $V_{\alpha i}$ ($i=1,2,3$) are presented in ref.~\cite{Zaglauer:1988gz}. See ref.~\cite{Kimura:2002wd} for the related work which allows us to construct the exact form of the oscillation probability. Due to the sum rule $\lambda_{1} + \lambda_{2} + \lambda_{3} = m^2_{1} + m^2_{2} + m^2_{3} + a$, the two out of the three eigenvalues are independent, reassuring the existence of (only) two independent oscillation modes.

With use of the rephasing $S_{\alpha \beta} \rightarrow e^{ i \frac{ \lambda_{1} }{2E} x} S_{\alpha \beta}$, one can define the amplitudes $S_{\alpha \beta}^{A}$ and $S_{\alpha \beta}^{S}$ as 
\begin{eqnarray} 
&& 
S_{\alpha \beta}^{A} 
\equiv  
V_{\alpha 3} V^{*}_{\beta 3} 
\left[ e^{ - i \frac{ ( \lambda_{3} - \lambda_{1} ) }{2E} x} - 1 \right], 
\nonumber \\
&& 
S_{\alpha \beta}^{S} \equiv  
V_{\alpha 2} V^{*}_{\beta 2} 
\left[ e^{ - i \frac{ ( \lambda_{2} - \lambda_{1} ) }{2E} x} - 1 \right], 
\label{ZS-decompose}
\end{eqnarray}
by which the $S$ matrix can be written, after the phase redefinition, as 
\begin{eqnarray} 
&& 
S_{\alpha \beta} 
= \delta_{\alpha \beta} + S_{\alpha \beta}^{A} + S_{\alpha \beta}^{S}. 
\end{eqnarray}
This procedure which leads to an amplitude decomposition in matter was called as the ZS decomposition in ref.~\cite{Minakata:2020ijz}. Thus, we know the general solution of the amplitude decomposition in matter under the uniform matter density approximation. 

\subsection{DMP decomposition} 
\label{sec:DMP-decomposition}

Then, one may ask: What is missing? The answer is lack of physical interpretation of the ZS decomposition. Or, in other word, it is very hard, if not impossible, to extract a clear physical picture out of the exact expressions of the eigenvalues and the $V$ matrix given in ref.~\cite{Zaglauer:1988gz}. Since we want to understand how the dynamical two modes in the general solution are connected to the physically motivated two modes, the matter-dressed ``atmospheric'' and the matter-dressed ``solar'' oscillations, we need a better way even though it is only an approximate solution. We argue in the rest of this paper that the framework proposed by Denton {\it et al.}~\cite{Denton:2016wmg} provides, to our knowledge, the best solution for this purpose. 

Despite a partial overlap between this and the previous papers~\cite{Minakata:2020ijz}, there is a clear difference between them about the setting of the problem and approach to it. In ref.~\cite{Minakata:2020ijz}, we have looked for the prescription for amplitude decomposition under the condition that the dynamical two modes allow interpretation of the matter-affected atmospheric and the solar oscillations. That is why we had to restrict our usage of the ZS construction to the guideline for the wave factor structure. 

Whereas in this paper, we start from the general principle of amplitude decomposition, and state that the ZS decomposition provides the exact solution in generic matter environment with a uniform density. But, because its physical interpretation is untenable, we replace the ZS decomposition by the DMP decomposition. As far as the numerical accuracy is concerned they are indistinguishable by eye~\cite{Parke:2019vbs}. More importantly, we illuminate by the explicit calculation how the above physical two modes are buried into the DMP amplitude decomposition. While the DMP solution is introduced by using the $V$ matrix elements~\cite{Minakata:2020ijz}, neither the explicit expressions of the decomposed amplitudes nor the decomposed probabilities are presented. In this paper we provide them in all the relevant oscillation channels. 

\section{The DMP perturbation theory in a nutshell} 
\label{sec:DMP-formulation}

The DMP perturbation theory~\cite{Denton:2016wmg} is the easiest possible framework to compute perturbative corrections, though this feature may not be appreciated so widely. It is not merely at a conceptual level, but is more true at the technical level, which is illuminated in this section. 

\subsection{The three-flavor neutrino evolution in matter} 
\label{sec:3nu-evolution}

We define the system of three-flavor neutrino evolution in matter. Though standard, we do it to define notations. 
The evolution of the three-flavor neutrinos in matter can be described by the Schr\"odinger equation in the flavor basis, $i \frac{d}{dx} \nu = H \nu$, with Hamiltonian 
\begin{eqnarray}
H= 
\frac{ 1 }{ 2E } \left\{ 
U \left[
\begin{array}{ccc}
0 & 0 & 0 \\
0 & \Delta m^2_{21}& 0 \\
0 & 0 & \Delta m^2_{31} 
\end{array}
\right] U^{\dagger}
+
\left[
\begin{array}{ccc}
a(x) & 0 & 0 \\
0 & 0 & 0 \\
0 & 0 & 0
\end{array}
\right] 
\right\}, 
\label{flavor-basis-hamiltonian}
\end{eqnarray}
where $E$ is neutrino energy and $\Delta m^2_{ji} \equiv m^2_{j} - m^2_{i}$. 
In (\ref{flavor-basis-hamiltonian}), $U \equiv U_{\text{\tiny MNS}}$ denotes the standard $3 \times 3$ lepton flavor mixing matrix~\cite{Maki:1962mu} which relates the flavor neutrino states to the vacuum mass eigenstates as $\nu_{\alpha} = U_{\alpha i} \nu_{i}$, where $\alpha$ runs over $e, \mu, \tau$, and the mass eigenstate indices $i, j$ run over $1,2,$ and $3$. 
We use the lepton flavor mixing matrix in the ``ATM'' convention in which $e^{ \pm i \delta}$ is attached to the ``atmospheric angle'' $s_{23}$: 
\begin{eqnarray}
U_{\text{\tiny ATM}} &=&
\left[
\begin{array}{ccc}
1 & 0 &  0  \\
0 & c_{23} & s_{23} e^{ i \delta} \\
0 & - s_{23} e^{- i \delta} & c_{23} \\
\end{array}
\right] 
\left[
\begin{array}{ccc}
c_{13}  & 0 &  s_{13} \\
0 & 1 & 0 \\
- s_{13} & 0 & c_{13}  \\
\end{array}
\right] 
\left[
\begin{array}{ccc}
c_{12} & s_{12}  &  0  \\
- s_{12} & c_{12} & 0 \\
0 & 0 & 1 \\
\end{array}
\right] 
\nonumber \\
&\equiv &
U_{23} (\theta_{23}, \delta) U_{13} (\theta_{13}) U_{12} (\theta_{12}). 
\label{MNS-ATM}
\end{eqnarray}
The choice \eqref{MNS-ATM} is most convenient given the fact that it is used in both refs~\cite{Denton:2016wmg} and \cite{Minakata:2015gra}, and is physically equivalent with the more familiar PDG convention \cite{Zyla:2020zbs}.\footnote{
%%%%%%%%%%%%%% footnote %%%%%%%%%%%%%%%
Among the three typical conventions of the $U_{\text{\tiny MNS}}$ matrix~\cite{Martinez-Soler:2018lcy}, the ATM convention is, in fact, the most rational choice from the theoretical point of view. It is well known that $\theta_{23}$ is irrelevant for dynamical evolution in matter, as $U_{23}$ is rotated away from the evolution equation in the ``propagation basis'', see e.g., refs.~\cite{Minakata:1999ze,Blennow:2013rca}. Thus, the CP phase $\delta$ does not complicate the neutrino evolution in the ATM convention. }

The functions $a(x)$ in (\ref{flavor-basis-hamiltonian}) denote the Wolfenstein matter potential \cite{Wolfenstein:1977ue} due to charged current reactions 
\begin{eqnarray} 
a &=&  
2 \sqrt{2} G_F N_e E \approx 1.52 \times 10^{-4} \left( \frac{Y_e \rho}{\rm g\,cm^{-3}} \right) \left( \frac{E}{\rm GeV} \right) {\rm eV}^2.  
\label{matt-potential}
\end{eqnarray}
Here, $G_F$ is the Fermi constant, $N_e$ is the electron number density in matter. $\rho$ and $Y_e$ denote, respectively, the matter density and number of electron per nucleon in matter. For simplicity and clarity we will work with the uniform matter density approximation throughout this paper. But, it is in principle possible to extend our treatment to varying matter density case if adiabaticity holds. 

\subsection{The DMP framework in brief} 
\label{sec:DMP-in-brief}

The DMP perturbation theory is a very natural framework as an approximate treatment of the three neutrino flavor transformation in matter. It has been well known that the matter effect significantly modifies the mixing angles $\theta_{12}$ and $\theta_{13}$, but not $\theta_{23}$ and CP phase $\delta$ \cite{Zaglauer:1988gz,Blennow:2013rca}. This feature is nicely incorporated into the DMP framework which utilizes the successive 1-3 and 1-2 space rotations with the mixing angles $\phi$ (matter-affected $\theta_{13}$) and $\psi$ (matter-affected $\theta_{12}$) to approximately diagonalize the Hamiltonian~\cite{Denton:2016wmg}.\footnote{
%%%%%%%%%%%%%% footnote %%%%%%%%%%%%%%%%%
This method for approximate diagonalization of the Hamiltonian has been known as the Jacobi method, and was first applied to the three-neutrino oscillation by the authors of ref.~\cite{Agarwalla:2013tza} who performed the two rotations in different order. 
}

In formulating perturbation theory one has to specify the basis in which one computes $S$ matrix elements perturbatively. Starting from the flavor eigenstate basis with the Hamiltonian $H$, we transform to the mass eigenstate basis in matter, which we call the ``check basis'' with the Hamiltonian $\check{H}$, 
\begin{eqnarray} 
&&
\check{H} 
=U^{\dagger}_{12} (\psi) U^{\dagger}_{13} (\phi) U^{\dagger}_{23} (\theta_{23}, \delta) H U_{23} (\theta_{23}, \delta) U_{13} (\phi) U_{12} (\psi).
\label{check-Hamiltonian}
\end{eqnarray}
The mixing angles in matter, $\phi$ and $\psi$, are determined by diagonalizing the Hamiltonians in each step \cite{Denton:2016wmg}, and their expressions are given in appendix~\ref{sec:DMP-variables}. 

The DMP perturbation theory utilizes the expansion parameter $\epsilon$ defined by 
\begin{eqnarray} 
&&
\epsilon \equiv \frac{ \Delta m^2_{21} }{ \Delta m^2_{ \text{ren} } }, 
\hspace{10mm}
\Delta m^2_{ \text{ren} } \equiv \Delta m^2_{31} - s^2_{12} \Delta m^2_{21},
\label{epsilon-Dm2-ren-def}
\end{eqnarray}
where $\Delta m^2_{ \text{ren} }$ is the ``renormalized'' atmospheric $\Delta m^2$ defined in ref.~\cite{Minakata:2015gra}. It should be remembered that throughout this paper $\Delta m^2_{ \text{ren} }$, $\epsilon$, $\Delta m^2_{31}$ etc. are the mass-ordering sign active. That is, $\Delta m^2_{31}$ (and $\Delta m^2_{ \text{ren} }$) is positive and negative for the normal mass ordering (NMO) and the inverted mass ordering (IMO), respectively. Note that $\epsilon <0$ for the IMO. 

The check basis Hamiltonian can be decomposed into the unperturbed and perturbed parts~\cite{Denton:2016wmg}, 
\begin{eqnarray} 
&&\check{H} = \check{H}_{0} + \check{H}_{1}, 
\nonumber 
\end{eqnarray}
\begin{eqnarray} 
&&
\check{H}_{0} = 
\left[
\begin{array}{ccc}
h_{1} & 0 & 0 \\
0 & h_{2} & 0 \\
0 & 0 & h_{3} \\
\end{array}
\right], 
\hspace{10mm}
\check{H}_{1} =
\epsilon c_{12} s_{12} s_{ (\phi - \theta_{13}) } \Delta_{ \text{ren} } 
\left[
\begin{array}{ccc}
0 & 0 & - s_{\psi} \\
0 & 0 & c_{\psi} \\
- s_{\psi} & c_{\psi} & 0 \\
\end{array}
\right]. 
\label{check-Hamiltonian-0th-1st}
\end{eqnarray}
In eq~\eqref{check-Hamiltonian-0th-1st}, $h_{i} \equiv \lambda_{i} / 2E$ ($i=1,2,3$) denote the eigenvalues of the unperturbed Hamiltonian. Hereafter we use the abbreviated notations $c_{12} \equiv \cos \theta_{12}$, $s_{\psi} \equiv  \sin \psi$, $c_{\phi} \equiv  \cos \phi$, $s_{(\phi - \theta_{13})} \equiv \sin (\phi - \theta_{13})$, etc. and $\Delta_{ \text{ren} } \equiv \Delta m^2_{ \text{ren} } / 2E$. 

The calculation of the $\check{S}$ matrix can be done routinely. Given the Schr\"odinger equation $i \frac{d}{dx} \check{S} = \check{H} (x) \check{S}$, we define 
\begin{eqnarray} 
\Omega(x) = e^{i \check{H}_{0} x} \check{S} (x).
\label{def-omega}
\end{eqnarray}
which obeys the evolution equation 
\begin{eqnarray} 
i \frac{d}{dx} \Omega(x) = H_{1} \Omega(x) 
\label{omega-evolution}
\end{eqnarray}
where
\begin{eqnarray} 
H_{1} \equiv e^{i \check{H}_{0} x} \check{H}_{1} e^{-i \check{H}_{0} x} .
\label{def-H1}
\end{eqnarray}
Then, $\Omega(x)$ can be computed perturbatively as
\begin{eqnarray} 
\Omega(x) &=& 1 + 
(-i) \int^{x}_{0} dx' H_{1} (x') + 
(-i)^2 \int^{x}_{0} dx' H_{1} (x') \int^{x'}_{0} dx'' H_{1} (x'') 
+ \cdot \cdot \cdot,
\label{Omega-expansion}
\end{eqnarray}
and the $\check{S}$ matrix is given by
\begin{eqnarray} 
\check{S} (x) =  
e^{-i \check{H}_{0} x} \Omega(x). 
\label{check-Smatrix}
\end{eqnarray}

Having obtained the check basis $\check{S}$ matrix it is straightforward to calculate the flavor basis $S$ matrix:
\begin{eqnarray} 
S = U_{23} (\theta_{23}, \delta) U_{13} (\phi) U_{12} (\psi) \check{S} U^{\dagger}_{12} (\psi) U^{\dagger}_{13} (\phi) U^{\dagger}_{23} (\theta_{23}, \delta).  
\label{flavor-Smatrix}
\end{eqnarray}
The explicit expressions of the zeroth and first order $S$ matrix elements are given in appendix~\ref{sec:S-matrix-0th-1st}. Then, the rest of the work is to compute the oscillation probability, $P(\nu_{\beta} \rightarrow \nu_{\alpha}) = \vert S_{\alpha \beta} \vert^2$. 

The descriptions of the framework in this section, if assisted by appendix~\ref{sec:DMP-variables} in which the expressions of the eigenvalues $\lambda_{i}$ or $h_{i}$, and the mixing angles $\phi$ and $\psi$ are given, must be sufficient for the readers to derive the formulas which will be presented in this paper. 
To discuss the antineutrino channels we reverse the signs of the matter potential $a$ and the CP phase $\delta$. This remark, since it is so well known,  will not be repeated in the discussion of each channel. 

\subsection{Simplified notations}
\label{sec:notations}

In this paper we use the following simplified notations $(i, j = 1,2,3)$. For quantities in vacuum 
\begin{eqnarray} 
&& 
\Delta_{ji} \equiv \frac{ \Delta m^2_{ji} }{ 2E }, 
\hspace{10mm}
\Delta_{ \text{ren} } \equiv \frac{ \Delta m^2_{ \text{ren} } }{ 2E }, 
\hspace{10mm}
\label{Delta-ij-def}
\end{eqnarray}
where $\Delta m^2_{ \text{ren} }$ is defined in eq.~\eqref{epsilon-Dm2-ren-def}. For the variables in matter  we use 
\begin{eqnarray} 
h_{i} \equiv \frac{ \lambda_{i} }{ 2E } ~~(i=1,2,3), 
\hspace{10mm}
h_{\pm, 0} \equiv \frac{ \lambda_{\pm, 0} }{ 2E },
\hspace{10mm}
\Delta_{ a } \equiv \frac{ a }{ 2E },
\label{hi-def}
\end{eqnarray}
where $\lambda_{\pm}$, $\lambda_{0}$, $h_{\pm}$, and $h_{0}$ denote the eigenvalues to be used in section~\ref{sec:helio-P}. 

\section{DMP amplitude decomposition} 
\label{sec:DMP-amplitude-decomposition} 

In this section we construct from scratch the DMP amplitude decomposition. We do it here by using the $S$ matrix method, which is in accord with the method for amplitude decomposition employed in the main body of ref.~\cite{Minakata:2020ijz}. We hope that the formulas we derive in this paper are much easier to reproduce for the readers who are not familiar with the $V$ matrix method \cite{Minakata:1998bf}. In anticipation of the physical interpretation of the decomposed amplitudes in section~\ref{sec:interpretation}, we denote them as $S_{\alpha \beta}^{ \text{atm} }$ and $S_{\alpha \beta}^{ \text{sol} }$. Our construction is valid in both the normal and inverted mass orderings, which we abbreviate as the NMO and IMO, respectively, since section~\ref{sec:DMP-in-brief}. In section~\ref{sec:interpretation-rephasing} we discuss a possibility that they are correlated with the choices of the $S$ matrix rephasing.  

We first present general structure of amplitude decomposition, in particular, the dual definitions of it. To focus on the conceptual questions, and for clear-cut exposition of the points, we concentrate on the $\nu_{\mu} - \nu_{e}$ channel in the main text of this paper. The amplitude decomposition in the other channels will be discussed in appendices~\ref{sec:amp-decompose-disapp} and \ref{sec:amp-decompose-mu-tau}.

\subsection{Two different definitions of the amplitude decomposition} 

Starting from the generic form of the $S$ matrix in matter, 
$S_{\alpha \beta} = V_{\alpha 1} V^{*}_{\beta 1} e^{ - i h_{1} x} + V_{\alpha 2} V^{*}_{\beta 2} e^{ - i h_{2} x} + V_{\alpha 3} V^{*}_{\beta 3} e^{ - i h_{3} x}$ (see \eqref{S-matrix-ZS}), there exist two ways of defining the amplitude decomposition. With use of the rephasing $S_{\alpha \beta} \rightarrow e^{ i h_{1} x} S_{\alpha \beta}$, one can define the amplitude decomposition $S_{\alpha \beta} = \delta_{\alpha \beta} + S_{\alpha \beta}^{ \text{atm} } + S_{\alpha \beta}^{ \text{sol} }$, where 
\begin{eqnarray} 
&& 
S_{\alpha \beta}^{ \text{atm} } 
\equiv  
V_{\alpha 3} V^{*}_{\beta 3} 
\left[ e^{ - i ( h_{3} - h_{1} ) x } - 1 \right], 
\nonumber \\
&& 
S_{\alpha \beta}^{ \text{sol} } \equiv  
V_{\alpha 2} V^{*}_{\beta 2} 
\left[ e^{ - i ( h_{2} - h_{1} ) x } - 1 \right]. 
\label{DMP-decompose-h1-rephase}
\end{eqnarray} 
If we use the different rephasing, $S_{\alpha \beta} \rightarrow e^{ i h_{2} x} S_{\alpha \beta}$, the decomposed amplitudes read\footnote{
%%%%%%%%%%%%% footnote %%%%%%%%%%%%%%
One might feel curious why the wave factor $[ e^{ i ( h_{2} - h_{1} ) x } - 1 ]$ appears with the $e^{ i h_{2} x}$ rephasing, in contrast to the $[ e^{ - i ( h_{2} - h_{1} ) x } - 1 ]$ factor in the case of $e^{ i h_{1} x}$ rephasing. It will be cleared up in section~\ref{sec:psi-symmetry}. 
}
\begin{eqnarray} 
&& 
S_{\alpha \beta}^{ \text{atm} } 
\equiv  
V_{\alpha 3} V^{*}_{\beta 3} 
\left[ e^{ - i ( h_{3} - h_{2} ) x } - 1 \right], 
\nonumber \\
&& 
S_{\alpha \beta}^{ \text{sol} } \equiv  
V_{\alpha 1} V^{*}_{\beta 1} 
\left[ e^{ i ( h_{2} - h_{1} ) x } - 1 \right].
\label{DMP-decompose-h2-rephase}
\end{eqnarray}
By being different only in the overall phase, of course, these two decomposed amplitudes lead to the same probability. 
For a simpler nomenclature for the two rephasing methods, $S_{\alpha \beta} \rightarrow e^{ i h_{1} x} S_{\alpha \beta}$ and $S_{\alpha \beta} \rightarrow e^{ i h_{2} x} S_{\alpha \beta}$, we denote them as the ``$e^{ i h_{1} x}$ rephasing'' and the ``$e^{ i h_{2} x}$ rephasing'', respectively. 
Later in section~\ref{sec:interpretation-rephasing}
we will present our physical interpretation of the two different decompositions in eqs.~\eqref{DMP-decompose-h1-rephase} and \eqref{DMP-decompose-h2-rephase}.

By using the decomposed amplitudes $S_{\alpha \beta} = \delta_{\alpha \beta} + S_{\alpha \beta}^{\text{atm}} + S_{\alpha \beta}^{\text{sol}}$, in either the $e^{ i h_{1} x}$ or $e^{ i h_{2} x}$ rephasing, the oscillation probability is also decomposed into the non-interference and interference terms as \cite{Huber:2019frh,Minakata:2020ijz} 
\begin{eqnarray} 
&& 
P(\nu_{\beta} \rightarrow \nu_{\alpha}) = 
P(\nu_{\beta} \rightarrow \nu_{\alpha})^{ \text {non-int-fer} } 
+ P(\nu_{\beta} \rightarrow \nu_{\alpha})^{ \text {int-fer} }.
\label{decomposed-probability}
\end{eqnarray}

\subsection{DMP $S$ matrix elements in the $\nu_{\mu} - \nu_{e}$ channel} 

The zeroth and the first order flavor basis amplitudes can be calculated by using the relevant formulas in appendix~\ref{sec:S-matrix-0th-1st}: 
\begin{eqnarray} 
&&
S_{e \mu}^{(0)} = 
c_{23} c_{\phi} c_{\psi} s_{\psi} 
\left( e^{ - i h_{2} x } - e^{ - i h_{1} x } \right) 
- s_{23} c_{\phi} s_{\phi} e^{ - i \delta} 
\left( c^2_{\psi} e^{ - i h_{1} x } + s^2_{\psi} e^{ - i h_{2} x } - e^{ - i h_{3} x } \right),
\nonumber \\ 
&& 
S_{e \mu}^{(1)} 
= s_{23} \cos 2\phi e^{ - i \delta} \hat{S}_{13}  
+ c_{23} s_{\phi} \hat{S}_{23} 
\nonumber \\
&=& 
\epsilon c_{12} s_{12} s_{ (\phi - \theta_{13}) } 
\biggl[
s_{23} \cos 2\phi c_{\psi} s_{\psi} e^{ - i \delta} 
\left\{ \frac{ \Delta_{ \text{ren} } }{ h_{3} - h_{2} } 
\left( e^{ - i h_{3} x } - e^{ - i h_{2} x } \right) 
- \frac{ \Delta_{ \text{ren} } }{ h_{3} - h_{1} } 
\left( e^{ - i h_{3} x } - e^{ - i h_{1} x } \right) 
\right\} 
\nonumber \\
&+& 
c_{23} s_{\phi} 
\left\{ c^2_{\psi} \frac{ \Delta_{ \text{ren} } }{ h_{3} - h_{2} } 
\left( e^{ - i h_{3} x } - e^{ - i h_{2} x } \right) 
+ s^2_{\psi} \frac{ \Delta_{ \text{ren} } }{ h_{3} - h_{1} } 
\left( e^{ - i h_{3} x } - e^{ - i h_{1} x } \right) \right\} 
\biggr].
\label{S-emu-0th-1st}
\end{eqnarray}
In eq.~\eqref{S-emu-0th-1st}, $\phi$ and $\psi$ denote, respectively, the matter-dressed mixing angles $\theta_{13}$ and $\theta_{12}$ \cite{Denton:2016wmg}, as mentioned in section~\ref{sec:DMP-in-brief}. 
We remark here that the superscripts $(0)$ and $(1)$ on the $S$ matrix elements and the probabilities imply the order of DMP perturbation throughout this paper.

We discuss the two rephasing methods, the $e^{ i h_{1} x}$ rephasing and $e^{ i h_{2} x}$ rephasing,  in parallel. For reasons we explain later we discuss the $e^{ i h_{2} x}$ rephasing first, and then, the case of $e^{ i h_{1} x}$ rephasing follows. 

\subsection{The decomposed amplitudes and probabilities in the $\nu_{\mu} - \nu_{e}$ channel: $e^{ i h_{2} x}$ rephasing}
\label{sec:decomposition-mue-h2}

With the $e^{ i h_{2} x}$ rephasing, the decomposed amplitudes in the zeroth- and first-order DMP expansion read 
\begin{eqnarray} 
&&
\left( S_{e \mu}^{\text{atm}} \right)^{(0)} 
= s_{23} c_{\phi} s_{\phi} e^{ - i \delta} \left( e^{ - i ( h_{3} - h_{2} ) x } - 1 \right), 
\nonumber \\ 
&& 
\left( S_{e \mu}^{\text{sol}} \right)^{(0)} 
= 
- c_{\phi} \left( c_{23} c_{\psi} s_{\psi} 
+ s_{23} s_{\phi} c^2_{\psi} e^{ - i \delta} \right) 
\left( e^{ i ( h_{2} - h_{1} ) x } - 1 \right).
\label{DMP-decompose-emu-0th-h2}
\end{eqnarray}
\begin{eqnarray} 
\left( S_{e \mu}^{ \text{atm} } \right)^{(1)} 
&=& 
\epsilon c_{12} s_{12} s_{ (\phi - \theta_{13}) } 
\left( c_{23} s_{\phi} c^2_{\psi} + s_{23} \cos 2\phi e^{ - i \delta} c_{\psi} s_{\psi} \right)
\frac{ \Delta_{ \text{ren} } }{ h_{3} - h_{2} } 
\left( e^{ - i ( h_{3} - h_{2} ) x } - 1 \right) 
\nonumber \\ 
&+& 
\epsilon c_{12} s_{12} s_{ (\phi - \theta_{13}) } 
\left( c_{23} s_{\phi} s^2_{\psi} - s_{23} \cos 2\phi e^{ - i \delta} c_{\psi} s_{\psi} \right) 
\frac{ \Delta_{ \text{ren} } }{ h_{3} - h_{1} } 
\left( e^{ - i ( h_{3} - h_{2} ) x } - 1 \right), 
\nonumber \\ 
\left( S_{e \mu}^{ \text{sol} } \right)^{(1)} 
&=& - \epsilon c_{12} s_{12} s_{ (\phi - \theta_{13}) } 
\left( c_{23} s_{\phi} s^2_{\psi} - s_{23} \cos 2\phi c_{\psi} s_{\psi} e^{ - i \delta} \right) 
\frac{ \Delta_{ \text{ren} } }{ h_{3} - h_{1} } 
\left( e^{ i ( h_{2} - h_{1} ) x } - 1 \right). 
\nonumber \\
\label{DMP-decompose-emu-1st-h2}
\end{eqnarray}

The decomposed oscillation probability, the non-interference and interference terms in eq.~\eqref{decomposed-probability} are given in the zeroth and first order as 
\begin{eqnarray} 
&& 
\left[ P(\nu_{\mu} \rightarrow \nu_{e})^{(0)} \right]^{ \text {non-int-fer} } 
= \biggl | \left( S_{e \mu}^{ \text{atm} } \right)^{(0)} \biggr |^2 
+ \biggl | \left( S_{e \mu}^{ \text{sol} } \right)^{(0)} \biggr |^2 
\nonumber \\ 
&=& 
4 s^2_{23} c^2_{\phi} s^2_{\phi}
\sin^2 \frac{ ( h_{3} - h_{2} ) x }{2} 
+ 4 c^2_{\phi} c^2_{\psi} 
\left( c^2_{23} s^2_{\psi} + s^2_{23} s^2_{\phi} c^2_{\psi} 
+ 2 c_{23} s_{23} s_{\phi} c_{\psi} s_{\psi} \cos \delta \right) 
\sin^2 \frac{ ( h_{2} - h_{1} ) x }{2}, 
\nonumber \\
&& 
\left[ P(\nu_{\mu} \rightarrow \nu_{e})^{(0)} \right]^{ \text {int-fer} } 
= 2 \mbox{Re} \left[ 
\left\{ \left( S_{e \mu}^{ \text{atm} } \right)^{(0)} \right\}^* 
\left( S_{e \mu}^{ \text{sol} } \right)^{(0)} \right] 
\nonumber \\ 
&=& 
- 4 s_{23} c^2_{\phi} s_{\phi} c_{\psi} 
\left( s_{23} s_{\phi} c_{\psi} + c_{23} s_{\psi} \cos \delta \right) 
%\nonumber \\ &\times& 
\left\{ - \sin^2 \frac{ ( h_{3} - h_{1} ) x }{2} 
+ \sin^2 \frac{ ( h_{3} - h_{2} ) x }{2} 
+ \sin^2 \frac{ ( h_{2} - h_{1} ) x }{2} \right\} 
\nonumber \\ 
&-& 
8 c_{23} s_{23} c^2_{\phi} s_{\phi} c_{\psi} s_{\psi} \sin \delta 
\sin \frac{ ( h_{3} - h_{1} ) x }{2} \sin \frac{ ( h_{2} - h_{1} ) x }{2} 
\sin \frac{ ( h_{3} - h_{2} ) x }{2}. 
\label{decomposed-P-mue-0th-h2}
\end{eqnarray}
\begin{eqnarray} 
&& 
\left[ P(\nu_{\mu} \rightarrow \nu_{e})^{(1)} \right]^{ \text {non-int-fer} } 
= 2 \mbox{Re} \left[ 
\left\{ \left( S_{e \mu}^{ \text{atm} } \right)^{(0)} \right\}^* 
\left( S_{e \mu}^{ \text{atm} } \right)^{(1)} \right] 
+ 
2 \mbox{Re} \left[ 
\left\{ \left( S_{e \mu}^{ \text{sol} } \right)^{(0)} \right\}^* 
\left( S_{e \mu}^{ \text{sol} } \right)^{(1)} \right] 
\nonumber \\ 
&=& 
8 \epsilon c_{12} s_{12} s_{ (\phi - \theta_{13}) } c_{\phi} s_{\phi} 
\left( c_{23} s_{23} s_{\phi} c_{\psi} \cos \delta + s^2_{23} \cos 2\phi s_{\psi} \right)
c_{\psi} \frac{ \Delta_{ \text{ren} } }{ h_{3} - h_{2} } 
\sin^2 \frac{ ( h_{3} - h_{2} ) x }{2} 
\nonumber \\ 
&+&
8 \epsilon c_{12} s_{12} s_{ (\phi - \theta_{13}) } c_{\phi} s_{\phi} 
\left( c_{23} s_{23} s_{\phi} s_{\psi} \cos \delta - s^2_{23} \cos 2\phi c_{\psi} 
\right) 
s_{\psi} \frac{ \Delta_{ \text{ren} } }{ h_{3} - h_{1} } 
\sin^2 \frac{ ( h_{3} - h_{2} ) x }{2} 
\nonumber \\ 
&+& 
8 \epsilon c_{12} s_{12} s_{ (\phi - \theta_{13}) } c_{\phi} c_{\psi} 
\left[ s_{\phi} \left( c^2_{23} s^2_{\psi} - s^2_{23} \cos 2\phi c^2_{\psi} \right) 
+ c_{23} s_{23} c_{\psi} s_{\psi} \cos \delta
\left( s^2_{\phi} - \cos 2\phi \right) \right]
\nonumber \\ 
&\times&
s_{\psi} \frac{ \Delta_{ \text{ren} } }{ h_{3} - h_{1} } 
\sin^2 \frac{ ( h_{2} - h_{1} ) x }{2}, 
\nonumber 
\end{eqnarray}
\begin{eqnarray} 
&& 
\left[ P(\nu_{\mu} \rightarrow \nu_{e})^{(1)} \right]^{ \text {int-fer} } 
= 2 \mbox{Re} \left[ 
\left\{ \left( S_{e \mu}^{ \text{atm} } \right)^{(0)} \right\}^* 
\left( S_{e \mu}^{ \text{sol} } \right)^{(1)} \right] 
+ 
2 \mbox{Re} \left[ 
\left\{ \left( S_{e \mu}^{ \text{sol} } \right)^{(0)} \right\}^* 
\left( S_{e \mu}^{ \text{atm} } \right)^{(1)} \right] 
\nonumber \\ 
&=& 
- 4 \epsilon c_{12} s_{12} s_{ (\phi - \theta_{13}) } c_{\phi} c_{\psi} 
\biggl\{ 
s_{\phi} c_{\psi} s_{\psi} 
\left( c^2_{23} + s^2_{23} \cos 2\phi \right) 
+ c_{23} s_{23} \cos \delta 
\left( \cos 2\phi s^2_{\psi} + s^2_{\phi} c^2_{\psi} \right) 
\biggr\}
\nonumber \\ 
&\times& 
c_{\psi} \frac{ \Delta_{ \text{ren} } }{ h_{3} - h_{2} } 
\left\{ - \sin^2 \frac{ ( h_{3} - h_{1} ) x }{2} + \sin^2 \frac{ ( h_{3} - h_{2} ) x }{2} + \sin^2 \frac{ ( h_{2} - h_{1} ) x }{2} \right\} 
\nonumber \\ 
&+& 
4 \epsilon c_{12} s_{12} s_{ (\phi - \theta_{13}) } c_{\phi} 
\biggl\{
s_{\phi} c_{\psi} 
\left[ s^2_{23} \cos 2\phi ( 1 + c^2_{\psi} ) - c^2_{23} s^2_{\psi} \right] 
- c_{23} s_{23} s_{\psi} \cos \delta 
\left[ s^2_{\phi} ( 1 + c^2_{\psi} ) - \cos 2\phi c^2_{\psi} \right] 
\biggr\}
\nonumber \\ 
&\times&
s_{\psi} \frac{ \Delta_{ \text{ren} } }{ h_{3} - h_{1} } 
\left\{ - \sin^2 \frac{ ( h_{3} - h_{1} ) x }{2} + \sin^2 \frac{ ( h_{3} - h_{2} ) x }{2} + \sin^2 \frac{ ( h_{2} - h_{1} ) x }{2} \right\} 
\nonumber \\ 
&+& 
8 \epsilon c_{23} s_{23} c_{\phi} s_{ (\phi - \theta_{13}) } c_{12} s_{12} \sin \delta 
\biggl\{
\left( s^2_{\phi} - c^2_{\phi} s^2_{\psi} \right) 
c^2_{\psi} \frac{ \Delta_{ \text{ren} } }{ h_{3} - h_{2} } 
- \left( s^2_{\phi} - c^2_{\phi} c^2_{\psi} \right) 
s^2_{\psi} \frac{ \Delta_{ \text{ren} } }{ h_{3} - h_{1} } 
\biggr\} 
\nonumber \\ 
&\times& 
\sin \frac{ ( h_{3} - h_{1} ) x }{2} \sin \frac{ ( h_{2} - h_{1} ) x }{2} 
\sin \frac{ ( h_{3} - h_{2} ) x }{2}.
\label{decomposed-P-mue-1st-h2}
\end{eqnarray}

\subsection{Amplitude decomposition in the $\nu_{\mu} - \nu_{e}$ channel with $e^{ i h_{1} x}$ rephasing}
\label{sec:decomposition-mue-h1}

We discuss next the amplitude decomposition with the $e^{ i h_{1} x}$ rephasing. The decomposed amplitudes in the zeroth- and first-order DMP expansion are similarly given by 
\begin{eqnarray} 
&&
\left( S_{e \mu}^{ \text{atm} } \right)^{(0)} 
= 
s_{23} c_{\phi} s_{\phi} e^{ - i \delta} 
\left( e^{ - i ( h_{3} - h_{1} ) x } - 1 \right), 
\nonumber \\
&& 
\left( S_{e \mu}^{ \text{sol} } \right)^{(0)} 
= 
c_{\phi} 
\left( c_{23} c_{\psi} s_{\psi} - s_{23} s_{\phi} s^2_{\psi} e^{ - i \delta} \right) 
\left( e^{ - i ( h_{2} - h_{1} ) x } - 1 \right), 
\label{DMP-decompose-emu-0th-h1}
\end{eqnarray}
\begin{eqnarray} 
\left( S_{e \mu}^{ \text{atm} } \right)^{(1)} 
&=&
\epsilon c_{12} s_{12} s_{ (\phi - \theta_{13}) }
\left( c_{23} s_{\phi} c^2_{\psi} + s_{23} \cos 2\phi c_{\psi} s_{\psi} e^{ - i \delta} \right) 
\frac{ \Delta_{ \text{ren} } }{ h_{3} - h_{2} } 
\left( e^{ - i ( h_{3} - h_{1} ) x } - 1 \right) 
\nonumber \\
&+& 
\epsilon c_{12} s_{12} s_{ (\phi - \theta_{13}) }
\left( c_{23} s_{\phi} s^2_{\psi} - s_{23} \cos 2\phi c_{\psi} s_{\psi} e^{ - i \delta} \right) 
\frac{ \Delta_{ \text{ren} } }{ h_{3} - h_{1} } 
\left( e^{ - i ( h_{3} - h_{1} ) x } - 1 \right), 
\nonumber \\
\left( S_{e \mu}^{ \text{sol} } \right)^{(1)} 
&=&
- \epsilon c_{12} s_{12} s_{ (\phi - \theta_{13}) } 
\left( c_{23} s_{\phi} c^2_{\psi} + s_{23} \cos 2\phi c_{\psi} s_{\psi} e^{ - i \delta} \right) 
\frac{ \Delta_{ \text{ren} } }{ h_{3} - h_{2} } 
\left( e^{ - i ( h_{2} - h_{1} ) x } - 1 \right).
\nonumber \\
\label{DMP-decompose-emu-1st-h1}
\end{eqnarray}
A comparison between the amplitudes in eqs.~\eqref{DMP-decompose-emu-0th-h1} and~\eqref{DMP-decompose-emu-1st-h1} and the ones in eqs.~\eqref{DMP-decompose-emu-0th-h2} and \eqref{DMP-decompose-emu-1st-h2} with the $e^{ i h_{2} x}$ rephasing tells us something new, which we need to discuss first. 

\subsection{$\psi$ symmetry}
\label{sec:psi-symmetry}

That is, the decomposed amplitudes obtained with the $e^{ i h_{1} x}$ rephasing and the ones with the $e^{ i h_{2} x}$ rephasing are connected with each other by the transformations~\cite{Denton:2016wmg}
\begin{eqnarray} 
&& 
h_{1} \rightarrow h_{2}, 
\hspace{10mm}
h_{2} \rightarrow h_{1}, 
\nonumber \\
&&
c_{\psi} \rightarrow - s_{\psi}, 
\hspace{6mm}
s_{\psi} \rightarrow + c_{\psi}, 
\hspace{6mm}
\cos 2\psi \rightarrow - \cos 2\psi, 
\hspace{6mm}
\sin 2\psi \rightarrow - \sin 2\psi.
\label{psi-transformation}
\end{eqnarray}
They may be summarized as\footnote{
%%%%%%%%%%%%%% 
In fact there is invariance under $\psi \rightarrow \psi \pm \frac{\pi}{2}$~\cite{Denton:2016wmg}, but we take only the plus sign for definiteness, because this degeneracy does not appear to be important. 
The nature of the symmetry which include $h_{1} \leftrightarrow h_{2}$ transformations explains appearance of the $[ e^{ i ( h_{2} - h_{1} ) x } - 1 ]$ wave factor with the $e^{ i h_{2} x}$ rephasing, in contrast to the $[ e^{ - i ( h_{2} - h_{1} ) x } - 1 ]$ with the $e^{ i h_{1} x}$ rephasing.
}
\begin{eqnarray} 
&& 
\psi \rightarrow \psi + \frac{\pi}{2}. 
\label{varphi-transformation-summary}
\end{eqnarray}

Now, the both $S$ matrix elements obtained with the $e^{ i h_{2} x}$ and $e^{ i h_{1} x}$ rephasing must give the same probability. But, they are connected by the transformation~\eqref{psi-transformation}. It means that the oscillation probability is invariant under the $\psi$ transformation~\eqref{psi-transformation}. It is nothing but the $\psi$ symmetry uncovered by Denton {\it et al.}~\cite{Denton:2016wmg}. It is easy to confirm that the symmetry structure prevails in all the other oscillation channels. 

Thus, we have identified the origin of the $\psi$ symmetry: It is due to the freedom of doing rephasing in the $S$ matrix, assuming the probabilistic nature of quantum mechanics. The observable, in this case the oscillation probability, must be invariant under the phase redefinition of the $S$ matrix. The $\psi$ symmetry utilizes the special two points in the continuous phase transformations which leave the oscillation probability invariant. We believe that the new characterization of the $\psi$ symmetry deepen our understanding of the symmetry. 
One can easily show that the similar understanding can be extended to the $\varphi \rightarrow \varphi + \frac{\pi}{2}$ symmetry in the ``solar-resonance perturbation theory''~\cite{Martinez-Soler:2019nhb}, and to the $\phi \rightarrow \phi + \frac{\pi}{2}$ symmetry in the ``renormalized helio-perturbation theory''~\cite{Minakata:2015gra}, where $\varphi$ and $\phi$ are the matter-dressed $\theta_{12}$ and $\theta_{13}$, respectively. The both symmetries are uncovered in ref.~\cite{Martinez-Soler:2019nhb}. 

Previously, we have characterized the symmetry as ``dynamical'' one~\cite{Martinez-Soler:2019nhb}, not a symmetry in the Hamiltonian. From our new understanding this feature arises because we do rephasing with use of the ``dynamical'' variables, the eigenvalues of the Hamiltonian. They are the complicated functions of the parameters in the Hamiltonian, in particular in the ZS construction, and no simple interpretation as a symmetry of the Hamiltonian is possible. 

\subsection{The decomposed probabilities with $e^{ i h_{1} x }$ rephasing}

Then, the amplitude decomposition with the $e^{ i h_{1} x }$ rephasing should lead to the expression of the non-interference and interference parts of the probability which is consistent with the $\psi$ symmetry. This can be verified by an explicit computation with the decomposed amplitudes eqs.~\eqref{DMP-decompose-emu-0th-h1} and \eqref{DMP-decompose-emu-1st-h1}, which produces the following expressions of $P(\nu_{\mu} \rightarrow \nu_{e})^{ \text{non-int-fer} }$ and $P(\nu_{\mu} \rightarrow \nu_{e})^{ \text{int-fer} }$ to first order in the DMP expansion: 
\begin{eqnarray} 
&& 
\left[ P(\nu_{\mu} \rightarrow \nu_{e})^{(0)} \right]^{ \text{non-int-fer} }
\nonumber \\
&=& 
4 s^2_{23} c^2_{\phi} s^2_{\phi} 
\sin^2 \frac{ ( h_{3} - h_{1} ) x }{2} 
+ 4 c^2_{\phi} s^2_{\psi} 
\left( c^2_{23} c^2_{\psi} + s^2_{23} s^2_{\phi} s^2_{\psi} 
- 2 c_{23} s_{23} s_{\phi} c_{\psi} s_{\psi} \cos \delta \right) 
\sin^2 \frac{ ( h_{2} - h_{1} ) x }{2},
\nonumber \\
&& 
\left[ P(\nu_{\mu} \rightarrow \nu_{e})^{(0)} \right]^{ \text{int-fer} }
\nonumber \\
&=& 
- 4 s_{23} c^2_{\phi} s_{\phi} s_{\psi} 
\left( s_{23} s_{\phi} s_{\psi} - c_{23} c_{\psi} \cos \delta  \right) 
\left\{
- \sin^2 \frac{ ( h_{3} - h_{2} ) x }{2} 
+ \sin^2 \frac{ ( h_{3} - h_{1} ) x }{2} 
+ \sin^2 \frac{ ( h_{2} - h_{1} ) x }{2} 
\right\} 
\nonumber \\
&-& 
8 c_{23} s_{23} c^2_{\phi} s_{\phi} c_{\psi} s_{\psi} \sin \delta 
\sin \frac{ ( h_{3} - h_{1} ) x }{2} \sin \frac{ ( h_{2} - h_{1} ) x }{2} 
\sin \frac{ ( h_{3} - h_{2} ) x }{2},
\label{decomposed-P-mue-0th-h1}
\end{eqnarray} 
\begin{eqnarray} 
&& 
\left[ P(\nu_{\mu} \rightarrow \nu_{e})^{(1)} \right]^{ \text{non-int-fer} } 
\nonumber \\
&=& 
8 \epsilon c_{12} s_{12} s_{ (\phi - \theta_{13}) }
c_{\phi} s_{\phi} 
\left( c_{23} s_{23} s_{\phi} c_{\psi} \cos \delta  + s^2_{23} \cos 2\phi s_{\psi} \right) 
c_{\psi} \frac{ \Delta_{ \text{ren} } }{ h_{3} - h_{2} } 
\sin^2 \frac{ ( h_{3} - h_{1} ) x }{2} 
\nonumber \\
&+& 
8 \epsilon c_{12} s_{12} s_{ (\phi - \theta_{13}) } c_{\phi} s_{\phi} 
\left( c_{23} s_{23} s_{\phi} s_{\psi} \cos \delta  - s^2_{23} \cos 2\phi c_{\psi} \right) 
s_{\psi} \frac{ \Delta_{ \text{ren} } }{ h_{3} - h_{1} } 
\sin^2 \frac{ ( h_{3} - h_{1} ) x }{2} 
\nonumber \\
&-& 
8 \epsilon c_{12} s_{12} s_{ (\phi - \theta_{13}) } 
c_{\phi} s_{\psi} 
\left[ s_{\phi} \left( c^2_{23} c^2_{\psi} - s^2_{23} \cos 2\phi s^2_{\psi} \right) 
+ c_{23} s_{23} c_{\psi} s_{\psi} \cos \delta 
\left( \cos 2\phi - s^2_{\phi} \right) 
\right]
\nonumber \\
&\times&
c_{\psi} \frac{ \Delta_{ \text{ren} } }{ h_{3} - h_{2} } 
\sin^2 \frac{ ( h_{2} - h_{1} ) x }{2}, 
\nonumber 
%\label{decomposed-P-mue-1st-h1}
\end{eqnarray} 
\begin{eqnarray} 
&& 
\left[ P(\nu_{\mu} \rightarrow \nu_{e})^{(1)} \right]^{ \text{int-fer} } 
\nonumber \\
&=& 
4 \epsilon c_{12} s_{12} s_{ (\phi - \theta_{13}) } c_{\phi} 
\left[ s_{\phi} s_{\psi} 
\left\{ c^2_{23} c^2_{\psi} - s^2_{23} \cos 2\phi ( 1 + s^2_{\psi} ) \right\} 
- c_{23} s_{23} c_{\psi} \cos \delta 
\left\{ s^2_{\phi} ( 1 + s^2_{\psi} ) - \cos 2\phi s^2_{\psi} \right\} \right] 
\nonumber \\
&\times&
c_{\psi} \frac{ \Delta_{ \text{ren} } }{ h_{3} - h_{2} } 
\left\{ - \sin^2 \frac{ ( h_{3} - h_{2} ) x }{2} 
+ \sin^2 \frac{ ( h_{3} - h_{1} ) x }{2} 
+ \sin^2 \frac{ ( h_{2} - h_{1} ) x }{2} \right\} 
\nonumber \\
&+& 
4 \epsilon c_{12} s_{12} s_{ (\phi - \theta_{13}) } c_{\phi} s_{\psi} 
\left[ s_{\phi} c_{\psi} s_{\psi} 
\left( c^2_{23} + s^2_{23} \cos 2\phi \right) 
- c_{23} s_{23} \cos \delta 
\left\{ s^2_{\phi} s^2_{\psi} + \cos 2\phi c^2_{\psi} \right\}  \right]
\nonumber \\
&\times&
s_{\psi} \frac{ \Delta_{ \text{ren} } }{ h_{3} - h_{1} } 
\left\{ - \sin^2 \frac{ ( h_{3} - h_{2} ) x }{2} 
+ \sin^2 \frac{ ( h_{3} - h_{1} ) x }{2} 
+ \sin^2 \frac{ ( h_{2} - h_{1} ) x }{2} \right\} 
\nonumber \\
&+& 
8 \epsilon c_{23} s_{23} c_{\phi} s_{ (\phi - \theta_{13}) } c_{12} s_{12} 
\sin \delta 
\left\{ 
c^2_{\psi} \left( s^2_{\phi} - c^2_{\phi} s^2_{\psi} \right) 
\frac{ \Delta_{ \text{ren} } }{ h_{3} - h_{2} } 
- s^2_{\psi} \left( s^2_{\phi} - c^2_{\phi} c^2_{\psi} \right) 
\frac{ \Delta_{ \text{ren} } }{ h_{3} - h_{1} } 
\right\} 
\nonumber \\
&\times&
\sin \frac{ ( h_{3} - h_{1} ) x }{2} \sin \frac{ ( h_{2} - h_{1} ) x }{2} 
\sin \frac{ ( h_{3} - h_{2} ) x }{2}. 
\label{decomposed-P-mue-1st-h1}
\end{eqnarray} 
It is now easy to see that the decomposed probabilities with the $e^{ i h_{1} x }$ rephasing, eqs.~\eqref{decomposed-P-mue-0th-h1} and \eqref{decomposed-P-mue-1st-h1}, and the ones with the $e^{ i h_{2} x }$ rephasing, eqs.~\eqref{decomposed-P-mue-0th-h2} and \eqref{decomposed-P-mue-1st-h2}, are connected with each other by the $\psi$ transformation~\eqref{psi-transformation}. 
Notice that the total probability 
$P(\nu_{\mu} \rightarrow \nu_{e}) = P(\nu_{\mu} \rightarrow \nu_{e})^{ \text{non-int-fer} } + P(\nu_{\mu} \rightarrow \nu_{e})^{ \text{int-fer} }$ calculated with both the $e^{ i h_{1} x }$ and $e^{ i h_{2} x }$ rephasing should be identical with each other. It is verified by explicitly showing that they both agree with the zeroth and first order probabilities, eqs.~\eqref{total-P-mue-0th} and ~\eqref{total-P-mue-1st} given in appendix~\ref{sec:DMP-probability-mue}, which are computed by using the conventional method of calculation using the $S$ matrix elements given in eq.~\eqref{S-emu-0th-1st}. By this way, consistency among the three ways of calculation, the $e^{ i h_{1} x }$ and $e^{ i h_{2} x }$ rephasing, and the $S$ matrix method without the amplitude decomposition, is verified. 
The consistency check for the decomposed probabilities in the other channels, which are computed in appendices~\ref{sec:amp-decompose-disapp} and \ref{sec:amp-decompose-mu-tau}, is carried out by the same way as above. 

Before completing our discussion of the DMP decomposition in this section, we discuss the two remaining problems, both of which we feel very relevant. They are (1) CP phase dependence of the oscillation probability, and (2) possible physical interpretation of the $e^{ i h_{1} x }$ and $e^{ i h_{2} x }$ rephasing.

\subsection{CP phase dependence} 
\label{sec:phase-dependence}

The $\delta$-dependent terms in the oscillation probability have some universal coefficients, and hence they may be of interest. In fact, we observe a new regularity in the structure of Jarlskog factors in matter in the DMP probabilities. Our following discussion applies to the both probabilities calculated with the $e^{ i h_{2} x }$ and $e^{ i h_{1} x }$ rephasing. 

In the both decomposed probabilities $P(\nu_{\mu} \rightarrow \nu_{e})^{ \text{non-int-fer} }$ and $P(\nu_{\mu} \rightarrow \nu_{e})^{ \text{int-fer} }$, the $\delta$ dependent terms have the coefficients $J_{ \text{m-1st} } \equiv c_{23} s_{23} c^2_{\phi} s_{\phi} c_{\psi} s_{\psi}$ in the zeroth-order and $J_{ \text{m-2nd} } \equiv c_{23} s_{23} c_{\phi} s_{ (\phi - \theta_{13}) } c_{12} s_{12}$ in the first-order terms. The same feature is shared by the probabilities in the $P(\nu_{\mu} \rightarrow \nu_{\tau})$ channel, see appendix~\ref{sec:amp-decompose-mu-tau}. We take the NMO not to worry about the $\pm$ signs. 
By using the formulas in appendix~\ref{sec:DMP-variables}, it is straightforward to show that these Jarlskog factors in matter can be written exactly as 
\begin{eqnarray} 
&& 
J_{ \text{m-1st} } \equiv c_{23} s_{23} c^2_{\phi} s_{\phi} c_{\psi} s_{\psi} 
\nonumber \\
&=&
\frac{ \epsilon J_{r} }{ \left[ 1 + r_{a}^2 - 2 r_{a} \cos 2\theta_{13} \right] } 
\frac{ F_{-} ( r_{a} ) }{ F_{+} ( r_{a} ) }
\frac{ 1 }{ 
\left[ 1 - 4 \epsilon \frac{ \cos 2\theta_{12} }{ F_{+} ( r_{a} ) } 
+ 4 \epsilon^2 \frac{ \cos^2 2\theta_{12} + \sin^2 2\theta_{12} c^2_{\phi - \theta_{13}} }{ [ F_{+} ( r_{a} ) ]^2 } \right]^{1/2} }, 
\nonumber \\
&&
J_{ \text{m-2nd} } \equiv c_{23} s_{23} c_{\phi} s_{ (\phi - \theta_{13}) } c_{12} s_{12} 
\nonumber \\
&=&
\frac{ J_{r} }{ \sqrt{ 1 + r_{a}^2 - 2 r_{a} \cos 2\theta_{13} } } 
\left[ 1 + 
\frac{ 2 s^2_{13}  }
{ ( \cos 2\theta_{13} - r_{a} ) - \sqrt{ 1 + r_{a}^2 - 2 r_{a} \cos 2\theta_{13} } } \right],
\label{matter-Jarlskog}
\end{eqnarray}
where $J_{r} \equiv c_{23} s_{23} c^2_{13} s_{13} c_{12} s_{12}$ denotes the Jarlskog factor in vacuum \cite{Jarlskog:1985ht}, and $r_{a} \equiv \frac{a}{ \Delta m^2_{ \text{ren} } }$. In eq.~\eqref{matter-Jarlskog} we have defined the $F_{+}$ and $F_{-}$ functions as
\begin{eqnarray} 
&&
F_{\pm} ( r_{a} )
\equiv 
\left[ 1 \pm r_{a} \mp \sqrt{ 1 + r_{a}^2 - 2 r_{a} \cos 2\theta_{13} } \right]. 
\label{F-function-def}
\end{eqnarray}
That is, the matter Jarlskog factors are proportional to the vacuum one. 
This result is what is predicted by the Naumov identity for the T-odd $\sin \delta$ term \cite{Naumov:1991ju}, and by the general property of the T-even $\cos \delta$ term discussed in ref.~\cite{Asano:2011nj}.\footnote{
%%%%%%%%%%%%%%%% footnote %%%%%%%%%%%%%%%%%
The general theorem given in ref.~\cite{Asano:2011nj} for the $\cos \delta$ term does not prove the $c^2_{13}$ factor, which is indeed missing in the probabilities in the $\nu_{\mu} - \nu_{\tau}$ sector. But, it is known empirically \cite{Kimura:2002wd} that the $c^2_{13}$ always appears in the $\nu_{\mu} \rightarrow \nu_{e}$ channel. For more examples, see e.g., ref.~\cite{Martinez-Soler:2019nhb} and sections 9 in ref.~\cite{Minakata:2020ijz}. 
}
A general discussion of T-odd terms in the DMP theory is also given in ref.~\cite{Denton:2016wmg}. 

\subsection{Physical interpretation of the $e^{ i h_{1} x }$ and $e^{ i h_{2} x }$ rephasing }
\label{sec:interpretation-rephasing}

We present here the two different views and usages of the freedom of the $e^{ i h_{1} x }$ and $e^{ i h_{2} x }$ rephasing. The decomposed probabilities $P(\nu_{\mu} \rightarrow \nu_{e})^{ \text{int-fer} }$ ($P(\nu_{\mu} \rightarrow \nu_{e})^{ \text{non-int-fer} }$ as well) with the $e^{ i h_{1} x }$ and $e^{ i h_{2} x }$ rephasing are different from each other in magnitudes, though they are related by the $\psi$ transformation. That is, the $\psi$ symmetry, which existed in the total probability $P(\nu_{\mu} \rightarrow \nu_{e})^{ \text{non-int-fer} } + P(\nu_{\mu} \rightarrow \nu_{e})^{ \text{int-fer} }$, is broken in each individual term in the decomposed probability. Then, the  test probability defined by introducing the $q$ parameter~\cite{Huber:2019frh,Minakata:2020ijz} 
\begin{eqnarray} 
P(\nu_{\beta} \rightarrow \nu_{\alpha}: q) 
&=& 
P(\nu_{\beta} \rightarrow \nu_{\alpha})^{ \text{non-int-fer} } 
+ q P(\nu_{\beta} \rightarrow \nu_{\alpha})^{ \text{int-fer} } 
\label{ansatz}
\end{eqnarray}
to quantify the statistical significance for observing the interference effect 
is different between the cases of $e^{ i h_{1} x }$ and $e^{ i h_{2} x }$ rephasing for $q \neq 1$. 

Using the difference in the test probability \eqref{ansatz} with the $e^{ i h_{1} x }$ and $e^{ i h_{2} x }$ rephasing at $q \neq 1$, the first usage of the two different rephasing is, in a given mass ordering, 
\begin{itemize}

\item 
In analyses of quantifying the effect of interference effect one can use the two different rephasing formulas to estimate the uncertainty due to the choice of the theoretical frameworks. 

\end{itemize}
\noindent 
It shares the similar spirit as the method we have employed in our JUNO analysis in ref.~\cite{Huber:2019frh}, in which we have examined the two cases of the atmospheric $\Delta m^2$, $\Delta m^2_{31}$ and $\Delta m^2_{32}$, which can be derived by using $e^{ i ( m^2_{1} / 2E ) x }$ and $e^{ i ( m^2_{2} / 2E ) x }$ rephasing, respectively. 

Toward the possible second usage, we argue that: 
\begin{itemize}

\item 
The $e^{ i h_{2} x }$ and $e^{ i h_{1} x }$ rephasing amplitudes correspond, respectively, to the IMO and NMO, the viewpoint we advertised in ref.~\cite{Minakata:2020ijz}. 

\end{itemize}
\noindent 
If one looks at the three-neutrino energy level crossing diagram, e.g., in Fig.~1 in ref.~\cite{Denton:2016wmg}, one recognizes that the atmospheric resonance is in the 2-3 and 1-3 level crossings in the NMO and IMO, respectively. We also note that with the $e^{ i h_{2} x }$ and $e^{ i h_{1} x }$ rephasing, the leading term of the probability in eq.~\eqref{decomposed-P-mue-0th-h2} and \eqref{decomposed-P-mue-0th-h1}, respectively, takes the form of $\sin^2 \frac{ ( h_{3} - h_{2} ) x }{2}$ and $\sin^2 \frac{ ( h_{3} - h_{1} ) x }{2}$. They describe the atmospheric resonance in the NMO and IMO, respectively. Therefore, the above interpretation is quite natural. Thus, we mean by the second usage of the two rephasing freedom use of the $e^{ i h_{2} x }$ rephasing in the analysis assuming the NMO, and $e^{ i h_{1} x }$ rephasing in the analysis assuming the IMO.\footnote{
%%%%%%%%%%%%%%%% footnote %%%%%%%%%%%%%%%%%
But, of course, one should remember that the total probabilities, or the test probabilities at $q=1$, with the $e^{ i h_{2} x }$ and $e^{ i h_{1} x }$ rephasing describe exactly the same physics. Therefore, our second usage of the two rephasing freedom only make sense when we talk about the interference analyses with use of the test probability \eqref{ansatz}. }

However, before one of the neutrino mass ordering is established, the analyses of experimental data will be done by assuming either the NMO or IMO one by one, or by marginalizing over the two mass orderings. Therefore, practically one may take effectively the first attitude above. But, when the mass ordering is established, one may want to decide whether one choose the possible second usage, $e^{ i h_{2} x }$-rephasing$-$NMO or $e^{ i h_{1} x }$-rephasing$-$IMO correspondence as an ansatz for the analysis, which we do recommend.\footnote{
%%%%%%%%%%%%%%%% footnote %%%%%%%%%%%%%%%%%
The era of established neutrino mass ordering may not be too far remote, given the result of analysis in ref.~\cite{Cabrera:2020own}. }

In either cases, we believe that the expressions of the decomposed oscillation probability eqs.~\eqref{decomposed-P-mue-0th-h2} and \eqref{decomposed-P-mue-1st-h2}, and/or eqs.~\eqref{decomposed-P-mue-0th-h1} and \eqref{decomposed-P-mue-1st-h1}, are ready for use in data analyses to quantify the observation of interference effect under the approximation of uniform matter density.

\section{Physical interpretation of the dynamical modes} 
\label{sec:interpretation} 

At the start of our discussion in section~\ref{sec:DMP-amplitude-decomposition}, we have foretold that the two dynamical modes in the decomposed amplitudes, denoted as the ``A'' and ``S'' modes, can be interpreted as the matter-affected atmospheric and the solar oscillation modes, respectively. Let us try to give some foundation on this statement. We do this by showing that the DMP decomposed amplitudes smoothly connect themselves into the ``atmospheric'' and ``solar'' waves in each appropriate kinematical phase spaces. 
Speaking more precisely, we take suitable limits of the DMP amplitude decomposition into the regions in which one of the two modes can be clearly identified either as the ``atmospheric'' or the ``solar'' waves. Each one of these modes is described by the suitable perturbative framework, as discussed below. The amplitude decomposition in the related frameworks is discussed in ref.~\cite{Minakata:2020ijz}. 

\subsection{Approaching to region of the enhanced atmospheric-scale oscillation} 
\label{sec:helio-P}

We discuss first the ``renormalized helio-perturbation theory'' \cite{Minakata:2015gra} limit of the DMP perturbation theory. For a simpler terminology we call the former as the MP model. It is the most suited one for this purpose among the similar frameworks so far developed which perturbs around the atmospheric resonance~\cite{Arafune:1996bt,Cervera:2000kp,Freund:2001pn,Akhmedov:2004ny,Minakata:2015gra}, because it is a ``half-way'' DMP and also due to its favorable properties such as inclusion of only the correct wave factors.\footnote{
%%%%%%%%%%%%%% footnote %%%%%%%%%%%%%%
One of the problem is that $( - i \Delta_{21} x)$ term could be interpreted as $\left( e^{ - i \Delta_{21} x } -1 \right)$ approximately. But, $\Delta_{21} x \approx \frac{\Delta_{21}}{\Delta_{31}} \simeq \epsilon$ can also be understood as the expansion parameter, which would introduce confusion in doing the decomposition. In the first-order treatment in ref.~\cite{Minakata:2020ijz} we have circumvented this issue, but the problem becomes more serious when we go to higher orders.  
}

To illuminate the point we discuss the case of NMO and IMO in parallel. The suitable limit to approach the atmospheric resonance region would be $\epsilon \ll 1$, keeping $\theta_{13}$ and $\phi$ finite, and 
\begin{eqnarray} 
&&
r_{a} \equiv \frac{a}{ \Delta m^2_{ \text{ren} } } 
\simeq 
\frac{ \Delta_{a} }{ \Delta_{31} } 
\left( 1 + s^2_{12} \frac{ \Delta_{21} }{ \Delta_{31} } \right) 
\sim \mathcal{O} (1).
\label{ra-def}
\end{eqnarray}
We remind the readers for the notations $\Delta_{a} \equiv \frac{a}{2E}$ and $\Delta_{ji} \equiv \frac{ \Delta m^2_{ji} }{2E}$, as defined in section~\ref{sec:notations}. Some aspects of this limit are discussed in appendix~\ref{sec:DMP-variables}. 

One can show, by using the expression in appendix~\ref{sec:DMP-variables}, that the DMP eigenvalues $h_{i} \equiv \frac{\lambda_{i}}{2E}$ ($i=1,2,3$) can be written by the MP eigenvalues $h_{\pm, 0} \equiv \frac{\lambda_{\pm, 0}}{2E}$ to order $\epsilon$: 
\begin{eqnarray} 
&& 
\text{NMO}: 
\hspace{8mm}
h_{3} - h_{2} = h_{+} - h_{-}, 
\hspace{8mm}
h_{3} - h_{1} = h_{+} - h_{0},  
\hspace{8mm}
h_{2} - h_{1} = h_{-} - h_{0}, 
\nonumber \\ 
&& 
\text{IMO}: 
\hspace{8mm}
h_{3} - h_{2} = h_{+} - h_{0}, 
\hspace{8mm}
h_{3} - h_{1} = h_{+} - h_{-},  
\hspace{8mm}
h_{2} - h_{1} = h_{0} - h_{-}.
\label{DMP-MP-eigenvalues}
\end{eqnarray}
See Fig.~3 in ref.~\cite{Minakata:2015gra}. 
Notice that every helio-perturbation theory has a drawback of the wrong solar-level crossing as discussed in ref.~\cite{Minakata:2015gra}, and the complexity of the correspondence of the DMP - MP eigenvalues in eq.~\eqref{DMP-MP-eigenvalues} reflects this drawback. But, it does not cause the problem in our task because our limit is toward the atmospheric resonance region, and the solar-level crossing is outside the region of validity of the helio--perturbation theory. 

The mixing angle $\phi$, the matter-dressed $\theta_{13}$, is given in eq.~\eqref{2phi}, the universal form which is valid independent of the NMO or IMO. $\psi$, the matter-dressed $\theta_{12}$ take the simple approximate forms $\cos 2 \psi = \mp 1 + \mathcal{O} (\epsilon^2)$ and 
\begin{eqnarray}
&& 
\sin 2 \psi 
= \pm \epsilon \sin 2\theta_{12} c_{\phi - \theta_{13}} 
\frac{ \Delta_{ \text{ren} } }{ h_{-} - h_{0} }.
\nonumber
\end{eqnarray}
That is, $\psi \approx \frac{\pi}{2}$ ($\psi \approx 0$) at around the atmospheric resonance in the NMO (IMO). See Fig.~1 in ref.~\cite{Denton:2016wmg}. This property guarantees a smooth connection to the MP model from the DMP.

Interestingly, in the limit to the atmospheric resonance region, the amplitude decomposition formulas in the zeroth and first order can be written in the universal form, i.e., independent of the NMO or IMO, as 
\begin{eqnarray} 
\left( S_{e \mu}^{\text{atm}} \right)^{(0)} 
&=&
s_{23} c_{\phi} s_{\phi} e^{ - i \delta} \left( e^{ - i ( h_{+} - h_{-} ) x } - 1 \right), 
\nonumber \\ 
\left( S_{e \mu}^{\text{sol}} \right)^{(0)} 
&=&
- \epsilon c_{23} c_{12} s_{12} c_{\phi} c_{\phi - \theta_{13}} 
\frac{ \Delta_{ \text{ren} } }{ h_{-} - h_{0} } 
\left( e^{ i ( h_{-} - h_{0} ) x } - 1 \right),
\nonumber \\ 
\left( S_{e \mu}^{ \text{atm} } \right)^{(1)} 
&=& 
\epsilon c_{23} c_{12} s_{12} s_{\phi} s_{ (\phi - \theta_{13}) } 
\frac{ \Delta_{ \text{ren} } }{ h_{+} - h_{0} } 
\left( e^{ - i ( h_{+} - h_{-} ) x } - 1 \right), 
\nonumber \\ 
\left( S_{e \mu}^{ \text{sol} } \right)^{(1)} 
&=& 
- \epsilon c_{23} c_{12} s_{12} s_{\phi} s_{ (\phi - \theta_{13}) } 
\frac{ \Delta_{ \text{ren} } }{ h_{+} - h_{0} } 
\left( e^{ i ( h_{-} - h_{0} ) x } - 1 \right). 
\label{MP-decomposition-emu-NMO-IMO}
\end{eqnarray}
Then, the non-interference and interference parts of the probability in the both mass orderings  can be written as 
\begin{eqnarray} 
P(\nu_{\mu} \rightarrow \nu_{e})^{ \text {non-int-fer} } 
&=&
4 s^2_{23} c^2_{\phi} s^2_{\phi} 
\sin^2 \frac{ ( h_{+} - h_{-} ) x }{2} 
\nonumber \\ 
&+& 
8 \epsilon c_{23} s_{23} c_{\phi} s^2_{\phi} c_{12} s_{12} s_{ (\phi - \theta_{13}) } \cos \delta 
\frac{ \Delta_{ \text{ren} } }{ h_{+} - h_{0} } 
\sin^2 \frac{ ( h_{+} - h_{-} ) x }{2},
\nonumber \\
P(\nu_{\mu} \rightarrow \nu_{e})^{ \text {int-fer} } 
&=& 
- 4 \epsilon c_{23} s_{23} c_{\phi} s_{\phi} c_{12} s_{12} \cos \delta 
\left\{ 
c_{\phi} c_{\phi - \theta_{13}} 
\frac{ \Delta_{ \text{ren} } }{ h_{-} - h_{0} } 
+ s_{\phi} s_{ (\phi - \theta_{13}) } 
\frac{ \Delta_{ \text{ren} } }{ h_{+} - h_{0} } 
\right\} 
\nonumber \\
&\times& 
\left\{ 
- \sin^2 \frac{ ( h_{+} - h_{0} ) x }{2}
+ \sin^2 \frac{ ( h_{+} - h_{-} ) x }{2} 
+ \sin^2 \frac{ ( h_{-} - h_{0} ) x }{2}
\right\} 
\nonumber \\ 
&-& 
8 \epsilon c_{23} s_{23} c_{\phi} s_{\phi} c_{12} s_{12} \sin \delta 
\left\{ 
c_{\phi} c_{\phi - \theta_{13}} 
\frac{ \Delta_{ \text{ren} } }{ h_{-} - h_{0} } 
+ s_{\phi} s_{ (\phi - \theta_{13}) } 
\frac{ \Delta_{ \text{ren} } }{ h_{+} - h_{0} }  
\right\} 
\nonumber \\
&\times& 
\sin \frac{ ( h_{+} - h_{0} ) x }{2} \sin \frac{ ( h_{-} - h_{0} ) x }{2} 
\sin \frac{ ( h_{+} - h_{-} ) x }{2}.  
\label{decomposed-P-mue-MP}
\end{eqnarray}
Notice that the MP first order terms in \eqref{decomposed-P-mue-MP} come from both the DMP leading and the next to leading order terms. Using the expressions of $V$ matrix elements given in ref.~\cite{Minakata:2015gra}, or using the formalism in our previous paper~\cite{Minakata:2020ijz}, one can easily work out the amplitude decomposition in the MP model. An explicit computation that is carried out confirms our results in eqs.~\eqref{MP-decomposition-emu-NMO-IMO} and \eqref{decomposed-P-mue-MP}. Of course, one can verify that $P(\nu_{\mu} \rightarrow \nu_{e}) = P(\nu_{\mu} \rightarrow \nu_{e})^{ \text {non-int-fer} } + P(\nu_{\mu} \rightarrow \nu_{e})^{ \text {int-fer} }$ reproduces (T-conjugate of) eq.~(B.2) in appendix~B in ref.~\cite{Minakata:2015gra}. 

For simplicity and clarity we have restricted our discussion in this section to the $\nu_{\mu} \rightarrow \nu_{e}$ channel. But, the generalization to the other channels is straightforward. The key point is that at around region of the atmospheric resonance the matter-dressed $\theta_{12}$ ``freezes'' into $\psi \approx \frac{\pi}{2}$ ($\psi \approx 0$) in the NMO (IMO).

\subsection{Approaching to region of the enhanced solar-scale oscillations}
\label{sec:solar-resonance-P} 

We discuss the limit toward the ``solar-resonance perturbation theory'' \cite{Martinez-Soler:2019nhb}. It is the perturbative framework whose region of validity is with enhanced solar-scale oscillations 
\begin{eqnarray} 
r_{a}^{ \text{sol} } \equiv \frac{a}{\Delta m^{2}_{21}} = \frac{ \Delta_{a} }{ \Delta_{21} } 
= \frac{a}{ \epsilon \Delta m^2_{ \text{ren} } } \sim \mathcal{O} (1).
\label{ra-sol-def}
\end{eqnarray}
The framework has an effective expansion parameter 
\begin{eqnarray}
A_{ \text{exp} } 
&\equiv& 
c_{13} s_{13} 
\biggl | \frac{ a }{ \Delta m^2_{31} } \biggr | %%expansion-parameter
%\nonumber \\&=&
%0.8875 \times 10^{-2} x 0.146 = 0.1296 \times 10^{-2} x (3/2.8) 2 (200MeV) = 0.278 \times 10^{-2} 
= 2.78 \times 10^{-3} 
\left(\frac{ \Delta m^2_{31} }{ 2.4 \times 10^{-3}~\mbox{eV}^2}\right)^{-1}
\left(\frac{\rho}{3.0 \,\text{g/cm}^3}\right) \left(\frac{E}{200~\mbox{MeV}}\right), 
\nonumber \\
\label{expansion-parameter}
\end{eqnarray}
which guarantees smallness of the perturbative corrections, as confirmed in ref.~\cite{Martinez-Soler:2019nhb}. 

Since we want to keep $A_{ \text{exp} } \propto \frac{ a }{ \Delta m^2_{31} } \simeq r_{a}$ small, we take the limit $r_{a} \ll1$ keeping $r_{a}^{ \text{sol} }$ finite to approach to the solar-resonance region. Noticing that $r_{a} = \epsilon r_{a}^{ \text{sol} }$, the limit $r_{a} \ll1$ is in harmony with the smallness of the DMP expansion parameter $\epsilon$. Therefore, the solar-resonance perturbation theory limit is feasible in the DMP framework. Our treatment below applies to both the NMO and IMO. 

We expand the eigenvalues and the mixing angles keeping $r_{a}^{ \text{sol} }$ finite. The eigenvalues read, to order $r_{a}$, as 
\begin{eqnarray} 
h_{-} 
&=& 
\Delta_{ \text{ren} } 
\left[ c^2_{13} r_{a} + \epsilon s^2_{12} \right], 
\nonumber \\
h_{0} 
&=& \Delta_{ \text{ren} } \epsilon c^2_{12}, 
\nonumber \\
h_{+} 
&=& \Delta_{ \text{ren} } 
\left[ 1 + s^2_{13} r_{a} + \epsilon s^2_{12} \right].
\label{lambda-pm-solarR}
\end{eqnarray}
For the mixing angle $\phi$ we refer section~\ref{sec:DMP-matterP-0th}
for the approximate formulas for the small $r_{a}$ approximation. Namely, the approximation on $\theta_{13}$ in eq.~\eqref{angles-matter-P} applies to the present case, but not the ones for $\psi$ which involve $r_{a}^{ \text{sol} }$. 
Using $\lambda_{0} + \lambda_{-} = \Delta m^2_{21} \left( 1 + c^2_{13} r_{a}^{ \text{sol} } \right)$ and the similar expansion inside the square root, the DMP eigenvalues are given by 
\begin{eqnarray} 
&& 
h_{1} 
= \frac{1}{2} \Delta_{21} 
\left[ \left( 1 + c^2_{13} r_{a}^{ \text{sol} } \right) 
- \left\{
1 - 2 c^2_{13} \cos 2 \theta_{12} r_{a}^{ \text{sol} }  
+ c^4_{13} \left( r_{a}^{ \text{sol} } \right)^2  
\right\}^{1/2}
\right], 
\nonumber \\
&& 
h_{2} 
= \frac{1}{2} \Delta_{21} 
\left[ \left( 1 + c^2_{13} r_{a}^{ \text{sol} } \right) 
+ \left\{
1 - 2 c^2_{13} \cos 2 \theta_{12} r_{a}^{ \text{sol} }  
+ c^4_{13} \left( r_{a}^{ \text{sol} } \right)^2  
\right\}^{1/2}
\right], 
\nonumber \\
&& 
h_{3} = 
\Delta_{31} + s^2_{13} \Delta_{a}. 
\label{lambda-123-solarR}
\end{eqnarray}
which reproduces the eigenvalues in eqs.~(16) and (17) in ref.~\cite{Martinez-Soler:2019nhb}. Notice that the $\epsilon$ correction in $\Delta_{ \text{ren} }$ and $\epsilon s^2_{12}$ term in $h_{+} $ cancel with each other. 
Similarly, the expressions of $\psi$ reads 
\begin{eqnarray}
&& 
\cos 2 \psi 
= - \frac{ - \cos 2\theta_{12} + c^2_{13} r_{a}^{ \text{sol} } }{ \sqrt{ \left( \cos 2\theta_{12} - c^2_{13} r_{a}^{ \text{sol} } \right)^2 + \sin^2 2\theta_{12} } },
\nonumber \\
&& 
\sin 2 \psi 
= \frac{ \sin 2\theta_{12} }{ \sqrt{ \left( \cos 2\theta_{12} - c^2_{13} r_{a}^{ \text{sol} } \right)^2 
+ \sin^2 2\theta_{12} } },
\label{2psi-solarR}
\end{eqnarray}
which again reproduces precisely the expressions in eq.~(14) in ref.~\cite{Martinez-Soler:2019nhb}. 

Thus, the zeroth order eigenvalues and the mixing angles are reproduced, which means that the solar-resonance perturbation theory at its leading order can be reached as the appropriate limit of the DMP theory. Since the prescription for amplitude decomposition we use in this and the previous papers~\cite{Minakata:2020ijz} is the same, it is obvious that the DMP amplitude decomposition smoothly tend to the decomposition using the solar-resonance perturbation theory discussed in ref.~\cite{Minakata:2020ijz}. Notice that our statement applies to all the oscillation channel, because the whole structure of the solar-resonance perturbation theory at the leading order is reproduced. Since the first order correction is small with the tiny $A_{ \text{exp} }$, we do not enter into the discussion of first order corrections. 

\subsection{Matter-dressed atmospheric and solar oscillations in entire terrestrial region?}
\label{sec:terrestrial-region} 

Based upon the results obtained in the previous sections~\ref{sec:helio-P} and \ref{sec:solar-resonance-P}, we argue that the dynamical two modes described by the DMP decomposition can be interpreted as the matter-dressed atmospheric and the matter-dressed solar neutrino oscillations in the entire ``terrestrial-experiments-covered'' region. By it we mean the energies and baselines of, roughly speaking, the neutrino energy-baseline covered by the atmospheric neutrino observation. For short it will be denoted as the ``terrestrial region'' hereafter. See the drawing e.g., in Fig.~1 of ref.~\cite{Minakata:2019gyw} with keeping in mind some extension to higher energy side. 

The validity of characterization of the dynamical two modes is of course true in the two resonance-enhanced regions that are explicitly treated in sections~\ref{sec:helio-P} and \ref{sec:solar-resonance-P}, but it is also likely to prevail in region between. In fact, the regions of validity of the both perturbation theories to first order, the solar-resonance perturbation theory at low-energy side and the one of the renormalized helio-perturbation theory at high-energy side, almost occupy the entire terrestrial region. See Figs.~2 and 3 in ref.~\cite{Martinez-Soler:2019nhb}, and Fig.~1 in ref.~\cite{Minakata:2015gra}. Therefore, there is not so much room for inserting a new dynamical mode between the regions of validity of the both perturbation theories valid at low and high energy regions. Then, it is very likely that the dynamical two modes in the terrestrial region can be regarded as the matter-dressed solar and the matter-dressed atmospheric neutrino oscillations.

As an outcome of the exercise we engaged in this section, we have obtained a new picture of the DMP perturbation theory as a unified perturbative framework for neutrino oscillation which covers the whole terrestrial neutrino oscillations. Certainly it is the unique framework among the perturbative schemes so far proposed that covers the two resonances, due to the atmospheric- and the solar-scale enhancements.\footnote{
%%%%%%%%%%%%% footnote %%%%%%%%%%%%%%%
A possible missing piece in our discussion might be the treatment of amplitude decomposition with use of the Agarwalla {\it et al.} Jacobi-method based framework~\cite{Agarwalla:2013tza}, another candidate for ``unified'' perturbative framework for terrestrial neutrino oscillations.
}

\section{DMP amplitude decomposition applied to the system with infinitesimal matter potential} 
\label{sec:decomposition-matter-P}

We have observed in ref.~\cite{Minakata:2020ijz} that a straightforward application of the vacuum definition of amplitude decomposition~\cite{Huber:2019frh} fails with an infinitesimal matter potential. It looks like really a contrived case because the three neutrino eigenvalues remain the same as in vacuum, and the strong matter modification of the two vacuum modes should not exist. 
Since we set up the framework of DMP amplitude decomposition, it should be possible to understand how this issue is treated by the DMP decomposition. It is the purpose of our discussion in this section to examine this point.

We first examine the matter perturbation theory limit of the DMP theory, deferring the discussion of amplitude decomposition later in this section. Since we are talking about an infinitesimal matter potential the limit we take is both $r_{a} \ll1$, and $r_{a}^{ \text{sol} } \ll 1$.

\subsection{The $\nu_{\mu} - \nu_{e}$ amplitude to first order in matter perturbation theory }

The oscillation $S$ matrix element in the $\nu_{\mu} - \nu_{e}$ channel valid to first order in matter perturbation theory which was derived in ref.~\cite{Minakata:2020ijz}, which we recapitulate here: 
\begin{eqnarray}
S_{e \mu} &=& 
c_{13} s_{12} 
\left( c_{23} c_{12} - s_{23} s_{13} s_{12} e^{ - i \delta} \right) 
\left( e^{ - i \Delta_{21} x } - 1 \right) 
+ s_{23} c_{13} s_{13} e^{ - i \delta} 
\left( e^{ - i \Delta_{31} x } -1 \right) 
\nonumber \\
&+& 
c_{12} s_{12} c^3_{13} 
\left( \cos 2\theta_{12} c_{23} - \sin 2\theta_{12} s_{13} s_{23} e^{ - i \delta} \right) 
\frac{ \Delta_{a} }{ \Delta_{21} } 
\left( e^{ - i \Delta_{21} x } - 1 \right) 
\nonumber \\
&+& 
c_{12} c_{13} s_{13} 
\left( - s_{12} s_{13} c_{23} + \cos 2\theta_{13} c_{12} s_{23} e^{ -i \delta }  \right) 
\frac{ \Delta_{a} }{ \Delta_{31} } 
\left( e^{ - i \Delta_{31} x } - 1 \right) 
\nonumber \\
&+& 
s_{12} c_{13} s_{13} 
\left( c_{12} s_{13} c_{23} + \cos 2\theta_{13} s_{12} s_{23} e^{ -i \delta } \right) 
\frac{ \Delta_{a} }{ \Delta_{31} - \Delta_{21} } 
\left( e^{ - i \Delta_{31} x } - e^{ - i \Delta_{21} x } \right) 
\nonumber \\
&+& 
( -i \Delta_{a} x) 
\biggl[ 
c^3_{12} c^3_{13} 
\left( - s_{12} c_{23} - c_{12} s_{13} s_{23} e^{ - i \delta} \right) 
\nonumber \\
&+& 
s^3_{12} c^3_{13} 
\left( c_{12} c_{23} - s_{12} s_{13} s_{23} e^{- i \delta}  \right)
e^{ - i \Delta_{21} x } 
+ c_{13} s^3_{13} s_{23} e^{-i\delta} 
e^{ - i \Delta_{31} x } 
\biggr]. 
\label{Semu-matter-P-1st} 
\end{eqnarray}
The matter perturbation theory used in ref.~\cite{Minakata:2020ijz} involves the unique expansion parameter $\frac{ a }{ \Delta m^2_{31} }$. But, since the DMP expansion involve the expansion parameter $\epsilon \equiv \frac{ \Delta m^2_{21} }{ \Delta m^2_{ \text{ren} } }$, we can reproduce the fourth line of \eqref{Semu-matter-P-1st} only in an expanded form 
\begin{eqnarray} 
c_{13} s_{13} s_{12} 
\left( c_{23} s_{13} c_{12} + s_{23} \cos 2\theta_{13} s_{12} e^{ -i \delta } \right) 
\frac{ \Delta_{a} }{ \Delta_{31} } \left( 1 + \frac{ \Delta_{21} }{ \Delta_{31} } \right) 
\left( e^{ - i \Delta_{31} x } - e^{ - i \Delta_{21} x } \right). 
\label{fourth-term}
\end{eqnarray}

\subsection{The DMP amplitude in the leading order} 
\label{sec:DMP-matterP-0th}

Let us start with $S_{e \mu}$ in the leading order DMP expansion, the first line in eq.~\eqref{S-emu-0th-1st}. We note that, as we are close to vacuum, our discussion in this section does not distinguish between the NMO and IMO. To obtain the approximate formula valid to first order in $r_{a}$ and $r_{a}^{ \text{sol} }$, we use the following formulas 
\begin{eqnarray} 
&& 
h_{1} = c^2_{13} c^2_{12} \Delta_{a}, 
\nonumber \\
&&
h_{2} = \Delta_{21} + c^2_{13} s^2_{12} r_{a} \Delta_{ \text{ren} },
\nonumber \\
&&
h_{3} = \left( 1 + \epsilon s^2_{12} + s^2_{13} r_{a} \right) \Delta_{ \text{ren} }.
\label{hi-matter-P}
\end{eqnarray}
\begin{eqnarray}
&&
c_{\phi} = c_{13} \left( 1 - s^2_{13} r_{a} \right), 
\hspace{16mm} 
s_{\phi} = s_{13} \left( 1 + c^2_{13} r_{a} \right), 
\nonumber \\
&& 
\cos ( \phi - \theta_{13} ) = 1, 
\hspace{23mm} 
\sin ( \phi - \theta_{13} ) = c_{13} s_{13} r_{a}, 
\nonumber \\
&& 
c_{\psi} = c_{12} \left( 1 - s^2_{12} c^2_{13} r_{a}^{ \text{sol} } \right), 
\hspace{8mm} 
s_{\psi} = s_{12} \left( 1 + c^2_{12} c^2_{13} r_{a}^{ \text{sol} } \right).
\label{angles-matter-P}
\end{eqnarray}
By using these formulas, one can obtain, to leading order in $r_{a}$ and $r_{a}^{ \text{sol} }$,  
\begin{eqnarray} 
S_{e \mu}^{(0)} 
&=& 
c_{13} s_{12} 
\left( c_{23} c_{12} - s_{23} s_{13} s_{12} e^{ - i \delta} \right) 
\left( e^{ - i \Delta_{21} x } - 1 \right) 
+ s_{23} c_{13} s_{13} e^{ - i \delta} 
\left( e^{ - i \Delta_{31} x } -1 \right) 
\nonumber \\
&+& 
c^3_{13} c_{12} s_{12} 
\left( c_{23} \cos 2\theta_{12} 
- s_{23} s_{13} \sin 2\theta_{12} e^{ - i \delta} \right) 
r_{a}^{ \text{sol} }  
\left( e^{ - i \Delta_{21} x } - 1 \right) 
\nonumber \\
&-& 
c_{13} s_{13} s_{12} 
\left( c_{23} s_{13} c_{12} 
+ s_{23} \cos 2\theta_{13} s_{12} e^{ - i \delta} \right) 
r_{a} 
\left( e^{ - i \Delta_{21} x } - 1 \right) 
\nonumber \\
&+& 
s_{23} c_{13} s_{13} e^{ - i \delta} 
\cos 2\theta_{13} 
r_{a} 
\left( e^{ - i \Delta_{31} x } -1 \right) 
\nonumber \\
&+& 
\biggl[
c^3_{13} c^3_{12} 
\left( - c_{23} s_{12} - s_{23} s_{13} c_{12} e^{ - i \delta} 
\right) 
+ c^3_{13} s^3_{12} 
\left( c_{23} c_{12} - s_{23} s_{13} s_{12} e^{ - i \delta} \right) 
e^{ - i \Delta_{21} x } 
\nonumber \\
&+& 
s_{23} c_{13} s^3_{13} e^{ - i \delta} e^{ - i \Delta_{31} x } 
\biggr] 
( - i \Delta_{a} x ).
\label{DMP-Semu-leading}
\end{eqnarray}
Noticing that by using the leading order property $r_{a} \simeq \frac{ \Delta_{a} }{ \Delta_{31} }$ and $r_{a}^{ \text{sol} } \simeq \frac{ \Delta_{a} }{ \Delta_{21} }$, the $\left( e^{ - i \Delta_{31} x } -1 \right) $ term in $S_{e \mu}^{(0)}$ in eq.~\eqref{DMP-Semu-leading} corresponds to the leading order term in the fourth line in eq.~\eqref{Semu-matter-P-1st}, ignoring the higher order $\frac{ \Delta_{a} }{ \Delta_{31} } \frac{ \Delta_{21} }{ \Delta_{31} }$ term in \eqref{fourth-term}. Therefore, the leading order term in $S_{e \mu}^{(0)}$ in \eqref{Semu-matter-P-1st} obtained by the first-order matter perturbation theory is reproduced by the DMP leading order amplitude. 

\subsection{$r_{a}$ correction and the DMP amplitude in the next to leading order}
\label{sec:DMP-matterP-1st}

Derivation of the next to leading order term involving $\frac{ \Delta_{a} }{ \Delta_{31} } \frac{ \Delta_{21} }{ \Delta_{31} }$ term in \eqref{fourth-term} is a little more complicated. It is easy to show that the first-order DMP correction in eq.~\eqref{S-emu-0th-1st} produces $c^2_{12}$ times the higher order $\frac{ \Delta_{a} }{ \Delta_{31} } \frac{ \Delta_{21} }{ \Delta_{31} }$ term in \eqref{fourth-term}. Then, we need the similar term, $s^2_{12}$ times the same term in \eqref{fourth-term} to reproduce, using $c^2_{12} + s^2_{12} = 1$, the desired higher order term in \eqref{fourth-term}. Though it may seem unlikely to happen, it indeed occurs. This is accomplished by adding $\frac{ \Delta_{21} }{ \Delta_{31} } \frac{ \Delta_{a} }{ \Delta_{31} }$ term that arises from the DMP's $r_{a}$ correction,  
$r_{a} \simeq \frac{ \Delta_{a} }{ \Delta_{31} } \left( 1 + s^2_{12} \frac{ \Delta_{21} }{ \Delta_{31} } \right)$, see eq.~\eqref{ra-def},\footnote{
%%%%%%%%%%%%%% footnote %%%%%%%%%%%%%%
Notice that there is no similar correction from $r_{a}^{ \text{sol} }$. It should also be noticed that $\epsilon r_{a}$ term which existed in $\lambda_{\pm}$ cancel against the similar term that comes from the $r_{a}$ correction. See appendix~\ref{sec:DMP-variables} for the definition of $\lambda_{\pm}$. As a consequence the wave factors have very simple expressions 
$e^{ - i h_{1} x } \simeq \left[ 1 - i c^2_{13} c^2_{12} \Delta_{a} x \right]$, 
$e^{ - i h_{2} x } \simeq e^{ - i \Delta_{21} x } \left[ 1 - i c^2_{13} s^2_{12} ( \Delta_{a} x ) \right]$, and 
$e^{ - i h_{3} x } \simeq e^{ - i \Delta_{31} x } \left[ 1 - i s^2_{13} ( \Delta_{a} x ) \right]$. 
}
in the third and fourth terms in eq.~\eqref{DMP-Semu-leading}, and the $\frac{1}{ h_{3} - h_{1} }$ term in the DMP first order correction:  
\begin{eqnarray} 
&& 
- c_{13} s_{13} s_{12} s^2_{12} 
\left( c_{23} s_{13} c_{12} + s_{23} \cos 2\theta_{13} s_{12} e^{ -i \delta } \right) 
\frac{ \Delta_{21} }{ \Delta_{31} } \frac{ \Delta_{a} }{ \Delta_{31} } 
\left( e^{ - i \Delta_{21} x } - 1 \right) 
\nonumber \\
&+&
c_{13} s_{13} s_{12} s^2_{12} 
\left( c_{23} s_{13} c_{12} + s_{23} \cos 2\theta_{13} s_{12} e^{ - i \delta} \right) 
\frac{ \Delta_{21} }{ \Delta_{31} } \frac{ \Delta_{a} }{ \Delta_{31} }
\left( e^{ - i \Delta_{31} x } - 1 \right) 
\nonumber \\
&=& 
c_{13} s_{13} s_{12} s^2_{12} 
\left( c_{23} s_{13} c_{12} + s_{23} \cos 2\theta_{13} s_{12} e^{ - i \delta} \right) 
\frac{ \Delta_{21} }{ \Delta_{31} } \frac{ \Delta_{a} }{ \Delta_{31} }
\left( e^{ - i \Delta_{31} x } - e^{ - i \Delta_{21} x } \right).
\label{s12square-term}
\end{eqnarray}
Notice that the first term of \eqref{s12square-term} is from the pure $r_{a}$ correction and the second term by combining the first order DMP correction and the $r_{a}$ correction. 

Thus, the $\nu_{\mu} - \nu_{e}$ amplitude obtained to first order in the matter perturbation theory is reproduced by the DMP framework to order $\frac{ \Delta_{21} }{ \Delta_{31} } \frac{ \Delta_{a} }{ \Delta_{31} }$. The success in the relatively simple manner above is due to a desirable property of the DMP framework that the effective expansion parameter involves $\sin ( \phi - \theta_{13} ) \simeq c_{13} s_{13} r_{a}$ which makes the near-vacuum limit transparent, as emphasized in ref.~\cite{Denton:2016wmg}. 

\subsection{DMP amplitude decomposition with infinitesimal matter potential}

We briefly discuss the DMP amplitude decomposition with the $e^{ i h_{1} x }$ rephasing. Since the near vacuum limit does not distinguish the mass orderings, we do not treat the $e^{ i h_{2} x }$ rephasing. But, they both lead to the same expressions of the probabilities, as they differ only by the $S$ matrix phase.

The amplitude decomposition formulas at the zeroth and the first order are given in eq.~\eqref{DMP-decompose-emu-1st-h1}. After taking the similar near vacuum limit the decomposed amplitudes read 
\begin{eqnarray} 
&&
\left( S_{e \mu}^{ \text{atm} } \right)^{(0)} 
= 
s_{23} c_{13} s_{13} e^{ - i \delta} \left( e^{ - i \Delta_{31} x } - 1 \right) 
\nonumber \\
&+& 
s_{23} c_{13} s_{13} \cos 2\theta_{13} e^{ - i \delta} r_{a} 
\left( e^{ - i \Delta_{31} x } - 1 \right) 
+ s_{23} c_{13} s_{13} e^{ - i \delta} 
( s^2_{13} - c^2_{13} c^2_{12} ) ( - i \Delta_{a} x ) e^{ - i \Delta_{31} x }, 
\nonumber \\
&&
\left( S_{e \mu}^{ \text{sol} } \right)^{(0)} 
= 
c_{13} s_{12} 
\left( c_{23} c_{12} - s_{23} s_{13} s_{12} e^{ - i \delta} \right) 
\left( e^{ - i \Delta_{21} x } - 1 \right) 
\nonumber \\
&& \hspace{-12mm} 
+ 
\left[ 
- c_{13} s_{13} s_{12} 
\left( c_{23} s_{13} c_{12} 
+ s_{23} \cos 2\theta_{13} s_{12} e^{ - i \delta} \right) r_{a} 
+ c^3_{13} c_{12} s_{12} 
\left( c_{23} \cos 2\theta_{12} 
- s_{23} s_{13} \sin 2\theta_{12} 
e^{ - i \delta} \right) r_{a}^{ \text{sol} } 
\right] 
\nonumber \\
&\times& 
\left( e^{ - i \Delta_{21} x } - 1 \right) 
\nonumber \\
&-& 
c^3_{13} s_{12} \cos 2\theta_{12} 
\left( c_{23} c_{12} - s_{23} s_{13} s_{12} e^{ - i \delta} \right) 
( - i \Delta_{a} x ) e^{ - i \Delta_{21} x }. 
\label{amp-decompose-emu-0th-matterP}
\end{eqnarray}
As we have learned in section~\ref{sec:DMP-matterP-0th}, we need to supply the order 
$\frac{ \Delta_{21} }{ \Delta_{31} } \frac{ \Delta_{a} }{ \Delta_{31} }$ term, but it can be done in exactly the same way as in section~\ref{sec:DMP-matterP-1st}. Namely the term comes from the $r_{a}$ correction and the first order DMP amplitudes. Then, we recover the order 
$\frac{ \Delta_{21} }{ \Delta_{31} } \frac{ \Delta_{a} }{ \Delta_{31} }$ term as seen in eq.~\eqref{fourth-term}, which can be distributed to the atmospheric and the solar amplitudes.\footnote{
%%%%%%%%%%%%%%% footnote %%%%%%%%%%%%%%%%
An apparent minor problem is that a part of the $(- i \Delta_{a} x )$ terms is missing. But, it is because the rephasing can produce such terms when expanded, and hence it will not affect the oscillation probability. 
}

The investigation reported in this section was motivated by the question of how we can understand a subtle feature of amplitude decomposition with infinitesimal matter potential. The problematic term of the genuine mixed atmospheric and solar waves that exists in the first-order matter perturbation theory is recovered as a $\frac{ \Delta_{21} }{ \Delta_{31} } \frac{ \Delta_{a} }{ \Delta_{31} }$ term in the DMP perturbation theory. 
We have learned that in the near-vacuum limit of the DMP decomposition, the term (i.e., perturbatively recovered fourth term in eq.~\eqref{Semu-matter-P-1st}) is distributed to the atmospheric and solar amplitudes according to the wave factors $\left( e^{ - i \Delta_{31} x } - 1 \right)$ and $\left( e^{ - i \Delta_{21} x } - 1 \right)$. Therefore, the answer to the question of how to understand the subtle term from the viewpoint of DMP decomposition is that the wave factor decomposition defined in sections~\ref{sec:principle} and \ref{sec:DMP-decomposition} transcends the energy denominator confusion.

\section{Some remarks toward phenomenology} 
\label{sec:remarks}

Though doing phenomenology is beyond the scope of this paper, it may be appropriate to add a few remarks toward experimental analyses of the interference effects in the ongoing and upcoming neutrino experiments. In section~\ref{sec:introduction}, we gave a tentative argument to indicate that detection of the atmospheric-solar interference might be easier than observing CP violation. It is extremely interesting to see if this expectation is true, for example, by doing the combined analysis of the T2K and NO$\nu$A data~\cite{Abe:2019vii,Acero:2019ksn}. 

Then, what is next? An interesting topic would be the diagnostics of the interference effect. The main ingredient in the interference term appears to be oscillation channel dependent: While the CP-phase effect is dominant in the $\nu_{\mu} \rightarrow \nu_{e}$ channel, CP-phase independent terms are the majority in the $\nu_{\mu} \rightarrow \nu_{\tau}$ and $\nu_{\mu} \rightarrow \nu_{\mu}$ channels~\cite{Minakata:2020ijz}. 
If one wants to understand possible interplay between the CP conserving and CP violating terms, an exploration of the $\nu_{\mu} \rightarrow \nu_{e}$ channel may be most feasible. In this case, given dominance of CP phase effect in the interference, precision measurement is required. In this sense we are facing with an interesting and exciting time, just at stepping into the precision era of neutrino experiments with the muon-neutrino superbeams, T2HK and DUNE~\cite{Abe:2018uyc,Abi:2020evt}. 
If the low-energy extensions of the astrophysical neutrino experiments \cite{TheIceCube-Gen2:2016cap,Adrian-Martinez:2016zzs,Aartsen:2020fwb} measure $P(\nu_{\mu} \rightarrow \nu_{\tau})$ in a reasonable accuracy, it could offer another interesting opportunity for diagnosing the atmospheric-solar interference effect. 

The disappearance channels play a unique role in discussion of the interference effect. In ref.~\cite{Huber:2019frh}, we have analyzed the medium-baseline reactor neutrino experiment JUNO~\cite{An:2015jdp} to uncover the sensitivity of detection of the interference term, which turned out to be excellent, higher than 4$\sigma$. It utilizes the $\bar{\nu}_{e} \rightarrow \bar{\nu}_{e}$ channel, and hence it is purely non-CP phase effect even in matter~\cite{Kuo:1987km,Minakata:1999ze}. The role of the other disappearance channel, $\nu_{\mu} \rightarrow \nu_{\mu}$, on the diagnostics of the interference term is not yet investigated, which may shed new light on this problem. The amplitude decomposition in the $\nu_{e} \rightarrow \nu_{e}$ and $\nu_{\mu} \rightarrow \nu_{\mu}$ channels will be discussed in appendix~\ref{sec:amp-decompose-disapp}. 

The $\nu_{\mu} \rightarrow \nu_{\tau}$ channel discussed in appendix~\ref{sec:amp-decompose-mu-tau} is a very interesting one with favorable property that the interference effect is dominated by the CP-phase free part of the probability, which is in sharp contrast to the feature in the $\nu_{\mu} \rightarrow \nu_{e}$ channel~\cite{Minakata:2020ijz}.\footnote{
%%%%%%%%%%%%%% footnote %%%%%%%%%%%%%%
We do not discuss the $\nu_{e} \rightarrow \nu_{\tau}$ channel because the oscillation probability can be obtained by using the $\theta_{23}$ transformation 
$P(\nu_e \rightarrow \nu_\tau) = P(\nu_e \rightarrow \nu_\mu: c_{23} \rightarrow - s_{23}, s_{23} \rightarrow c_{23})$~\cite{Akhmedov:2004ny}, where $P(\nu_e \rightarrow \nu_\mu)$ is given by T-conjugate of $P(\nu_{\mu} \rightarrow \nu_{e})$.
}
Though it may not be so easy, statistical separation of $\nu_{\tau}$ in the atmospheric neutrinos is already successful at Super-Kamiokande (SK) to show $\nu_{\tau}$ appearance at significance level of 4.6$\sigma$~\cite{Li:2017dbe}. The experience would help the IceCube-PINGU, and KM3NeT/ORCA to detect $\nu_{\tau}$ in the atmospheric neutrinos~\cite{TheIceCube-Gen2:2016cap,Adrian-Martinez:2016zzs}. 

The long-baseline beam experiments may be more powerful for precision measurement. The $\nu_{\tau}$ detection requires relatively high-energy beam, and the DUNE experiment is likely to be the best candidate for such measurement. See refs.~\cite{deGouvea:2019ozk,Machado:2020yxl,Ghoshal:2019pab} and the references cited therein. 
But, even assuming perfect knowledge of the neutrino flux, the observable quantity is the product of the oscillation probability and the cross sections. Therefore, only a single measurement cannot determine the both. Certainly we need experimentalists' help to go forward. 

A completely different aspect of application of the DMP decomposition to JUNO is a possible role played by the earth matter effect. Though it is small in JUNO~\cite{Li:2016txk}, at a level of 1\%, it must be taken into account when the accuracy of measurement becomes a percent level. Since JUNO observes both the solar- and the atmospheric-scale oscillations in the same energy region, most of the frameworks treated in ref.~\cite{Minakata:2020ijz} cannot do the job. Therefore, the best framework would be the DMP decomposition, or possibly, the solar-resonance perturbation theory~\cite{Martinez-Soler:2019nhb}.

\section{Concluding remarks}
\label{sec:conclusion}

In this paper, we have discussed the amplitude decomposition in matter started from the first principle. We found that the Denton {\it et al.} (DMP) perturbation theory~\cite{Denton:2016wmg} provides the ideal foundation for this purpose. It not only possesses the appropriate wave factor structure as the Zaglauer-Schwarzer (ZS) construction~\cite{Zaglauer:1988gz} dictates, but also allows physical interpretation of the dynamically independent two modes. Speaking more precisely, 
\begin{itemize}

\item 
The DMP decomposition is a very good approximation to the exact ZS decomposition. The numerical accuracy of the oscillation probability formulas has been checked and found to be the best among all the perturbative schemes available to date~\cite{Parke:2019vbs}.

\item 
Physical picture of the dynamically independent two modes can be drawn by explicitly showing the smooth continuity of the DMP decomposition to the regions of the atmospheric- and the solar-scale enhanced oscillations (section~\ref{sec:interpretation}). 

\end{itemize}
\noindent
Moreover, the DMP decomposition illuminates how the subtle case uncovered in ref.~\cite{Minakata:2020ijz} should be treated in the light of the general principle of amplitude decomposition, as discussed in section~\ref{sec:decomposition-matter-P}. It is the case with failure of the vacuum prescription for the system with infinitesimal matter potential mentioned in section~\ref{sec:overview}. 

In a process of formulating the DMP decomposition we have identified the cause of the $\psi$ symmetry~\cite{Denton:2016wmg} in the oscillation probability, see section~\ref{sec:psi-symmetry}, as due to the $S$ matrix rephasing invariance. The same understanding can be extended to the similar symmetries in the renormalized helio-perturbation theory and the solar-resonance perturbation theory. 

In passing, it may be appropriate to remark on a new picture for the DMP framework as a unified perturbation theory for neutrino oscillation in terrestrial matter. It has the various favorable properties such as having the effective expansion parameter $\epsilon c_{12} s_{12} s_{ (\phi - \theta_{13}) } \simeq \epsilon c_{12} s_{12} c_{13} s_{13} r_{a}$ which makes the near-vacuum limit particularly transparent~\cite{Denton:2016wmg}. In section~\ref{sec:interpretation} we have checked that the DMP perturbation theory reproduces the frameworks known as the solar-resonance perturbation theory~\cite{Martinez-Soler:2019nhb}, and the atmospheric-resonance perturbation theory~\cite{Arafune:1996bt,Cervera:2000kp,Freund:2001pn,Akhmedov:2004ny,Minakata:2015gra}, by explicitly analyzing the limits to the respective regions of their validity. This property strongly suggests the picture of DMP perturbation theory as a unified perturbative framework whose region of validity spans the whole region covered by the terrestrial experiments. Here, the terrestrial experiments' region implies the neutrino energy-baseline covered by the atmospheric neutrino observation e.g., in Super-Kamiokande \cite{Jiang:2019xwn}. For a pictorial view of such terrestrial experiments' region, see e.g., Fig 1 in ref.~\cite{Minakata:2019gyw}.

Turning back to the amplitude decomposition in matter, upon setting up the prescription for the DMP decomposition we have derived all the expressions of the decomposed oscillation probabilities in all the relevant oscillation channels: 
For the $\nu_{\mu} \rightarrow \nu_{e}$ channel, see section~\ref{sec:DMP-amplitude-decomposition}, and for the $\nu_{e} \rightarrow \nu_{e}$ and $\nu_{\mu} \rightarrow \nu_{\mu}$ channels appendix~\ref{sec:amp-decompose-disapp}, and for the $\nu_{\mu} \rightarrow \nu_{\tau}$ channel appendix~\ref{sec:amp-decompose-mu-tau}. These expressions must be sufficient to perform the quantitative analyses of the experimental data to uncover the atmospheric-solar wave interference. 

Some of the readers may be reluctant by seeing no detailed phenomenological discussion in this paper. Though this is true, we must emphasize that we are making progress in a step by step manner. We are about to complete the first and second stages of the analysis strategy which consists of the following three stages: 
(1) To establish the principle and the prescriptions for the amplitude decomposition, 
(2) To derive all the necessary formulas which are required in the analyses in all the oscillation channels, and (3) To perform phenomenological analyses, and the data analyses in parallel, if possible. We believe that the real progress can only be made by taking the solid steps in each stage. 

Yet, there exists many unanswered theoretical questions. We lack discussions for physical properties in the various channels, for example, in the disappearance channels except for the $\bar{\nu}_{e} - \bar{\nu}_{e}$ channel treated in vacuum~\cite{Huber:2019frh}. In this paper we did not address the physical picture of the two dynamical modes outside the terrestrial-experiments favored kinematical phase space, e.g., super-high energy or super-dense matter. Since the system is effectively two-flavor we expect that the physics there is likely to be simple. We also did not enter into the problem of varying matter density. We hope that we can come back to these questions in the near future. 

Finally, we speculate that observation of neutrino interference in more generic sense might add a new page in quantum mechanics. Since the neutrino wave packet has macroscopic size, see the estimate e.g., in ref.~\cite{Minakata:2012kg}, it might contribute to deeper understanding of quantum mechanics. Even though it may be regarded as ``everyday physics'' nowadays, we  should bear in mind that quantum mechanics still offers a cradle of a new fundamental and technological advances \cite{Q-computation}.

\acknowledgments

The author thanks Patrick Huber and Rebekah Pestes for fruitful collaborations and stimulating discussions, with whom the project for detecting the interference effect has been started while I was visiting Virginia Tech, Blacksburg, VA for a year. He is grateful to Pilar Coloma for triggering his interest in the ZS amplitude decomposition. 
After start working on this project, we have received the numerous encouraging comments from people including Alexei Smirnov, Samoil Bilenky, Sandhya Choubey, Peter Denton, John Learned, Stephen Parke, Orlando Peres, Serguey Petcov, and Masashi Yokoyama. They continued to be the generous motive force for this work done during the Covid-19 isolation. 

\appendix 

\section{DMP-variables: Summary}
\label{sec:DMP-variables}

We summarize here the formulas for the eigenvalues and the mixing angles in matter. As a building block we need the expressions of $\lambda_{\mp}$ and $\lambda_{0}$, the eigenvalues of $2E$ times the zeroth-order Hamiltonian, that result from $\phi$ rotation which covers both mass orderings, the NMO ($\Delta m^2_{ \text{ren} } > 0$) and IMO ($\Delta m^2_{ \text{ren} } < 0$): 
\begin{eqnarray} 
\lambda_{-} &=& 
\frac{ 1 }{ 2 } \left[
\left( \Delta m^2_{ \text{ren} } + a \right) - {\rm sign}(\Delta m^2_{ \text{ren} }) \sqrt{ \left( \Delta m^2_{ \text{ren} } - a \right)^2 + 4 s^2_{13} a \Delta m^2_{ \text{ren} } }
\right] 
+ \epsilon \Delta m^2_{ \text{ren} } s^2_{12},
\nonumber\\
\lambda_{0} &=&  c^2_{12} ~\epsilon ~\Delta m^2_{ \text{ren} },
\label{lambda-pm0}  
\\
\lambda_{+} &=& 
\frac{ 1 }{ 2 } \left[
\left( \Delta m^2_{ \text{ren} } + a \right) + {\rm sign}(\Delta m^2_{ \text{ren} }) \sqrt{ \left( \Delta m^2_{ \text{ren} } - a \right)^2 + 4 s^2_{13} a \Delta m^2_{ \text{ren} } }
\right] 
+ \epsilon \Delta m^2_{ \text{ren} } s^2_{12}.
\nonumber 
\end{eqnarray}
This rotation is common in both refs.~\cite{Minakata:2015gra} and \cite{Denton:2016wmg}. With use of eq.~\eqref{lambda-pm0} and $r_{a} = \frac{a}{ \Delta m^2_{ \text{ren} } }$, the eigenvalue differences and the angle $\phi$ has the same form for the both mass ordering: 
\begin{eqnarray} 
&& 
\lambda_{+} - \lambda_{-} 
= \Delta m^2_{ \text{ren} } 
\sqrt{ 1 + r_{a}^2 - 2 r_{a} \cos 2\theta_{13} }, 
\nonumber\\
&&
\lambda_{-} - \lambda_{0} 
= \frac{ \Delta m^2_{ \text{ren} } }{ 2 } 
\left[ 1 + r_{a} - \sqrt{ 1 + r_{a}^2 - 2 r_{a} \cos 2\theta_{13} } \right] 
- \epsilon \Delta m^2_{ \text{ren} } \cos 2\theta_{12}, 
\nonumber\\
&&
\lambda_{+} - \lambda_{0} 
= \frac{ \Delta m^2_{ \text{ren} } }{ 2 } 
\left[ 1 + r_{a} + \sqrt{ 1 + r_{a}^2 - 2 r_{a} \cos 2\theta_{13} } \right] 
- \epsilon \Delta m^2_{ \text{ren} } \cos 2\theta_{12}. 
\label{eigenvalue-diff}
\end{eqnarray} 
\begin{eqnarray} 
\cos 2 \phi &=& 
\frac{ \Delta m^2_{ \text{ren} } \cos 2\theta_{13} - a }{ \lambda_{+} - \lambda_{-} } 
= \frac{ \cos 2\theta_{13} - r_{a} }{ 
\sqrt{ 1 + r_{a}^2 - 2 r_{a} \cos 2\theta_{13} } }, 
\nonumber \\
\sin 2 \phi &=& \frac{ \Delta m^2_{ \text{ren} } \sin 2\theta_{13} }{ \lambda_{+} - \lambda_{-} } 
= \frac{ \sin 2\theta_{13} }{ 
\sqrt{ 1 + r_{a}^2 - 2 r_{a} \cos 2\theta_{13} } }. 
\label{2phi}
\end{eqnarray}
When $a$ is varied from $- \infty$ to $+ \infty$, $\cos 2 \phi $ varies from -1 to +1 in the NMO and +1 to -1 in the IMO. In the both extremes $a \sim \pm \infty$ $\sin 2 \phi \sim 0$. Then, $\phi$ varies from 0 to $\frac{\pi}{2}$ (from $\frac{\pi}{2}$ to 0) when $a$ is varied from $- \infty$ to $+ \infty$ in the case of NMO (IMO). It agrees with Fig.~1 in ref.~\cite{Denton:2016wmg}.

After $\psi$ rotation we obtain the expressions of the DMP eigenvalues 
\begin{eqnarray} 
&& 
\lambda_{1}, \lambda_{2} 
= 
\frac{1}{2} 
\left[ ( \lambda_{-} + \lambda_{0} ) 
\mp \sqrt{ ( \lambda_{-} - \lambda_{0} )^2 
+ 4 A^2 } 
\right], 
\nonumber \\
&& \lambda_{3} = \lambda_{+},
\label{DMP-eigenvalues}
\end{eqnarray}
where $A \equiv \epsilon c_{12} s_{12} c_{\phi - \theta_{13}} \Delta m^2_{ \text{ren} }$, 
and the $\psi$ angle 
\begin{eqnarray}
&& 
\sin 2 \psi = 
\frac{ 2A }{ \sqrt{ ( \lambda_{-} - \lambda_{0} )^2 + 4 A^2 } } 
\simeq
\frac{ \pm 2 \epsilon \sin 2\theta_{12} c_{\phi - \theta_{13}} }
{ \left[ 1 + r_{a} - \sqrt{ 1 + r_{a}^2 - 2 r_{a} \cos 2\theta_{13} } \right] } + \mathcal{O} (\epsilon^2),
\nonumber \\
&& 
\cos 2 \psi = 
- \frac{ ( \lambda_{-} - \lambda_{0} ) }{ \sqrt{ ( \lambda_{-} - \lambda_{0} )^2 + 4 A^2 } } 
\simeq \mp 1 + \mathcal{O} (\epsilon^2), 
\label{2psi}
\end{eqnarray}
where the $\epsilon$ expanded values are quoted after $\simeq$ equality, and the upper and lower signs are for the case of NMO and IMO, respectively. Therefore, $\psi \simeq \frac{\pi}{2}$ ($\psi \simeq 0$)  around the atmospheric resonance in the NMO (IMO). This is again consistent with Fig.~1 in \cite{Denton:2016wmg}. 

\section{Zeroth and first order $S$ matrix elements}
\label{sec:S-matrix-0th-1st}

\subsection{Zeroth-order $S$ matrix elements}
\label{sec:S-matrix-0th}

With the zeroth-order $\check{S}$ matrix $\check{S}^{(0)} = e^{-i \check{H}_{0} x}$, the zeroth-order flavor basis $S$ matrix elements read 
\begin{eqnarray} 
S_{ee}^{(0)} &=& 
c^2_{\phi} \left( c^2_{\psi} e^{ - i h_{1} x } + s^2_{\psi} e^{ - i h_{2} x } \right)  + s^2_{\phi} e^{ - i h_{3} x },
\nonumber \\
S_{e \mu}^{(0)} &=& 
c_{23} c_{\phi} c_{\psi} s_{\psi} 
\left( e^{ - i h_{2} x } - e^{ - i h_{1} x } \right) 
- s_{23} c_{\phi} s_{\phi} e^{ - i \delta} 
\left( c^2_{\psi} e^{ - i h_{1} x } + s^2_{\psi} e^{ - i h_{2} x } - e^{ - i h_{3} x } \right), 
\nonumber \\ 
S_{e \tau}^{(0)} &=& 
- c_{23} c_{\phi} s_{\phi} 
\left( c^2_{\psi} e^{ - i h_{1} x } + s^2_{\psi} e^{ - i h_{2} x } - e^{ - i h_{3} x } \right) 
- s_{23} c_{\phi} c_{\psi} s_{\psi} e^{ i \delta} 
\left( e^{ - i h_{2} x } - e^{ - i h_{1} x } \right),
\nonumber \\
S_{\mu e}^{(0)} &=& 
c_{23} c_{\phi} c_{\psi} s_{\psi} 
\left( e^{ - i h_{2} x } - e^{ - i h_{1} x } \right) 
- s_{23} c_{\phi} s_{\phi} e^{ i \delta} 
\left( c^2_{\psi} e^{ - i h_{1} x } + s^2_{\psi} e^{ - i h_{2} x } 
- e^{ - i h_{3} x } \right),
\nonumber \\
S_{\mu \mu}^{(0)} &=& 
c^2_{23} 
\left( s^2_{\psi} e^{ - i h_{1} x } + c^2_{\psi} e^{ - i h_{2} x } \right) 
+ s^2_{23} 
\left\{ s^2_{\phi} \left( c^2_{\psi} e^{ - i h_{1} x } + s^2_{\psi} e^{ - i h_{2} x } \right) 
+ c^2_{\phi} e^{ - i h_{3} x } \right\} 
\nonumber \\
&-&
2 c_{23} s_{23} s_{\phi} c_{\psi} s_{\psi} \cos \delta 
\left( e^{ - i h_{2} x } - e^{ - i h_{1} x } \right),
\nonumber \\
S_{\mu \tau}^{(0)} &=& 
- \left( c^2_{23} - s^2_{23} e^{ 2 i \delta} \right) 
s_{\phi} c_{\psi} s_{\psi} \left( e^{ - i h_{2} x } - e^{ - i h_{1} x } \right) 
\nonumber \\
&+&
c_{23} s_{23} e^{ i \delta} 
\left[ 
s^2_{\phi} \left( c^2_{\psi} e^{ - i h_{1} x } + s^2_{\psi} e^{ - i h_{2} x } \right) 
+ c^2_{\phi} e^{ - i h_{3} x } 
- \left( s^2_{\psi} e^{ - i h_{1} x } + c^2_{\psi} e^{ - i h_{2} x } \right) 
\right],
\nonumber \\ 
S_{\tau e}^{(0)} &=& 
- c_{23} c_{\phi} s_{\phi} 
\left( c^2_{\psi} e^{ - i h_{1} x } + s^2_{\psi} e^{ - i h_{2} x } 
- e^{ - i h_{3} x } \right) 
- s_{23} c_{\phi} c_{\psi} s_{\psi} e^{ - i \delta} 
\left( e^{ - i h_{2} x } - e^{ - i h_{1} x } \right),
\nonumber \\ 
S_{\tau \mu}^{(0)} &=& 
- \left( c^2_{23} - s^2_{23} e^{ - 2 i \delta} \right) 
s_{\phi} c_{\psi} s_{\psi} \left( e^{ - i h_{2} x } - e^{ - i h_{1} x } \right) 
\nonumber \\
&+&
c_{23} s_{23} e^{ - i \delta} 
\left[ 
s^2_{\phi} \left( c^2_{\psi} e^{ - i h_{1} x } + s^2_{\psi} e^{ - i h_{2} x } \right) 
+ c^2_{\phi} e^{ - i h_{3} x } 
- \left( s^2_{\psi} e^{ - i h_{1} x } + c^2_{\psi} e^{ - i h_{2} x } \right) 
\right],
\nonumber \\
S_{\tau \tau}^{(0)} &=& 
s^2_{23} \left( s^2_{\psi} e^{ - i h_{1} x } + c^2_{\psi} e^{ - i h_{2} x } \right) 
+ c^2_{23} \left\{ s^2_{\phi} \left( c^2_{\psi} e^{ - i h_{1} x } + s^2_{\psi} e^{ - i h_{2} x } \right) 
+ c^2_{\phi} e^{ - i h_{3} x } \right\}
\nonumber \\
&+&
2 c_{23} s_{23} \cos \delta 
s_{\phi} c_{\psi} s_{\psi} \left( e^{ - i h_{2} x } - e^{ - i h_{1} x } \right). 
\label{flavor-S-0th}
\end{eqnarray}

\subsection{First-order $S$ matrix elements}
\label{sec:S-matrix-1st}

What is good in the DMP framework is the extreme simplicity of the perturbed Hamiltonian in eq.~\eqref{check-Hamiltonian-0th-1st}. Because of it the computed $S$ matrix elements only contains the two functions $\hat{S}_{13}$ and $\hat{S}_{23}$ of showing the wave property, 
\begin{eqnarray} 
&& 
\hat{S}_{13} 
= 
\epsilon c_{12} s_{12} s_{ (\phi - \theta_{13}) }
c_{\psi} s_{\psi} 
\left[ \frac{ \Delta_{ \text{ren} } }{ h_{3} - h_{2} } 
\left( e^{ - i h_{3} x } - e^{ - i h_{2} x } \right) 
- \frac{ \Delta_{ \text{ren} } }{ h_{3} - h_{1} } 
\left( e^{ - i h_{3} x } - e^{ - i h_{1} x } \right) 
\right], 
\nonumber \\
&& 
\hat{S}_{23} 
= 
\epsilon c_{12} s_{12} s_{ (\phi - \theta_{13}) } 
\left[
c^2_{\psi} \frac{ \Delta_{ \text{ren} } }{ h_{3} - h_{2} } 
\left( e^{ - i h_{3} x } - e^{ - i h_{2} x } \right) 
+ s^2_{\psi} \frac{ \Delta_{ \text{ren} } }{ h_{3} - h_{1} } 
\left( e^{ - i h_{3} x } - e^{ - i h_{1} x } \right) 
\right]. 
\nonumber \\
\label{hat-S-functions}
\end{eqnarray}
The first order flavor-basis $S$ matrix elements have the following simple expressions with $\hat{S}_{13}$ and $\hat{S}_{23}$ as 
\begin{eqnarray} 
S_{ee}^{(1)} 
&=& \sin 2\phi \hat{S}_{13}
\nonumber \\
S_{e \mu}^{(1)}  
&=& 
s_{23} \cos 2\phi e^{ - i \delta} \hat{S}_{13}  
+ c_{23} s_{\phi} \hat{S}_{23} 
\nonumber \\ 
S_{e \tau}^{(1)}  
&=& 
c_{23} \cos 2\phi \hat{S}_{13} 
- s_{23} s_{\phi} e^{ i \delta} \hat{S}_{23} 
\nonumber \\
S_{\mu e}^{(1)}  
&=& 
s_{23} \cos 2\phi e^{ i \delta} \hat{S}_{13} 
+ c_{23} s_{\phi} \hat{S}_{23} 
\nonumber \\
S_{\mu \mu}^{(1)}  
&=& 
- s^2_{23} \sin 2\phi \hat{S}_{13} 
+ \sin 2 \theta_{23} c_{\phi} \cos \delta \hat{S}_{23} 
\nonumber \\
S_{\mu \tau}^{(1)}  
&=& 
- c_{23} s_{23} \sin 2\phi e^{ i \delta} \hat{S}_{13} 
+ c_{\phi} \left( c^2_{23} - s^2_{23} e^{ 2 i \delta} \right) \hat{S}_{23} 
\nonumber \\ 
S_{\tau e}^{(1)}  
&=& 
c_{23} \cos 2\phi \hat{S}_{13} 
- s_{23} s_{\phi} e^{ - i \delta} \hat{S}_{23} 
\nonumber \\ 
S_{\tau \mu}^{(1)}  
&=& 
- c_{23} s_{23} \sin 2\phi e^{ - i \delta} \hat{S}_{13} 
+ c_{\phi} \left( c^2_{23} - s^2_{23} e^{ - 2 i \delta} \right) \hat{S}_{23} 
\nonumber \\
S_{\tau \tau}^{(1)} 
&=& 
- c^2_{23} \sin 2\phi \hat{S}_{13} 
- \sin 2\theta_{23} c_{\phi} \cos \delta \hat{S}_{23} 
\label{Flavor-S-elements-1st}
\end{eqnarray}

\section{The oscillation probability $P(\nu_{\mu} \rightarrow \nu_{e})$ by the $S$ matrix method}
\label{sec:DMP-probability-mue} 

Given the formulas for the DMP $S$ matrix elements in the $\nu_{\mu} \rightarrow \nu_{e}$ channel in the zeroth and first-order in eq.~\eqref{S-emu-0th-1st}, it is straightforward to derive the zeroth and first-order oscillation probabilities in the traditional way, i.e., without performing amplitude decomposition:
\begin{eqnarray} 
&&
P(\nu_{\mu} \rightarrow \nu_{e})^{(0)} 
= \vert S_{e \mu}^{(0)} \vert^2
\nonumber \\ 
&=& 
4 c^2_{\phi} c^2_{\psi} s^2_{\psi} 
\left( c^2_{23} - s^2_{23} s^2_{\phi} \right) 
\sin^2 \frac{ ( h_{2} - h_{1} ) x }{2} 
+ 4 s^2_{23} c^2_{\phi} s^2_{\phi} \left\{ c^2_{\psi} \sin^2 \frac{ ( h_{3} - h_{1} ) x }{2} 
+ s^2_{\psi} \sin^2 \frac{ ( h_{3} - h_{2} ) x }{2} \right\} 
\nonumber \\ 
&-& 
4 c_{23} s_{23} c^2_{\phi} s_{\phi} c_{\psi} s_{\psi} \cos \delta 
\left\{ 
\sin^2 \frac{ ( h_{3} - h_{2} ) x }{2} 
- \sin^2 \frac{ ( h_{3} - h_{1} ) x }{2} 
- \cos 2\psi \sin^2 \frac{ ( h_{2} - h_{1} ) x }{2} 
\right\} 
\nonumber \\ 
&-& 
8 c_{23} s_{23} c^2_{\phi} s_{\phi} c_{\psi} s_{\psi} \sin \delta 
\sin \frac{ ( h_{3} - h_{1} ) x }{2} \sin \frac{ ( h_{2} - h_{1} ) x }{2} 
\sin \frac{ ( h_{3} - h_{2} ) x }{2}. 
\label{total-P-mue-0th}
\end{eqnarray}
\begin{eqnarray} 
&&
P(\nu_{\mu} \rightarrow \nu_{e})^{(1)} 
= 2 \mbox{Re} \left[ \left( S_{e \mu}^{(0)} \right)^* S_{e \mu}^{(1)} \right] 
\nonumber \\ 
&=& 
- \frac{ \Delta_{ \text{ren} } }{ h_{3} - h_{2} } 
4 \epsilon c_{23} c_{12} s_{12} s_{ (\phi - \theta_{13}) } c_{\phi} c_{\psi} s_{\psi} 
\left( c_{23} s_{\phi} c^2_{\psi} + s_{23} \cos 2\phi c_{\psi} s_{\psi} \cos \delta \right) 
\nonumber \\ 
&\times& 
\left\{ \sin^2 \frac{ ( h_{3} - h_{2} ) x }{2} 
- \sin^2 \frac{ ( h_{3} - h_{1} ) x }{2} 
+ \sin^2 \frac{ ( h_{2} - h_{1} ) x }{2} \right\} 
\nonumber \\ 
&+& 
\frac{ \Delta_{ \text{ren} } }{ h_{3} - h_{1} } 
4 \epsilon c_{23} c_{12} s_{12} s_{ (\phi - \theta_{13}) } c_{\phi} c_{\psi} s_{\psi} 
\left( c_{23} s_{\phi} s^2_{\psi} - s_{23} \cos 2\phi c_{\psi} s_{\psi} \cos \delta \right) 
\nonumber \\ 
&\times&
\left\{ - \sin^2 \frac{ ( h_{3} - h_{2} ) x }{2} 
+ \sin^2 \frac{ ( h_{3} - h_{1} ) x }{2} 
+ \sin^2 \frac{ ( h_{2} - h_{1} ) x }{2} \right\} 
\nonumber \\ 
&+& 
\frac{ \Delta_{ \text{ren} } }{ h_{3} - h_{2} } 
4 \epsilon s_{23} c_{12} s_{12} s_{ (\phi - \theta_{13}) } c_{\phi} s_{\phi} 
\left( c_{23} s_{\phi} c^2_{\psi} \cos \delta + s_{23} \cos 2\phi c_{\psi} s_{\psi} \right) 
\nonumber \\ 
&\times&
\left\{ ( 1 + s^2_{\psi} ) \sin^2 \frac{ ( h_{3} - h_{2} ) x }{2} 
+ c^2_{\psi} \sin^2 \frac{ ( h_{3} - h_{1} ) x }{2} 
- c^2_{\psi} \sin^2 \frac{ ( h_{2} - h_{1} ) x }{2} \right\} 
\nonumber \\ 
&+& 
\frac{ \Delta_{ \text{ren} } }{ h_{3} - h_{1} } 
4 \epsilon s_{23} c_{12} s_{12} s_{ (\phi - \theta_{13}) } c_{\phi} s_{\phi} 
\left( c_{23} s_{\phi} s^2_{\psi} \cos \delta - s_{23} \cos 2\phi c_{\psi} s_{\psi} \right) 
\nonumber \\ 
&\times&
\left\{ s^2_{\psi} \sin^2 \frac{ ( h_{3} - h_{2} ) x }{2} 
+ ( 1 + c^2_{\psi} ) \sin^2 \frac{ ( h_{3} - h_{1} ) x }{2} 
- s^2_{\psi} \sin^2 \frac{ ( h_{2} - h_{1} ) x }{2} \right\} 
\nonumber \\ 
&+& 
8 \epsilon c_{23} s_{23} c_{12} s_{12} s_{ (\phi - \theta_{13}) } c_{\phi} \sin \delta 
\left\{
c^2_{\psi} \left( s^2_{\phi} - c^2_{\phi} s^2_{\psi} \right) 
\frac{ \Delta_{ \text{ren} } }{ h_{3} - h_{2} } 
- 
s^2_{\psi} \left( s^2_{\phi} - c^2_{\phi} c^2_{\psi} \right) 
\frac{ \Delta_{ \text{ren} } }{ h_{3} - h_{1} } 
\right\}
\nonumber \\ 
&\times&
\sin \frac{ ( h_{3} - h_{1} ) x }{2} \sin \frac{ ( h_{2} - h_{1} ) x }{2} 
\sin \frac{ ( h_{3} - h_{2} ) x }{2}.
\label{total-P-mue-1st}
\end{eqnarray}
We note that the $\psi$ symmetry, see section~\ref{sec:psi-symmetry}, holds in eqs.~\eqref{total-P-mue-0th} and \eqref{total-P-mue-1st}. 

\section{Amplitude decomposition in the disappearance channels}
\label{sec:amp-decompose-disapp} 

In the main body of the text in this paper we have concentrated on the $\nu_{\mu} \rightarrow \nu_{e}$ channel to focus on the conceptual issue related to the question, ``what is the correct way of performing the amplitude decomposition in matter?''. In this and the following appendices, we discuss the amplitude decomposition in some of the remaining channels, the $\nu_{e} \rightarrow \nu_{e}$ and $\nu_{\mu} \rightarrow \nu_{\mu}$ channels (appendix~\ref{sec:amp-decompose-disapp}), and the $\nu_{\mu} \rightarrow \nu_{\tau}$ channel (appendix~\ref{sec:amp-decompose-mu-tau}). 

Among them, the $\nu_{\mu} \rightarrow \nu_{\mu}$ channel may be most feasible experimentally, apart from the $\bar{\nu}_{e} \rightarrow \bar{\nu}_{e}$ measurement in JUNO. Once a long-baseline neutrino experiment start to operate, it will make measurement in both the $\nu_{\mu} \rightarrow \nu_{\mu}$ and $\nu_{\mu} \rightarrow \nu_{e}$ channels simultaneously. But, since the $\nu_{e} - \nu_{e}$ channel is the simplest one to discuss, we start by treating this channel. The $\bar{\nu}_{e} \rightarrow \bar{\nu}_{e}$ probability can be obtained from the $P(\nu_{e} \rightarrow \nu_{e})$ by just flipping the sign of the matter potential $a$, as CP phase $\delta$ is absent. 
Measurement of $P(\nu_{e} \rightarrow \nu_{e})$ in sizable matter effect might be possible by using the atmospheric neutrino observation 
\cite{Jiang:2019xwn,Abe:2018uyc,Abi:2020evt,TheIceCube-Gen2:2016cap,Adrian-Martinez:2016zzs,Aartsen:2020fwb}, or a neutrino factory~\cite{Geer:1997iz,DeRujula:1998umv}, assuming electron detection capability in the detector. 
Some of these apparatus, including the others e.g., T2HK and DUNE~\cite{Abe:2018uyc,Abi:2020evt}, have capabilities to measure the $\nu_{\mu} \rightarrow \nu_{\mu}$, $\nu_{\mu} \rightarrow \nu_{e}$, and its T conjugate channels.

\subsection{General formula for the disappearance probability with use of the decomposed amplitudes}
\label{sec:general-formula}

In the disappearance channels the amplitude decomposition has a ``unity'' in $S_{\alpha \alpha} = 1 + S_{\alpha \alpha}^{ \text{atm} } + S_{\alpha \alpha}^{ \text{sol} }$. As a consequence the expressions of the oscillation probability with use of the decomposed amplitudes are different from those in the appearance channel exhibited in section~\ref{sec:decomposition-mue-h2}. Therefore, it may be useful to present the general formula for the probability. They read in the zeroth and the first orders as 
\begin{eqnarray} 
P(\nu_{\alpha} \rightarrow \nu_{\alpha})^{(0)} 
&=&
1 + 2 \mbox{Re} \left[ \left( S_{\alpha \alpha}^{ \text{atm} } \right)^{(0)} \right] 
+ 2 \mbox{Re} \left[ \left( S_{\alpha \alpha}^{ \text{sol} } \right)^{(0)} \right] 
\nonumber \\
&+& 
\biggl | \left( S_{\alpha \alpha}^{ \text{atm} } \right)^{(0)} \biggr |^2 
+ \biggl | \left( S_{\alpha \alpha}^{ \text{sol} } \right)^{(0)} \biggr |^2
+ 2 \mbox{Re} \left[ 
\left\{ \left( S_{\alpha \alpha}^{ \text{atm} } \right)^{(0)} \right\}^* 
\left( S_{\alpha \alpha}^{ \text{sol} } \right)^{(0)} \right], 
\nonumber 
\end{eqnarray}
\begin{eqnarray} 
&& 
P(\nu_{\alpha} \rightarrow \nu_{\alpha})^{(1)} 
\nonumber \\
&=& 
2 \mbox{Re} \left[ 
\left( S_{\alpha \alpha}^{ \text{atm} } \right)^{(1)} \right] 
+ 2 \mbox{Re} \left[ 
\left( S_{\alpha \alpha}^{ \text{sol} } \right)^{(1)} \right] 
+ 2 \mbox{Re} \left[ 
\left\{ \left( S_{\alpha \alpha}^{ \text{atm} } \right)^{(0)} \right\}^* 
\left( S_{\alpha \alpha}^{ \text{atm} } \right)^{(1)} \right] 
+ 2 \mbox{Re} \left[ 
\left\{ \left( S_{\alpha \alpha}^{ \text{sol} } \right)^{(0)} \right\}^* 
\left( S_{\alpha \alpha}^{ \text{sol} } \right)^{(1)} \right] 
\nonumber \\
&+&
2 \mbox{Re} \left[ 
\left\{ \left( S_{\alpha \alpha}^{ \text{atm} } \right)^{(0)} \right\}^* 
\left( S_{\alpha \alpha}^{ \text{sol} } \right)^{(1)} \right] 
+ 2 \mbox{Re} \left[ 
\left\{ \left( S_{\alpha \alpha}^{ \text{sol} } \right)^{(0)} \right\}^* 
\left( S_{\alpha \alpha}^{ \text{atm} } \right)^{(1)} \right]. 
\label{general-formula-disapp}
\end{eqnarray}

\subsection{The decomposed amplitudes and probabilities in the $\nu_{e} - \nu_{e}$ channel}
\label{sec:decomposition-ee}

With the $e^{ i h_{2} x}$ rephasing, the decomposed amplitudes in the $\nu_{e} - \nu_{e}$ channel in the zeroth- and first-order DMP expansion read 
\begin{eqnarray} 
&&
\left( S_{ee}^{ \text{atm} } \right)^{(0)} 
= s^2_{\phi} \left( e^{ - i ( h_{3} - h_{2} ) x } - 1 \right), 
\nonumber \\
&& 
\left( S_{ee}^{ \text{sol} } \right)^{(0)} 
= c^2_{\phi} c^2_{\psi} \left( e^{ i ( h_{2} - h_{1} ) x } - 1 \right), 
\label{DMP-decompose-ee-0th-h2}
\end{eqnarray}
\begin{eqnarray} 
&&
\left( S_{ee}^{ \text{atm} } \right)^{(1)} 
= 
\epsilon c_{12} s_{12} s_{ (\phi - \theta_{13}) } c_{\psi} s_{\psi} 
\left( \frac{ \Delta_{ \text{ren} } }{ h_{3} - h_{2} } 
- \frac{ \Delta_{ \text{ren} } }{ h_{3} - h_{1} } \right) 
\left( e^{ - i ( h_{3} - h_{2} ) x } - 1 \right), 
\nonumber \\
&& 
\left( S_{ee}^{ \text{sol} } \right)^{(1)} 
= 
\epsilon c_{12} s_{12} s_{ (\phi - \theta_{13}) } c_{\psi} s_{\psi} 
\frac{ \Delta_{ \text{ren} } }{ h_{3} - h_{1} } \left( e^{ i ( h_{2} - h_{1} ) x } - 1 \right). 
\label{DMP-decompose-ee-1st-h2}
\end{eqnarray}
Similarly, the decomposed amplitudes with the $e^{ i h_{1} x}$ rephasing read 
\begin{eqnarray} 
&&
\left( S_{ee}^{ \text{atm} } \right)^{(0)} 
= s^2_{\phi} \left( e^{ - i ( h_{3} - h_{1} ) x } - 1 \right) 
\nonumber \\
&& 
\left( S_{ee}^{ \text{sol} } \right)^{(0)} 
= c^2_{\phi} s^2_{\psi} \left( e^{ - i ( h_{2} - h_{1} ) x } - 1 \right) 
\label{DMP-decompose-ee-0th-h1}
\end{eqnarray}
\begin{eqnarray} 
&&
\left( S_{ee}^{ \text{atm} } \right)^{(1)} 
= 
\epsilon c_{12} s_{12} s_{ (\phi - \theta_{13}) } c_{\psi} s_{\psi} 
\left( \frac{ \Delta_{ \text{ren} } }{ h_{3} - h_{2} } 
- \frac{ \Delta_{ \text{ren} } }{ h_{3} - h_{1} } \right) 
\left( e^{ - i ( h_{3} - h_{1} ) x } - 1 \right), 
\nonumber \\
&& 
\left( S_{ee}^{ \text{sol} } \right)^{(1)} 
= 
- \epsilon c_{12} s_{12} s_{ (\phi - \theta_{13}) } c_{\psi} s_{\psi} 
\frac{ \Delta_{ \text{ren} } }{ h_{3} - h_{2} } 
\left( e^{ - i ( h_{2} - h_{1} ) x } - 1 \right).
\label{DMP-decompose-ee-1st-h1}
\end{eqnarray}
It is obvious that the amplitudes $S_{ee}^{ \text{atm} }$ and $S_{ee}^{ \text{sol} }$ with the $e^{ i h_{2} x}$ and $e^{ i h_{1} x}$ rephasing are related with each other by the $\psi$ transformation (see section~\ref{sec:psi-symmetry}) in the both zeroth and first order. 

Therefore, we present below the expressions of the decomposed oscillation probabilities with the $e^{ i h_{2} x}$ rephasing only. The expressions with the $e^{ i h_{1} x}$ rephasing can be obtained by the $\psi$ transformation~\eqref{psi-transformation}. They read in the zeroth order 
\begin{eqnarray} 
&& 
\left[ P(\nu_{e} \rightarrow \nu_{e})^{(0)} \right]^{ \text{non-int-fer} } 
= 
1 - 4 c^2_{\phi} s^2_{\phi} \sin^2 \frac{ ( h_{3} - h_{2} ) x }{2} 
- 4 c^2_{\phi} c^2_{\psi} ( s^2_{\phi} + c^2_{\phi} s^2_{\psi} ) \sin^2 \frac{ ( h_{2} - h_{1} ) x }{2}, 
\nonumber \\
&& 
\left[ P(\nu_{e} \rightarrow \nu_{e})^{(0)} \right]^{ \text{int-fer} } 
= 
4 c^2_{\phi} s^2_{\phi} c^2_{\psi} 
\left\{ 
\sin^2 \frac{ ( h_{3} - h_{2} ) x }{2}
- \sin^2 \frac{ ( h_{3} - h_{1} ) x }{2}
+ \sin^2 \frac{ ( h_{2} - h_{1} ) x }{2}
\right\}, 
\nonumber \\
\label{decomposed-P-ee-0th-h2}
\end{eqnarray}
and in the first order\footnote{
%%%%%%%%%%%%%% footnote %%%%%%%%%%%%%%%
In the published version of this paper, ref.~\cite{Minakata:2020oxb}, an overall factor $\sin 2 \phi$ has been missed in the first order expressions of the non-interference and interference parts of $P(\nu_{e} \rightarrow \nu_{e})^{(1)}$ in eq.~\eqref{decomposed-P-ee-1st-h2}, which is corrected here.  }
\begin{eqnarray} 
&&
\left[ P(\nu_{e} \rightarrow \nu_{e})^{(1)} \right]^{ \text{non-int-fer} } 
\nonumber \\
&=&
- 4 \epsilon c_{12} s_{12} s_{ (\phi - \theta_{13}) } \cos 2\phi \sin 2\phi c_{\psi} s_{\psi} 
\left( \frac{ \Delta_{ \text{ren} } }{ h_{3} - h_{2} } 
- \frac{ \Delta_{ \text{ren} } }{ h_{3} - h_{1} } \right) 
\sin^2 \frac{ ( h_{3} - h_{2} ) x }{2} 
\nonumber \\
&-& 
4 \epsilon c_{12} s_{12} s_{ (\phi - \theta_{13}) } c_{\psi} s_{\psi} 
\left( s^2_{\psi} - \cos 2\phi c^2_{\psi} \right) 
\frac{ \Delta_{ \text{ren} } }{ h_{3} - h_{1} } 
\sin^2 \frac{ ( h_{2} - h_{1} ) x }{2}, 
\nonumber \\
&&
\left[ P(\nu_{e} \rightarrow \nu_{e})^{(1)} \right]^{ \text{int-fer} } 
\nonumber \\
&=&
4 \epsilon c_{12} s_{12} s_{ (\phi - \theta_{13}) } \sin 2\phi c_{\psi} s_{\psi} 
\left\{ 
c^2_{\phi} c^2_{\psi} \frac{ \Delta_{ \text{ren} } }{ h_{3} - h_{2} } 
+ ( s^2_{\phi} - c^2_{\phi} c^2_{\psi} ) \frac{ \Delta_{ \text{ren} } }{ h_{3} - h_{1} } 
\right\} 
\nonumber \\
&\times&
\left\{ 
\sin^2 \frac{ ( h_{3} - h_{2} ) x }{2}
- \sin^2 \frac{ ( h_{3} - h_{1} ) x }{2}
+ \sin^2 \frac{ ( h_{2} - h_{1} ) x }{2}
\right\}. 
\label{decomposed-P-ee-1st-h2}
\end{eqnarray}
It is verified by an explicit computation of the probability in the decomposition with the $e^{ i h_{1} x}$ rephasing that the expressions of $P(\nu_{e} \rightarrow \nu_{e})^{ \text{non-int-fer} }$ and $P(\nu_{e} \rightarrow \nu_{e})^{ \text{int-fer} }$ to first order agree with the ones obtained by using the $\psi$ transformation of eqs.~\eqref{decomposed-P-ee-0th-h2} and \eqref{decomposed-P-ee-1st-h2}.

\subsection{The decomposed amplitudes and probabilities in the $\nu_{\mu} - \nu_{\mu}$ channel}
\label{sec:decomposition-mumu}

As in the $\nu_{e} - \nu_{e}$ channel we first compare the decomposed amplitudes in the $\nu_{\mu} - \nu_{\mu}$ channel between the $e^{ i h_{2} x}$ and $e^{ i h_{2} x}$ rephasing in the zeroth- and first-order DMP expansion. With the $e^{ i h_{2} x}$ rephasing they read 
\begin{eqnarray} 
&&
\left( S_{\mu \mu}^{ \text{atm} } \right)^{(0)} 
= s^2_{23} c^2_{\phi} \left( e^{ - i ( h_{3} - h_{2} ) x } - 1 \right), 
\nonumber \\
&& 
\left( S_{\mu \mu}^{ \text{sol} } \right)^{(0)} 
= 
\left( c^2_{23} s^2_{\psi} + s^2_{23} s^2_{\phi} c^2_{\psi} 
+ 2 c_{23} s_{23} s_{\phi} c_{\psi} s_{\psi} \cos \delta \right) 
\left( e^{ i ( h_{2} - h_{1} ) x } - 1 \right), 
\label{DMP-decompose-mumu-0th-h2}
\end{eqnarray}
\begin{eqnarray} 
\left( S_{\mu \mu}^{ \text{atm} } \right)^{(1)} 
&=& 
\epsilon c_{12} s_{12} s_{ (\phi - \theta_{13}) } 
\left( - s^2_{23} \sin 2\phi c_{\psi} s_{\psi} + \sin 2 \theta_{23} c_{\phi} c^2_{\psi} \cos \delta \right) 
\frac{ \Delta_{ \text{ren} } }{ h_{3} - h_{2} } 
\left( e^{ - i ( h_{3} - h_{2} ) x } - 1 \right) 
\nonumber \\
&+& 
\epsilon c_{12} s_{12} s_{ (\phi - \theta_{13}) } 
\left( s^2_{23} \sin 2\phi c_{\psi} s_{\psi} + \sin 2 \theta_{23} c_{\phi} s^2_{\psi} \cos \delta \right) 
\frac{ \Delta_{ \text{ren} } }{ h_{3} - h_{1} } \left( e^{ - i ( h_{3} - h_{2} ) x } - 1 \right), 
\nonumber \\
\left( S_{\mu \mu}^{ \text{sol} } \right)^{(1)} 
&=& 
- \epsilon c_{12} s_{12} s_{ (\phi - \theta_{13}) } 
\left( s^2_{23} \sin 2\phi c_{\psi} s_{\psi} + \sin 2 \theta_{23} c_{\phi} s^2_{\psi} \cos \delta \right) 
\frac{ \Delta_{ \text{ren} } }{ h_{3} - h_{1} } 
\left( e^{ i ( h_{2} - h_{1} ) x } - 1 \right). 
\nonumber \\
\label{DMP-decompose-mumu-1st-h2}
\end{eqnarray}
With the $e^{ i h_{1} x}$ rephasing they have the forms 
\begin{eqnarray} 
&&
\left( S_{\mu \mu}^{ \text{atm} } \right)^{(0)} 
= s^2_{23} c^2_{\phi} \left( e^{ - i ( h_{3} - h_{1} ) x } - 1 \right), 
\nonumber \\
&& 
\left( S_{\mu \mu}^{ \text{sol} } \right)^{(0)} 
= 
\left( c^2_{23} c^2_{\psi} + s^2_{23} s^2_{\phi} s^2_{\psi} - 2 c_{23} s_{23} s_{\phi} c_{\psi} s_{\psi} \cos \delta \right) 
\left( e^{ - i ( h_{2} - h_{1} ) x } - 1 \right), 
\label{DMP-decompose-mumu-0th-h1}
\end{eqnarray}
\begin{eqnarray} 
\left( S_{\mu \mu}^{ \text{atm} } \right)^{(1)} 
&=&
\epsilon c_{12} s_{12} s_{ (\phi - \theta_{13}) } 
\left( - s^2_{23} \sin 2\phi c_{\psi} s_{\psi} + \sin 2 \theta_{23} c_{\phi} c^2_{\psi} \cos \delta 
\right) 
\frac{ \Delta_{ \text{ren} } }{ h_{3} - h_{2} } 
\left( e^{ - i ( h_{3} - h_{1} ) x } - 1 \right) 
\nonumber \\
&+& 
\epsilon c_{12} s_{12} s_{ (\phi - \theta_{13}) } 
\left( s^2_{23} \sin 2\phi c_{\psi} s_{\psi} + \sin 2 \theta_{23} c_{\phi} s^2_{\psi} \cos \delta 
\right) 
\frac{ \Delta_{ \text{ren} } }{ h_{3} - h_{1} } 
\left( e^{ - i ( h_{3} - h_{1} ) x } - 1 \right), 
\nonumber \\
\left( S_{\mu \mu}^{ \text{sol} } \right)^{(1)} 
&=&
\epsilon c_{12} s_{12} s_{ (\phi - \theta_{13}) } 
\left( s^2_{23} \sin 2\phi c_{\psi} s_{\psi} - \sin 2 \theta_{23} c_{\phi} c^2_{\psi} \cos \delta \right) 
\frac{ \Delta_{ \text{ren} } }{ h_{3} - h_{2} } 
\left( e^{ - i ( h_{2} - h_{1} ) x } - 1 \right). 
\nonumber \\
\label{DMP-decompose-mumu-1st-h1}
\end{eqnarray}
We notice again that the decomposed amplitudes with the $e^{ i h_{1} x}$ rephasing can be obtained by the $\psi$ transformation from the ones with the $e^{ i h_{2} x}$ rephasing in the both zeroth and first order. Therefore, as in the $\nu_{e} - \nu_{e}$ channel, we present here only the expressions of the decomposed probabilities obtained with the $e^{ i h_{2} x}$ rephasing. They read in the zeroth and first order as 
\begin{eqnarray} 
\left[ P(\nu_{\mu} \rightarrow \nu_{\mu})^{(0)} \right]^{ \text{non-int-fer} } 
&=& 
1 - 4 s^2_{23} c^2_{\phi} \left( c^2_{23} + s^2_{23} s^2_{\phi} \right) 
\sin^2 \frac{ ( h_{3} - h_{2} ) x }{2} 
\nonumber \\
&-& 
4 \left( c^2_{23} s^2_{\psi} + s^2_{23} s^2_{\phi} c^2_{\psi} 
+ 2 c_{23} s_{23} s_{\phi} c_{\psi} s_{\psi} \cos \delta \right) 
\nonumber \\&\times& 
\left\{ c^2_{23} c^2_{\psi} + s^2_{23} ( c^2_{\phi} + s^2_{\phi} s^2_{\psi} )
- 2 c_{23} s_{23} s_{\phi} c_{\psi} s_{\psi} \cos \delta \right\}
\sin^2 \frac{ ( h_{2} - h_{1} ) x }{2}, 
\nonumber \\
\left[ P(\nu_{\mu} \rightarrow \nu_{\mu})^{(0)} \right]^{ \text{int-fer} } 
&=& 
4 s^2_{23} c^2_{\phi} 
\left( c^2_{23} s^2_{\psi} + s^2_{23} s^2_{\phi} c^2_{\psi} 
+ 2 c_{23} s_{23} s_{\phi} c_{\psi} s_{\psi} \cos \delta \right) 
\nonumber \\
&\times& 
\left\{ 
\sin^2 \frac{ ( h_{3} - h_{2} ) x }{2}
- \sin^2 \frac{ ( h_{3} - h_{1} ) x }{2}
+ \sin^2 \frac{ ( h_{2} - h_{1} ) x }{2} 
\right\}, 
\label{decomposed-P-mumu-0th-h2}
\end{eqnarray}

\begin{eqnarray} 
&& 
\left[ P(\nu_{\mu} \rightarrow \nu_{\mu})^{(1)} \right]^{ \text{non-int-fer} } 
\nonumber \\
&=& 
4 \epsilon c_{12} s_{12} s_{ (\phi - \theta_{13}) } 
\left( c^2_{23} - s^2_{23} \cos 2\phi \right) 
\left( s^2_{23} \sin 2\phi s_{\psi} - \sin 2 \theta_{23} c_{\phi} c_{\psi} \cos \delta \right) 
c_{\psi} \frac{ \Delta_{ \text{ren} } }{ h_{3} - h_{2} } 
\sin^2 \frac{ ( h_{3} - h_{2} ) x }{2} 
\nonumber \\
&-& 
4 \epsilon c_{12} s_{12} s_{ (\phi - \theta_{13}) } 
\left( c^2_{23} - s^2_{23} \cos 2\phi \right) 
\left( s^2_{23} \sin 2\phi c_{\psi} + \sin 2 \theta_{23} c_{\phi} s_{\psi} \cos \delta \right)
s_{\psi} \frac{ \Delta_{ \text{ren} } }{ h_{3} - h_{1} } 
\sin^2 \frac{ ( h_{3} - h_{2} ) x }{2} 
\nonumber \\
&+& 
4 \epsilon c_{12} s_{12} s_{ (\phi - \theta_{13}) } 
\left( s^2_{23} \sin 2\phi c_{\psi} + \sin 2 \theta_{23} c_{\phi} s_{\psi} \cos \delta \right) 
\nonumber \\
&\times& 
\left[ 
s^2_{23} c^2_{\phi} 
+ \left( c^2_{23} - s^2_{23} s^2_{\phi} \right) \cos 2\psi 
- 4 c_{23} s_{23} s_{\phi} c_{\psi} s_{\psi} \cos \delta 
\right] 
s_{\psi} \frac{ \Delta_{ \text{ren} } }{ h_{3} - h_{1} } \sin^2 \frac{ ( h_{2} - h_{1} ) x }{2},
\nonumber \\
&& 
\left[ P(\nu_{\mu} \rightarrow \nu_{\mu})^{(1)} \right]^{ \text{int-fer} } 
\nonumber \\
&=& 
4 \epsilon c_{12} s_{12} s_{ (\phi - \theta_{13}) } 
\left( - s^2_{23} \sin 2\phi s_{\psi} + \sin 2 \theta_{23} c_{\phi} c_{\psi} \cos \delta \right) 
\left( c^2_{23} s^2_{\psi} + s^2_{23} s^2_{\phi} c^2_{\psi} 
+ 2 c_{23} s_{23} s_{\phi} c_{\psi} s_{\psi} \cos \delta \right) 
\nonumber \\
&\times& 
c_{\psi} \frac{ \Delta_{ \text{ren} } }{ h_{3} - h_{2} } 
\left\{ - \sin^2 \frac{ ( h_{3} - h_{1} ) x }{2} 
+ \sin^2 \frac{ ( h_{3} - h_{2} ) x }{2} 
+ \sin^2 \frac{ ( h_{2} - h_{1} ) x }{2} 
\right\} 
\nonumber \\
&+& 
4 \epsilon c_{12} s_{12} s_{ (\phi - \theta_{13}) } 
\left( s^2_{23} \sin 2\phi c_{\psi} + \sin 2 \theta_{23} c_{\phi} s_{\psi} \cos \delta \right) 
\left\{ c^2_{23} s^2_{\psi} 
+ s^2_{23} ( s^2_{\phi} c^2_{\psi} - c^2_{\phi} ) 
+ 2 c_{23} s_{23} s_{\phi} c_{\psi} s_{\psi} \cos \delta 
\right\} 
\nonumber \\
&\times& 
s_{\psi} \frac{ \Delta_{ \text{ren} } }{ h_{3} - h_{1} } 
\left\{ - \sin^2 \frac{ ( h_{3} - h_{1} ) x }{2} 
+ \sin^2 \frac{ ( h_{3} - h_{2} ) x }{2} 
+ \sin^2 \frac{ ( h_{2} - h_{1} ) x }{2} \right\}. 
\label{decomposed-P-mumu-1st-h2}
\end{eqnarray} 
Again it is explicitly checked that the $\psi$ transformation on eqs.~\eqref{decomposed-P-mumu-0th-h2} and \eqref{decomposed-P-mumu-1st-h2} agree with the explicitly computed  decomposed probabilities using the amplitudes with the $e^{ i h_{1} x}$ rephasing, as they should. 

\section{Amplitude decomposition in the $\nu_{\mu} \rightarrow \nu_{\tau}$ channel}
\label{sec:amp-decompose-mu-tau} 

Finally, we discuss the $\nu_{\mu} \rightarrow \nu_{\tau}$ channel. Though the measurement of the oscillation probability would be tough this is very interesting channel for the diagnostics of the interference effect, as mentioned in section~\ref{sec:remarks}. 

\subsection{The decomposed amplitudes and probabilities in the $\nu_{\mu} - \nu_{\tau}$ channel}
\label{sec:decomposition-mutau}

We first show that the decomposed amplitudes in the $\nu_{\mu} - \nu_{\tau}$ channel with the $e^{ i h_{2} x }$ rephasing and the ones with the $e^{ i h_{1} x }$ rephasing are related with each other by the $\psi$ transformation. The former read as follows:
\begin{eqnarray} 
&&
\left( S_{\tau \mu}^{ \text{atm} } \right)^{(0)} 
= c_{23} s_{23} c^2_{\phi} e^{ - i \delta} 
\left( e^{ - i ( h_{3} - h_{2} ) x } - 1 \right), 
\nonumber \\
&&
\left( S_{\tau \mu}^{ \text{sol} } \right)^{(0)} 
= e^{ - i \delta} \left\{
\left( c^2_{23} e^{ i \delta} - s^2_{23} e^{ - i \delta} \right) s_{\phi} c_{\psi} s_{\psi} 
+ c_{23} s_{23} \left( s^2_{\phi} c^2_{\psi} - s^2_{\psi} \right) \right\} 
\left( e^{ i ( h_{2} - h_{1} ) x } - 1 \right), 
\nonumber \\
\label{DMP-decompose-mutau-0th-h2}
\end{eqnarray}
\begin{eqnarray} 
&&
\left( S_{\tau \mu}^{ \text{atm} } \right)^{(1)} 
\nonumber \\
&=& 
\epsilon c_{12} s_{12} s_{ (\phi - \theta_{13}) } e^{ - i \delta} 
\left\{
- c_{23} s_{23} \sin 2\phi c_{\psi} s_{\psi} 
+ \left( c^2_{23} e^{ i \delta} - s^2_{23} e^{ - i \delta} \right) c_{\phi} c^2_{\psi} \right\} 
\frac{ \Delta_{ \text{ren} } }{ h_{3} - h_{2} } 
\left( e^{ - i ( h_{3} - h_{2} ) x } - 1 \right) 
\nonumber \\
&+& 
\epsilon c_{12} s_{12} s_{ (\phi - \theta_{13}) } e^{ - i \delta}
\left\{ 
c_{23} s_{23} \sin 2\phi c_{\psi} s_{\psi} 
+ \left( c^2_{23} e^{ i \delta} - s^2_{23} e^{ - i \delta} \right) c_{\phi} s^2_{\psi} 
\right\} 
\frac{ \Delta_{ \text{ren} } }{ h_{3} - h_{1} } 
\left( e^{ - i ( h_{3} - h_{2} ) x } - 1 \right), 
\nonumber \\
&& 
\left( S_{\tau \mu}^{ \text{sol} } \right)^{(1)} 
\nonumber \\
&=& 
- \epsilon c_{12} s_{12} s_{ (\phi - \theta_{13}) } e^{ - i \delta} 
\left\{
c_{23} s_{23} 
\sin 2\phi c_{\psi} s_{\psi} 
+ \left( c^2_{23} e^{ i \delta} - s^2_{23} e^{ - i \delta} \right) 
c_{\phi} s^2_{\psi} 
\right\} 
\frac{ \Delta_{ \text{ren} } }{ h_{3} - h_{1} } 
\left( e^{ i ( h_{2} - h_{1} ) x } - 1 \right). 
\nonumber \\
\label{DMP-decompose-mutau-1st-h2}
\end{eqnarray}
While the amplitudes with the $e^{ i h_{1} x }$ rephasing read 
\begin{eqnarray} 
&&
\left( S_{\tau \mu}^{ \text{atm} } \right)^{(0)} 
=
c_{23} s_{23} c^2_{\phi} e^{ - i \delta} \left( e^{ - i ( h_{3} - h_{1} ) x } - 1 \right), 
\nonumber \\
&&
\left( S_{\tau \mu}^{ \text{sol} } \right)^{(0)} 
=
e^{ - i \delta} \left\{ 
- \left( c^2_{23} e^{ i \delta} - s^2_{23} e^{ - i \delta} \right) 
s_{\phi} c_{\psi} s_{\psi} 
+ c_{23} s_{23} \left( s^2_{\phi} s^2_{\psi} - c^2_{\psi} \right) 
\right\} 
\left( e^{ - i ( h_{2} - h_{1} ) x } - 1 \right), 
\nonumber \\
\label{DMP-decompose-mutau-0th-h1}
\end{eqnarray} 
\begin{eqnarray} 
&&
\left( S_{\tau \mu}^{ \text{atm} } \right)^{(1)} 
\nonumber \\
&=& 
\epsilon c_{12} s_{12} s_{ (\phi - \theta_{13}) } e^{ - i \delta} 
\left\{ 
- c_{23} s_{23} \sin 2\phi c_{\psi} s_{\psi} 
+ \left( c^2_{23} e^{ i \delta} - s^2_{23} e^{ - i \delta} \right) c_{\phi} c^2_{\psi} 
\right\} 
\frac{ \Delta_{ \text{ren} } }{ h_{3} - h_{2} } 
\left( e^{ - i ( h_{3} - h_{1} ) x } - 1 \right) 
\nonumber \\
&+& 
\epsilon c_{12} s_{12} s_{ (\phi - \theta_{13}) } e^{ - i \delta} 
\left\{ 
c_{23} s_{23} \sin 2\phi c_{\psi} s_{\psi} 
+ \left( c^2_{23} e^{ i \delta} - s^2_{23} e^{ - i \delta} \right) c_{\phi} s^2_{\psi} 
\right\} 
\frac{ \Delta_{ \text{ren} } }{ h_{3} - h_{1} } 
\left( e^{ - i ( h_{3} - h_{1} ) x } - 1 \right), 
\nonumber \\
&&
\left( S_{\tau \mu}^{ \text{sol} } \right)^{(1)} 
\nonumber \\
&=&
\epsilon c_{12} s_{12} s_{ (\phi - \theta_{13}) } e^{ - i \delta} 
\left\{ 
c_{23} s_{23} \sin 2\phi c_{\psi} s_{\psi} 
- \left( c^2_{23} e^{ i \delta} - s^2_{23} e^{ - i \delta} \right) c_{\phi} c^2_{\psi} 
\right\} 
\frac{ \Delta_{ \text{ren} } }{ h_{3} - h_{2} } 
\left( e^{ - i ( h_{2} - h_{1} ) x } - 1 \right). 
\nonumber \\
\label{DMP-decompose-mutau-1st-h1}
\end{eqnarray} 
One can easily verify that they are related by the $\psi$ transformation. Then, we just present the expressions of the decomposed probabilities with the $e^{ i h_{2} x }$ rephasing: 
\begin{eqnarray} 
&& 
\left[ P(\nu_{\mu} \rightarrow \nu_{\tau})^{(0)} \right]^{ \text{non-int-fer} }
\nonumber \\
&=& 
4 c^2_{23} s^2_{23} c^4_{\phi} 
\sin^2 \frac{ ( h_{3} - h_{2} ) x }{2} 
\nonumber \\
&+& 
\biggl\{ 
\sin^2 2\theta_{23} ( s^4_{\psi} + s^4_{\phi} c^4_{\psi} ) 
+ 4 \cos^2 2\theta_{23} s^2_{\phi} c^2_{\psi} s^2_{\psi} 
\nonumber \\
&+& 
8 c_{23} s_{23} s_{\phi} c_{\psi} s_{\psi} \cos 2\theta_{23} 
\left( s^2_{\phi} c^2_{\psi} - s^2_{\psi} \right) \cos \delta 
- 8 c^2_{23} s^2_{23} s^2_{\phi} c^2_{\psi} s^2_{\psi} \cos 2\delta 
\biggr\}
\sin^2 \frac{ ( h_{2} - h_{1} ) x }{2}, 
\nonumber \\
&& 
\left[ P(\nu_{\mu} \rightarrow \nu_{\tau})^{(0)} \right]^{ \text{int-fer} }
\nonumber \\
&=& 
\left\{ 4 c_{23} s_{23} c^2_{\phi} c_{23} s_{23} \left( s^2_{\phi} c^2_{\psi} - s^2_{\psi} \right) 
+ 4 c_{23} s_{23} c^2_{\phi} s_{\phi} c_{\psi} s_{\psi} \cos 2\theta_{23} \cos \delta \right\} 
\nonumber \\
&\times&
\left\{ 
\sin^2 \frac{ ( h_{3} - h_{2} ) x }{2}
- \sin^2 \frac{ ( h_{3} - h_{1} ) x }{2} 
+ \sin^2 \frac{ ( h_{2} - h_{1} ) x }{2} 
\right\} 
\nonumber \\
&+& 
8 c_{23} s_{23} c^2_{\phi} s_{\phi} c_{\psi} s_{\psi} \sin \delta 
\sin \frac{ ( h_{3} - h_{1} ) x }{2} \sin \frac{ ( h_{2} - h_{1} ) x }{2} 
\sin \frac{ ( h_{3} - h_{2} ) x }{2}, 
\label{decomposed-P-mutau-0th-h2}
\end{eqnarray}
\begin{eqnarray} 
&& 
\left[ P(\nu_{\mu} \rightarrow \nu_{\tau})^{(1)} \right]^{ \text{non-int-fer} } 
\nonumber \\
&=& 
8 \epsilon c_{12} s_{12} s_{ (\phi - \theta_{13}) } c_{23} s_{23} c^2_{\phi} 
\biggl[ 
\left\{
- c_{23} s_{23} \sin 2\phi s_{\psi} + \cos 2\theta_{23} c_{\phi} c_{\psi} \cos \delta \right\} 
c_{\psi} \frac{ \Delta_{ \text{ren} } }{ h_{3} - h_{2} } 
\nonumber \\
&+&
\left\{ 
c_{23} s_{23} \sin 2\phi c_{\psi} + \cos 2\theta_{23} c_{\phi} s_{\psi} \cos \delta 
\right\} 
s_{\psi} \frac{ \Delta_{ \text{ren} } }{ h_{3} - h_{1} } 
\biggr] 
\sin^2 \frac{ ( h_{3} - h_{2} ) x }{2} 
\nonumber \\
&-& 
8 \epsilon c_{12} s_{12} s_{ (\phi - \theta_{13}) } 
%\nonumber \\&\times& 
\biggl\{ 
\left[ 1 - 2 c^2_{23} s^2_{23} ( 1 + \cos 2\delta ) \right] 
c_{\phi} s_{\phi} c_{\psi} s^2_{\psi} 
+ c^2_{23} s^2_{23} \sin 2\phi \left( s^2_{\phi} c^2_{\psi} - s^2_{\psi} \right) c_{\psi} 
\nonumber \\
&+& 
c_{23} s_{23} \cos 2\theta_{23} s_{\psi} \cos \delta 
\left[ s_{\phi} \sin 2\phi c^2_{\psi} 
+ \left( s^2_{\phi} c^2_{\psi} - s^2_{\psi} \right) c_{\phi} \right] 
\biggr\} 
s_{\psi} \frac{ \Delta_{ \text{ren} } }{ h_{3} - h_{1} } 
\sin^2 \frac{ ( h_{2} - h_{1} ) x }{2}, 
\nonumber 
%\label{decomposed-P-mutau-1st-h2}
\end{eqnarray}
\begin{eqnarray} 
&& 
\left[ P(\nu_{\mu} \rightarrow \nu_{\tau})^{(1)} \right]^{ \text{int-fer} } 
\nonumber \\
&=& 
4 \epsilon c_{12} s_{12} s_{ (\phi - \theta_{13}) } 
%\nonumber \\&\times&
\biggl\{ 
\left[ 1 - 2 c^2_{23} s^2_{23} ( 1 + \cos 2\delta ) \right] 
c_{\phi} s_{\phi} c_{\psi} s^2_{\psi} 
- c^2_{23} s^2_{23} \sin 2\phi c_{\psi} 
\left[ \cos 2\phi + ( 1 + s^2_{\phi} ) s^2_{\psi} \right] 
\nonumber \\
&+&
c_{23} s_{23} \cos 2\theta_{23} s_{\psi} \cos \delta 
\left[ s_{\phi} \sin 2\phi c^2_{\psi} - c_{\phi} \left\{ \cos 2\phi + ( 1 + s^2_{\phi} ) s^2_{\psi} \right\} 
\right] 
\biggr\} 
\nonumber \\
&\times& 
s_{\psi} \frac{ \Delta_{ \text{ren} } }{ h_{3} - h_{1} } 
\left\{
- \sin^2 \frac{ ( h_{3} - h_{1} ) x }{2} 
+ \sin^2 \frac{ ( h_{3} - h_{2} ) x }{2} 
+ \sin^2 \frac{ ( h_{2} - h_{1} ) x }{2} 
\right\} 
\nonumber \\
&+& 
4 \epsilon c_{12} s_{12} s_{ (\phi - \theta_{13}) } 
%\nonumber \\&\times& 
\biggl\{ 
\left[ 1 - 2 c^2_{23} s^2_{23} ( 1 + \cos 2\delta ) \right] 
c_{\phi} s_{\phi} c^2_{\psi} s_{\psi} 
- c^2_{23} s^2_{23} \sin 2\phi s_{\psi} \left( s^2_{\phi} c^2_{\psi} - s^2_{\psi} \right) 
\nonumber \\
&+& 
c_{23} s_{23} \cos 2\theta_{23} c_{\psi} \cos \delta 
\left[ c_{\phi} \left( s^2_{\phi} c^2_{\psi} - s^2_{\psi} \right) 
- s_{\phi} \sin 2\phi s^2_{\psi} \right] 
\biggr\} 
\nonumber \\
&\times&
c_{\psi} \frac{ \Delta_{ \text{ren} } }{ h_{3} - h_{2} } 
\left\{
- \sin^2 \frac{ ( h_{3} - h_{1} ) x }{2} 
+ \sin^2 \frac{ ( h_{3} - h_{2} ) x }{2} 
+ \sin^2 \frac{ ( h_{2} - h_{1} ) x }{2} 
\right\} 
\nonumber \\
&-& 
8 \epsilon c_{23} s_{23} c_{\phi} s_{ (\phi - \theta_{13}) } c_{12} s_{12} \sin \delta 
\left[
\left( s^2_{\phi} - c^2_{\phi}s^2_{\psi} \right) c^2_{\psi} 
\frac{ \Delta_{ \text{ren} } }{ h_{3} - h_{2} } 
- \left( s^2_{\phi} - c^2_{\phi} c^2_{\psi} \right) s^2_{\psi} 
\frac{ \Delta_{ \text{ren} } }{ h_{3} - h_{1} } 
\right]
\nonumber \\
&\times&
\sin \frac{ ( h_{3} - h_{1} ) x }{2} \sin \frac{ ( h_{2} - h_{1} ) x }{2} 
\sin \frac{ ( h_{3} - h_{2} ) x }{2}. 
\label{decomposed-P-mutau-1st-h2}
\end{eqnarray}

The decomposed probabilities $P(\nu_{\beta} \rightarrow \nu_{\alpha})^{ \text{non-int-fer} }$ and $P(\nu_{\beta} \rightarrow \nu_{\alpha})^{ \text{int-fer} }$, if added, must equal to the oscillation probability calculated by the conventional $S$ matrix method, giving a way of checking the consistency of the calculations. In the $\nu_{\mu} - \nu_{e}$ channel, the latter is given in appendix~\ref{sec:DMP-probability-mue}, and the above consistency check is executed explicitly in both of the $e^{ i h_{2} x }$ and $e^{ i h_{1} x }$ rephasing, as mentioned in section~\ref{sec:DMP-amplitude-decomposition}. The similar consistency checks for the decomposed probabilities in the $\nu_{e} - \nu_{e}$, $\nu_{\mu} - \nu_{\mu}$, and the $\nu_{\mu} - \nu_{\tau}$ channels are executed as well with the both $e^{ i h_{2} x }$ and $e^{ i h_{1} x }$ rephasing to first order in the DMP perturbation. 

\section{Addendum: Some miscellaneous formulas}  
\label{sec:miscellaneous} 

This appendix~\ref{sec:miscellaneous}, which is added only for the arXiv version of this paper, contains a missing pieces of the formulas in the published version of this paper. Though they are not difficult to derive, we hope that it may be of some help for the readers who want to verify all the formulas for the decomposed $S$ matrix elements and the probabilities in this paper. 

The only lacking information in appendices~\ref{sec:amp-decompose-disapp} and \ref{sec:amp-decompose-mu-tau} are the expressions of the decomposed probabilities in the $e^{ - i h_{1} x }$ re-phasing, and the oscillation probability $P(\nu_{e} \rightarrow \nu_{e})$ to first order in the DMP expansion. The latter is calculated by using the conventional $S$ matrix method as explained in section~\ref{sec:DMP-formulation}, and it serves for consistency check of the calculation. Notice that verifying 
$P(\nu_{\beta} \rightarrow \nu_{\alpha}) 
= P(\nu_{\beta} \rightarrow \nu_{\alpha})^{ \text{non-int-fer} } 
+ P(\nu_{\beta} \rightarrow \nu_{\alpha})^{ \text{int-fer} }$ is not always trivial in the $\nu_{\mu} - \nu_{\tau}$ sector. 

\subsection{$\nu_{e}$ disappearance channel}
\label{sec:nue-nue}

The non-interference and interference terms in the probability $P(\nu_{e} \rightarrow \nu_{e})$ in the $e^{ - i h_{1} x }$ re-phasing are given in the zeroth and first order by\footnote{
%%%%%%%%%%%%%% footnote %%%%%%%%%%%%%%%
In arXiv version 2, the first order expressions of the non-interference and interference parts of $P(\nu_{e} \rightarrow \nu_{e})^{(1)}$ with the $e^{ - i h_{1} x }$ re-phasing given in eq.~\eqref{decomposed-P-ee-1st-h1}, and their sum in eq.~\eqref{P-ee-1st} lacked an overall factor $\sin 2 \phi$, which is corrected here. }
\begin{eqnarray} 
&& 
\left[ P(\nu_{e} \rightarrow \nu_{e})^{(0)} \right]^{ \text{non-int-fer} } 
= 
1 - 4 c^2_{\phi} s^2_{\phi} \sin^2 \frac{ ( h_{3} - h_{1} ) x }{2} 
- 4 c^2_{\phi} s^2_{\psi} ( s^2_{\phi} + c^2_{\phi} c^2_{\psi} ) 
\sin^2 \frac{ ( h_{2} - h_{1} ) x }{2}, 
\nonumber \\
&& 
\left[ P(\nu_{e} \rightarrow \nu_{e})^{(0)} \right]^{ \text{int-fer} } 
= 
4 c^2_{\phi} s^2_{\phi} s^2_{\psi} 
\left\{
- \sin^2 \frac{ ( h_{3} - h_{2} ) x }{2} 
+ \sin^2 \frac{ ( h_{3} - h_{1} ) x }{2} 
+ \sin^2 \frac{ ( h_{2} - h_{1} ) x }{2} 
\right\},
\nonumber \\
\label{decomposed-P-ee-0th-h1}
\end{eqnarray}
\begin{eqnarray} 
&&
\left[ P(\nu_{e} \rightarrow \nu_{e})^{(1)} \right]^{ \text{non-int-fer} } 
\nonumber \\
&=&
- 4 \epsilon c_{12} s_{12} s_{ (\phi - \theta_{13}) } \cos 2\phi \sin 2\phi c_{\psi} s_{\psi} 
\left( \frac{ \Delta_{ \text{ren} } }{ h_{3} - h_{2} } 
- \frac{ \Delta_{ \text{ren} } }{ h_{3} - h_{1} } \right) 
\sin^2 \frac{ ( h_{3} - h_{1} ) x }{2} 
\nonumber \\
&+&
4 \epsilon c_{12} s_{12} s_{ (\phi - \theta_{13}) } \sin 2\phi 
\left( c^2_{\psi} - \cos 2\phi s^2_{\psi} \right) c_{\psi} s_{\psi} 
\frac{ \Delta_{ \text{ren} } }{ h_{3} - h_{2} } 
\sin^2 \frac{ ( h_{2} - h_{1} ) x }{2}, 
\nonumber \\
&&
\left[ P(\nu_{e} \rightarrow \nu_{e})^{(1)} \right]^{ \text{int-fer} } 
\nonumber \\
&=&
- 4 \epsilon c_{12} s_{12} s_{ (\phi - \theta_{13}) } \sin 2\phi c_{\psi} s_{\psi} 
\left\{ 
\left( s^2_{\phi} - c^2_{\phi} s^2_{\psi} \right) 
\frac{ \Delta_{ \text{ren} } }{ h_{3} - h_{2} } 
+ c^2_{\phi} s^2_{\psi}  
\frac{ \Delta_{ \text{ren} } }{ h_{3} - h_{1} } 
\right\} 
\nonumber \\
&\times& 
\left\{ - \sin^2 \frac{ ( h_{3} - h_{2} ) x }{2} 
+ \sin^2 \frac{ ( h_{3} - h_{1} ) x }{2} 
+ \sin^2 \frac{ ( h_{2} - h_{1} ) x }{2} \right\}.
\label{decomposed-P-ee-1st-h1}
\end{eqnarray}

The oscillation probability $P(\nu_{e} \rightarrow \nu_{e})$ calculated by using the $S$ matrix method to first order in the DMP expansion is given by
\begin{eqnarray} 
&& 
P(\nu_{e} \rightarrow \nu_{e})^{(0)} 
\nonumber \\
&=&
1 
- 4 c^4_{\phi} c^2_{\psi} s^2_{\psi} \sin^2 \frac{ ( h_{2} - h_{1} ) x }{2} 
- 4 c^2_{\phi} s^2_{\phi} \left( c^2_{\psi} \sin^2 \frac{ ( h_{3} - h_{1} ) x }{2} 
+ s^2_{\psi} \sin^2 \frac{ ( h_{3} - h_{2} ) x }{2} \right), 
\nonumber \\
\label{P-ee-0th}
\end{eqnarray}
\begin{eqnarray} 
&& 
P(\nu_{e} \rightarrow \nu_{e})^{(1)}  
= 
- 4 \epsilon c_{12} s_{12} s_{ (\phi - \theta_{13}) } c^2_{\phi} \sin 2\phi 
c_{\psi} s_{\psi} 
\nonumber \\
&\times& 
\biggl[
\frac{ \Delta_{ \text{ren} } }{ h_{3} - h_{2} } 
\biggl\{ 
s^2_{\psi} \sin^2 \frac{ ( h_{3} - h_{2} ) x }{2} 
+ c^2_{\psi} \sin^2 \frac{ ( h_{3} - h_{1} ) x }{2}
- c^2_{\psi} \sin^2 \frac{ ( h_{2} - h_{1} ) x }{2}
\biggr\}
\nonumber \\
&-&
\frac{ \Delta_{ \text{ren} } }{ h_{3} - h_{1} } 
\biggl\{ 
s^2_{\psi} \sin^2 \frac{ ( h_{3} - h_{2} ) x }{2}
+ c^2_{\psi} \sin^2 \frac{ ( h_{3} - h_{1} ) x }{2}
- s^2_{\psi} \sin^2 \frac{ ( h_{2} - h_{1} ) x }{2}
\biggr\}
\biggr]
\nonumber \\
&+& 
4 \epsilon c_{12} s_{12} s_{ (\phi - \theta_{13}) } s^2_{\phi} \sin 2\phi c_{\psi} s_{\psi} 
\biggl[
\frac{ \Delta_{ \text{ren} } }{ h_{3} - h_{2} } 
\sin^2 \frac{ ( h_{3} - h_{2} ) x }{2} 
- \frac{ \Delta_{ \text{ren} } }{ h_{3} - h_{1} } 
\sin^2 \frac{ ( h_{3} - h_{1} ) x }{2} 
\biggr]
\nonumber \\
\label{P-ee-1st}
\end{eqnarray}

\subsection{$\nu_{\mu}$ disappearance channel}
\label{sec:numu-numu}

Similarly we present the formulas of the decomposed probabilities with the $e^{ - i h_{1} x }$ re-phasing, and the expression of the probability obtained by the $S$ matrix method. The former read in the zeroth order 
\begin{eqnarray} 
&&
\left[ P(\nu_{\mu} \rightarrow \nu_{\mu})^{(0)} \right]^{ \text{non-int-fer} } 
=
1 - 4 s^2_{23} c^2_{\phi} ( c^2_{23} + s^2_{23} s^2_{\phi} ) 
\sin^2 \frac{ ( h_{3} - h_{1} ) x }{2} 
\nonumber \\
&-& 
4 \left( c^2_{23} c^2_{\psi} + s^2_{23} s^2_{\phi} s^2_{\psi} - 2 c_{23} s_{23} s_{\phi} c_{\psi} s_{\psi} \cos \delta \right) 
\nonumber \\
&\times& 
\left[ c^2_{23} s^2_{\psi} 
+ s^2_{23} ( c^2_{\phi} + s^2_{\phi} c^2_{\psi} ) 
+ 2 c_{23} s_{23} s_{\phi} c_{\psi} s_{\psi} \cos \delta 
\right]
\sin^2 \frac{ ( h_{2} - h_{1} ) x }{2}, 
\nonumber \\
&&
\left[ P(\nu_{\mu} \rightarrow \nu_{\mu})^{(0)} \right]^{ \text{int-fer} } 
=
4 s^2_{23} c^2_{\phi} 
\left( c^2_{23} c^2_{\psi} + s^2_{23} s^2_{\phi} s^2_{\psi} - 2 c_{23} s_{23} s_{\phi} c_{\psi} s_{\psi} \cos \delta \right) 
\nonumber \\
&\times&
\left\{ 
- \sin^2 \frac{ ( h_{3} - h_{2} ) x }{2} 
+ \sin^2 \frac{ ( h_{3} - h_{1} ) x }{2} 
+ \sin^2 \frac{ ( h_{2} - h_{1} ) x }{2} 
\right\}, 
\end{eqnarray}
and in the first order 
\begin{eqnarray} 
&& 
\left[ P(\nu_{\mu} \rightarrow \nu_{\mu})^{(1)} \right]^{ \text{non-int-fer} } 
\nonumber \\
&=& 
4 \epsilon c_{12} s_{12} s_{ (\phi - \theta_{13}) } 
\left( c^2_{23} - s^2_{23} \cos 2\phi \right) 
\left( s^2_{23} \sin 2\phi s_{\psi} - \sin 2 \theta_{23} c_{\phi} c_{\psi} \cos \delta 
\right) 
c_{\psi} \frac{ \Delta_{ \text{ren} } }{ h_{3} - h_{2} } 
\sin^2 \frac{ ( h_{3} - h_{1} ) x }{2} 
\nonumber \\
&-& 
4 \epsilon c_{12} s_{12} s_{ (\phi - \theta_{13}) } 
\left( c^2_{23} - s^2_{23} \cos 2\phi \right) 
\left( s^2_{23} \sin 2\phi c_{\psi} + \sin 2 \theta_{23} c_{\phi} s_{\psi} \cos \delta 
\right) 
s_{\psi} \frac{ \Delta_{ \text{ren} } }{ h_{3} - h_{1} } 
\sin^2 \frac{ ( h_{3} - h_{1} ) x }{2}
\nonumber \\
&-& 
4 \epsilon c_{12} s_{12} s_{ (\phi - \theta_{13}) } 
\left( s^2_{23} \sin 2\phi s_{\psi} - \sin 2 \theta_{23} c_{\phi} c_{\psi} \cos \delta \right) 
\nonumber \\
&\times&
\left[ 
s^2_{23} c^2_{\phi} - \left( c^2_{23} - s^2_{23} s^2_{\phi} \right) \cos 2\psi 
+ 4 c_{23} s_{23} s_{\phi} c_{\psi} s_{\psi} \cos \delta 
\right] 
c_{\psi} \frac{ \Delta_{ \text{ren} } }{ h_{3} - h_{2} } 
\sin^2 \frac{ ( h_{2} - h_{1} ) x }{2}, 
\nonumber \\
&& 
\left[ P(\nu_{\mu} \rightarrow \nu_{\mu})^{(1)} \right]^{ \text{int-fer} } 
\nonumber \\
&=& 
- 4 \epsilon c_{12} s_{12} s_{ (\phi - \theta_{13}) } 
\left( s^2_{23} \sin 2\phi s_{\psi} - \sin 2 \theta_{23} c_{\phi} c_{\psi} \cos \delta 
\right) 
\left[ c^2_{23} c^2_{\psi} + s^2_{23} ( s^2_{\phi} s^2_{\psi} - c^2_{\phi} ) 
- 2 c_{23} s_{23} s_{\phi} c_{\psi} s_{\psi} \cos \delta \right] 
\nonumber \\
&\times&
c_{\psi} \frac{ \Delta_{ \text{ren} } }{ h_{3} - h_{2} } 
\left\{ 
- \sin^2 \frac{ ( h_{3} - h_{2} ) x }{2} 
+ \sin^2 \frac{ ( h_{3} - h_{1} ) x }{2} 
+ \sin^2 \frac{ ( h_{2} - h_{1} ) x }{2} \right\} 
\nonumber \\
&+& 
4 \epsilon c_{12} s_{12} s_{ (\phi - \theta_{13}) } 
\left( s^2_{23} \sin 2\phi c_{\psi} + \sin 2 \theta_{23} c_{\phi} s_{\psi} \cos \delta 
\right) 
\left( c^2_{23} c^2_{\psi} + s^2_{23} s^2_{\phi} s^2_{\psi} - 2 c_{23} s_{23} s_{\phi} c_{\psi} s_{\psi} \cos \delta \right) 
\nonumber \\
&\times&
s_{\psi} \frac{ \Delta_{ \text{ren} } }{ h_{3} - h_{1} } 
\left\{ 
- \sin^2 \frac{ ( h_{3} - h_{2} ) x }{2} 
+ \sin^2 \frac{ ( h_{3} - h_{1} ) x }{2} 
+ \sin^2 \frac{ ( h_{2} - h_{1} ) x }{2} 
\right\}. 
\end{eqnarray}

The oscillation probability $P(\nu_{\mu} \rightarrow \nu_{\mu})$ from the $S$ matrix calculation is given by the following formulas. Notice that the corresponding formulas in the $\nu_{\mu} \rightarrow \nu_{e}$ channel are given in appendix~\ref{sec:DMP-probability-mue}. 
\begin{eqnarray} 
&& 
P(\nu_{\mu} \rightarrow \nu_{\mu})^{(0)} 
= 
1 - 4 \left( c^2_{23} s^2_{\psi} + s^2_{23} s^2_{\phi} c^2_{\psi} \right) 
\left( c^2_{23} c^2_{\psi} + s^2_{23} s^2_{\phi} s^2_{\psi} \right) 
\sin^2 \frac{ ( h_{2} - h_{1} ) x }{2} 
\nonumber \\
&-& 
4 s^2_{23} c^2_{\phi}
\biggl\{
\left( c^2_{23} s^2_{\psi} + s^2_{23} s^2_{\phi} c^2_{\psi} \right) 
\sin^2 \frac{ ( h_{3} - h_{1} ) x }{2} 
+ \left( c^2_{23} c^2_{\psi} + s^2_{23} s^2_{\phi} s^2_{\psi} \right) 
\sin^2 \frac{ ( h_{3} - h_{2} ) x }{2} 
\biggr\} 
\nonumber \\
&-&
8 c_{23} s_{23} s_{\phi} c_{\psi} s_{\psi} \cos \delta 
\nonumber \\
&\times& 
\left[ 
\left( c^2_{23} - s^2_{23} s^2_{\phi} \right) \cos 2\psi 
\sin^2 \frac{ ( h_{2} - h_{1} ) x }{2} 
- s^2_{23} c^2_{\phi} 
\left\{ \sin^2 \frac{ ( h_{3} - h_{2} ) x }{2} 
- \sin^2 \frac{ ( h_{3} - h_{1} ) x }{2} 
\right\} 
\right] 
\nonumber \\
&+& 
16 c^2_{23} s^2_{23} s^2_{\phi} c^2_{\psi} s^2_{\psi} \cos^2 \delta 
\sin^2 \frac{ ( h_{2} - h_{1} ) x }{2}, 
\end{eqnarray}
\begin{eqnarray} 
&& 
P(\nu_{\mu} \rightarrow \nu_{\mu})^{(1)} 
\nonumber \\&=& 
- 4 \epsilon c_{12} s_{12} s_{ (\phi - \theta_{13}) } c^2_{23} 
\left( - s^2_{23} \sin 2\phi s_{\psi} + \sin 2 \theta_{23} c_{\phi} c_{\psi} \cos \delta \right) 
c_{\psi} \frac{ \Delta_{ \text{ren} } }{ h_{3} - h_{2} } 
\nonumber \\
&\times& 
\left\{ 
c^2_{\psi} \sin^2 \frac{ ( h_{3} - h_{2} ) x }{2} 
+ s^2_{\psi} \sin^2 \frac{ ( h_{3} - h_{1} ) x }{2} 
- s^2_{\psi} \sin^2 \frac{ ( h_{2} - h_{1} ) x }{2} 
\right\} 
\nonumber \\
&-& 
4 \epsilon c_{12} s_{12} s_{ (\phi - \theta_{13}) } c^2_{23} 
\left( s^2_{23} \sin 2\phi c_{\psi} + \sin 2 \theta_{23} c_{\phi} s_{\psi} \cos \delta \right) 
s_{\psi} \frac{ \Delta_{ \text{ren} } }{ h_{3} - h_{1} } 
\nonumber \\
&\times& 
\left\{ 
c^2_{\psi} \sin^2 \frac{ ( h_{3} - h_{2} ) x }{2} 
+ s^2_{\psi} \sin^2 \frac{ ( h_{3} - h_{1} ) x }{2} 
- c^2_{\psi} \sin^2 \frac{ ( h_{2} - h_{1} ) x }{2} 
\right\} 
\nonumber \\
&+& 
4 \epsilon c_{12} s_{12} s_{ (\phi - \theta_{13}) } s^2_{23} 
\left( - s^2_{23} \sin 2\phi s_{\psi} + \sin 2 \theta_{23} c_{\phi} c_{\psi} \cos \delta \right) 
c_{\psi} \frac{ \Delta_{ \text{ren} } }{ h_{3} - h_{2} } 
\nonumber \\
&\times& 
\left\{ 
( c^2_{\phi} - s^2_{\phi} s^2_{\psi} ) \sin^2 \frac{ ( h_{3} - h_{2} ) x }{2} 
- s^2_{\phi} c^2_{\psi} \sin^2 \frac{ ( h_{3} - h_{1} ) x }{2} 
+ s^2_{\phi} c^2_{\psi} \sin^2 \frac{ ( h_{2} - h_{1} ) x }{2} 
\right\} 
\nonumber \\
&+& 
4 \epsilon c_{12} s_{12} s_{ (\phi - \theta_{13}) } s^2_{23} 
\left( s^2_{23} \sin 2\phi c_{\psi} + \sin 2 \theta_{23} c_{\phi} s_{\psi} \cos \delta \right) 
s_{\psi} \frac{ \Delta_{ \text{ren} } }{ h_{3} - h_{1} } 
\nonumber \\
&\times& 
\left\{
- s^2_{\phi} s^2_{\psi} \sin^2 \frac{ ( h_{3} - h_{2} ) x }{2}
+ ( c^2_{\phi} - s^2_{\phi} c^2_{\psi} ) \sin^2 \frac{ ( h_{3} - h_{1} ) x }{2} 
+ s^2_{\phi} s^2_{\psi} \sin^2 \frac{ ( h_{2} - h_{1} ) x }{2} 
\right\} 
\nonumber \\
&-& 
8 \epsilon c_{12} s_{12} s_{ (\phi - \theta_{13}) } c_{23} s_{23} s_{\phi} c_{\psi} s_{\psi} \cos \delta 
\left( s^2_{23} \sin 2\phi s_{\psi} - \sin 2 \theta_{23} c_{\phi} c_{\psi} \cos \delta \right) 
c_{\psi} \frac{ \Delta_{ \text{ren} } }{ h_{3} - h_{2} } 
\nonumber \\
&\times& 
\left\{ 
\sin^2 \frac{ ( h_{3} - h_{2} ) x }{2}
- \sin^2 \frac{ ( h_{3} - h_{1} ) x }{2} 
+ \sin^2 \frac{ ( h_{2} - h_{1} ) x }{2} 
\right\} 
\nonumber \\
&-& 
8 \epsilon c_{12} s_{12} s_{ (\phi - \theta_{13}) } c_{23} s_{23} s_{\phi} c_{\psi} s_{\psi} \cos \delta 
\left( s^2_{23} \sin 2\phi c_{\psi} + \sin 2 \theta_{23} c_{\phi} s_{\psi} \cos \delta \right) 
s_{\psi} \frac{ \Delta_{ \text{ren} } }{ h_{3} - h_{1} } 
\nonumber \\
&\times& 
\left\{ 
- \sin^2 \frac{ ( h_{3} - h_{2} ) x }{2} 
+ \sin^2 \frac{ ( h_{3} - h_{1} ) x }{2} 
+ \sin^2 \frac{ ( h_{2} - h_{1} ) x }{2} 
\right\}.
\end{eqnarray}

\subsection{$\nu_{\mu} \rightarrow \nu_{\tau}$ appearance channel}
\label{sec:numu-tau}

Our presentation follows exactly the styles in the previous sections. The decomposed probabilities with the $e^{ - i h_{1} x }$ re-phasing read in the zeroth order 
\begin{eqnarray} 
&& 
\left[ P(\nu_{\mu} \rightarrow \nu_{\tau})^{(0)} \right]^{ \text{non-int-fer} } 
= 4 c^2_{23} s^2_{23} c^4_{\phi} 
\sin^2 \frac{ ( h_{3} - h_{1} ) x }{2} 
\nonumber \\
&+& 
%4 
\biggl\{ 
\sin^2 2\theta_{23} 
\left( c^4_{\psi} + s^4_{\phi} s^4_{\psi} \right) 
+ 4 \cos^2 2\theta_{23} s^2_{\phi} c^2_{\psi} s^2_{\psi} 
\nonumber \\
&-& 
8 c_{23} s_{23} s_{\phi} c_{\psi} s_{\psi} \cos 2\theta_{23} 
\left( s^2_{\phi} s^2_{\psi} - c^2_{\psi} \right) \cos \delta 
- 8 c^2_{23} s^2_{23} s^2_{\phi} c^2_{\psi} s^2_{\psi} \cos 2\delta 
\biggr\} 
\sin^2 \frac{ ( h_{2} - h_{1} ) x }{2}, 
\nonumber \\
&& 
\left[ P(\nu_{\mu} \rightarrow \nu_{\tau})^{(0)} \right]^{ \text{int-fer} }
=  
4 \left\{ 
c^2_{23} s^2_{23} c^2_{\phi} \left( s^2_{\phi} s^2_{\psi} - c^2_{\psi} \right) 
- c_{23} s_{23} c^2_{\phi} s_{\phi} c_{\psi} s_{\psi} \cos 2\theta_{23} \cos \delta 
\right\} 
\nonumber \\
&\times&
\left\{ 
- \sin^2 \frac{ ( h_{3} - h_{2} ) x }{2} 
+ \sin^2 \frac{ ( h_{3} - h_{1} ) x }{2} 
+ \sin^2 \frac{ ( h_{2} - h_{1} ) x }{2} 
\right\} 
\nonumber \\
&+& 
8 c_{23} s_{23} c^2_{\phi} s_{\phi} c_{\psi} s_{\psi} \sin \delta 
\sin \frac{ ( h_{3} - h_{1} ) x }{2} \sin \frac{ ( h_{2} - h_{1} ) x }{2} 
\sin \frac{ ( h_{3} - h_{2} ) x }{2}. 
\end{eqnarray} 
and in the first order 

\begin{eqnarray} 
&& 
\left[ P(\nu_{\mu} \rightarrow \nu_{\tau})^{(1)} \right]^{ \text{non-int-fer} }
= 
8 \epsilon c_{12} s_{12} s_{ (\phi - \theta_{13}) } 
c_{23} s_{23} c^2_{\phi} 
\nonumber \\
&\times&
\biggl[ 
\left\{ 
- c_{23} s_{23} \sin 2\phi s_{\psi} 
+ \cos 2\theta_{23} c_{\phi} c_{\psi} \cos \delta 
\right\} 
c_{\psi} \frac{ \Delta_{ \text{ren} } }{ h_{3} - h_{2} } 
\sin^2 \frac{ ( h_{3} - h_{1} ) x }{2}
\nonumber \\
&+& 
\left\{ 
c_{23} s_{23} \sin 2\phi c_{\psi} 
+ \cos 2\theta_{23} c_{\phi} s_{\psi} \cos \delta 
\right\} 
s_{\psi} \frac{ \Delta_{ \text{ren} } }{ h_{3} - h_{1} } 
\sin^2 \frac{ ( h_{3} - h_{1} ) x }{2}
\biggr] 
\nonumber \\
&+& 
8 \epsilon c_{12} s_{12} s_{ (\phi - \theta_{13}) } 
%\nonumber \\&\times&
\biggl\{ 
c^2_{23} s^2_{23} \sin 2\phi s_{\psi} 
\left( s^2_{\phi} s^2_{\psi} - c^2_{\psi} \right) 
+ 
\left[ 1 - 2 c^2_{23} s^2_{23} ( 1 + \cos 2\delta ) \right]
c_{\phi} s_{\phi} c^2_{\psi} s_{\psi} 
\nonumber \\
&-& 
c_{23} s_{23} \cos 2\theta_{23} c_{\psi} \cos \delta 
\left[ 
\sin 2\phi s_{\phi} s^2_{\psi}
+ c_{\phi} \left( s^2_{\phi} s^2_{\psi} - c^2_{\psi} \right) 
\right] 
\biggr\}
c_{\psi} \frac{ \Delta_{ \text{ren} } }{ h_{3} - h_{2} } 
\sin^2 \frac{ ( h_{2} - h_{1} ) x }{2}, 
\nonumber \\
&& 
\left[ P(\nu_{\mu} \rightarrow \nu_{\tau})^{(1)} \right]^{ \text{int-fer} }
\nonumber \\
&=& 
4 \epsilon c_{12} s_{12} s_{ (\phi - \theta_{13}) } 
\biggl\{
- \left[ 1 - 2 c^2_{23} s^2_{23} ( 1 + \cos 2\delta ) \right] 
c_{\phi} s_{\phi} c^2_{\psi} s_{\psi} 
+ c^2_{23} s^2_{23} \sin 2\phi s_{\psi} 
\left[ \cos 2\phi + ( 1 + s^2_{\phi} ) c^2_{\psi} \right] 
\nonumber \\
&+& 
c_{23} s_{23} \cos 2\theta_{23} c_{\psi} \cos \delta 
\left[ 
s_{\phi} \sin 2\phi s^2_{\psi} 
- c_{\phi} 
\left\{ \cos 2\phi + ( 1 + s^2_{\phi} ) c^2_{\psi} \right\} 
\right] 
\biggr\} 
\nonumber \\
&\times&
c_{\psi} \frac{ \Delta_{ \text{ren} } }{ h_{3} - h_{2} } 
\left\{ 
- \sin^2 \frac{ ( h_{3} - h_{2} ) x }{2} 
+ \sin^2 \frac{ ( h_{3} - h_{1} ) x }{2} 
+ \sin^2 \frac{ ( h_{2} - h_{1} ) x }{2} 
\right\} 
\nonumber \\
&+& 
4 \epsilon c_{12} s_{12} s_{ (\phi - \theta_{13}) } 
\biggl\{
- \left[ 1 - 2 c^2_{23} s^2_{23} ( 1 + \cos 2\delta ) \right] 
c_{\phi} s_{\phi} c_{\psi} s^2_{\psi}  
+ c^2_{23} s^2_{23} \sin 2\phi c_{\psi} 
\left( s^2_{\phi} s^2_{\psi} - c^2_{\psi} \right) 
\nonumber \\
&+& 
c_{23} s_{23} \cos 2\theta_{23} s_{\psi} \cos \delta 
\left[ c_{\phi} \left( s^2_{\phi} s^2_{\psi} - c^2_{\psi} \right) 
- s_{\phi} \sin 2\phi c^2_{\psi} \right] 
\biggr\} 
\nonumber \\
&\times&
s_{\psi} \frac{ \Delta_{ \text{ren} } }{ h_{3} - h_{1} } 
\left\{ 
- \sin^2 \frac{ ( h_{3} - h_{2} ) x }{2} 
+ \sin^2 \frac{ ( h_{3} - h_{1} ) x }{2} 
+ \sin^2 \frac{ ( h_{2} - h_{1} ) x }{2} 
\right\} 
\nonumber \\
&-& 
8 \epsilon c_{12} s_{12} s_{ (\phi - \theta_{13}) } c_{23} s_{23} c_{\phi} \sin \delta 
\left\{
\left( s^2_{\phi} - c^2_{\phi} s^2_{\psi} \right) 
c^2_{\psi} \frac{ \Delta_{ \text{ren} } }{ h_{3} - h_{2} } 
- \left( s^2_{\phi} - c^2_{\phi} c^2_{\psi} \right) 
s^2_{\psi} \frac{ \Delta_{ \text{ren} } }{ h_{3} - h_{1} } 
\right\} 
\nonumber \\
&\times& 
\sin \frac{ ( h_{3} - h_{1} ) x }{2} \sin \frac{ ( h_{2} - h_{1} ) x }{2} 
\sin \frac{ ( h_{3} - h_{2} ) x }{2}.
\end{eqnarray}

The probability $P(\nu_{\mu} \rightarrow \nu_{\tau})$ from $S$ matrix method reads in each order as 
\begin{eqnarray} 
&& 
P(\nu_{\mu} \rightarrow \nu_{\tau})^{(0)} 
%\nonumber \\&=& 
= \left\{ 1 - 2 c^2_{23} s^2_{23} ( 1 + \cos 2\delta ) \right\} 
s^2_{\phi} \sin^2 2\psi 
\sin^2 \frac{ ( h_{2} - h_{1} ) x }{2}
\nonumber \\
&-& 
4 c^2_{23} s^2_{23} 
\biggl[ 
\left\{ c^2_{\psi} s^2_{\psi} ( 1 + s^2_{\phi} )^2 - s^2_{\phi} \right\}
\sin^2 \frac{ ( h_{2} - h_{1} ) x }{2} 
\nonumber \\
&+& 
c^2_{\phi} \left( s^2_{\phi} c^2_{\psi} - s^2_{\psi} \right) 
\sin^2 \frac{ ( h_{3} - h_{1} ) x }{2} 
+ c^2_{\phi} \left( s^2_{\phi} s^2_{\psi} - c^2_{\psi} \right) 
\sin^2 \frac{ ( h_{3} - h_{2} ) x }{2} 
\biggr]
\nonumber \\
&+& 
4 c_{23} s_{23} s_{\phi} c_{\psi} s_{\psi} \cos 2\theta_{23} \cos \delta 
\nonumber \\
&\times& 
\biggl\{ 
( 1 + s^2_{\phi} ) \cos 2\psi \sin^2 \frac{ ( h_{2} - h_{1} ) x }{2} 
- c^2_{\phi} \sin^2 \frac{ ( h_{3} - h_{1} ) x }{2} 
+ c^2_{\phi} \sin^2 \frac{ ( h_{3} - h_{2} ) x }{2} 
\biggr\} 
\nonumber \\
&+& 
8 c_{23} s_{23} c^2_{\phi} s_{\phi} c_{\psi} s_{\psi} \sin \delta 
\sin \frac{ ( h_{3} - h_{1} ) x }{2} \sin \frac{ ( h_{2} - h_{1} ) x }{2} 
\sin \frac{ ( h_{3} - h_{2} ) x }{2}, 
\end{eqnarray}
and 
\begin{eqnarray} 
&& 
P(\nu_{\mu} \rightarrow \nu_{\tau})^{(1)} 
\nonumber \\
&=& 
- 4 \epsilon c_{12} s_{12} s_{ (\phi - \theta_{13}) } s_{\phi} c_{\psi} s_{\psi} 
\biggl[
c_{23} s_{23} \sin 2\phi s_{\psi} \cos 2\theta_{23} \cos \delta 
- \left\{ 1 - 2 c^2_{23} s^2_{23} ( 1 + \cos 2\delta ) \right\} 
c_{\phi} c_{\psi} 
\biggr] 
\nonumber \\
&\times& 
c_{\psi} \frac{ \Delta_{ \text{ren} } }{ h_{3} - h_{2} } 
\biggl\{
\sin^2 \frac{ ( h_{3} - h_{2} ) x }{2} 
- \sin^2 \frac{ ( h_{3} - h_{1} ) x }{2} 
+ \sin^2 \frac{ ( h_{2} - h_{1} ) x }{2} 
\biggr\} 
\nonumber \\
&-& 
4 \epsilon c_{12} s_{12} s_{ (\phi - \theta_{13}) } s_{\phi} c_{\psi} s_{\psi} 
\biggl[
c_{23} s_{23} \sin 2\phi c_{\psi} \cos 2\theta_{23} \cos \delta 
+ \left\{ 1 - 2 c^2_{23} s^2_{23} ( 1 + \cos 2\delta ) \right\} 
c_{\phi} s_{\psi} 
\biggr] 
\nonumber \\
&\times& 
s_{\psi} \frac{ \Delta_{ \text{ren} } }{ h_{3} - h_{1} } 
\biggl\{
- \sin^2 \frac{ ( h_{3} - h_{2} ) x }{2} 
+ \sin^2 \frac{ ( h_{3} - h_{1} ) x }{2} 
+ \sin^2 \frac{ ( h_{2} - h_{1} ) x }{2} 
\biggr\} 
\nonumber \\
&-& 
4 c_{23} s_{23} \epsilon c_{12} s_{12} s_{ (\phi - \theta_{13}) } 
\left( c_{23} s_{23} \sin 2\phi s_{\psi} - c_{\phi} c_{\psi} \cos 2\theta_{23} \cos \delta \right) 
c_{\psi} \frac{ \Delta_{ \text{ren} } }{ h_{3} - h_{2} } 
\nonumber \\
&\times&
\biggl[ 
\left( s^2_{\phi} c^2_{\psi} - s^2_{\psi} \right) 
\biggl\{ - \sin^2 \frac{ ( h_{3} - h_{1} ) x }{2} 
+ \sin^2 \frac{ ( h_{2} - h_{1} ) x }{2} \biggr\} 
+ \left\{ \cos 2\phi + ( 1 + s^2_{\phi} ) c^2_{\psi} \right\} 
\sin^2 \frac{ ( h_{3} - h_{2} ) x }{2}  
\biggr] 
\nonumber \\
&+& 
4 c_{23} s_{23} 
\epsilon c_{12} s_{12} s_{ (\phi - \theta_{13}) } 
\left( c_{23} s_{23} \sin 2\phi c_{\psi} + c_{\phi} s_{\psi} \cos 2\theta_{23} \cos \delta \right) 
s_{\psi} \frac{ \Delta_{ \text{ren} } }{ h_{3} - h_{1} } 
\nonumber \\
&\times&
\biggl[
\left( s^2_{\phi} s^2_{\psi} - c^2_{\psi} \right) 
\biggl\{ - \sin^2 \frac{ ( h_{3} - h_{2} ) x }{2} 
+ \sin^2 \frac{ ( h_{2} - h_{1} ) x }{2} \biggr\} 
+ \left\{ \cos 2\phi + ( 1 + s^2_{\phi} ) s^2_{\psi} \right\} 
\sin^2 \frac{ ( h_{3} - h_{1} ) x }{2} 
\biggr] 
\nonumber \\
&-& 
8 \epsilon c_{23} s_{23} c_{12} s_{12} s_{ (\phi - \theta_{13}) } c_{\phi} \sin \delta 
\left\{
c^2_{\psi} \left( s^2_{\phi} - c^2_{\phi} s^2_{\psi} \right) 
\frac{ \Delta_{ \text{ren} } }{ h_{3} - h_{2} } 
- s^2_{\psi} \left( s^2_{\phi} - c^2_{\phi} c^2_{\psi} \right) 
\frac{ \Delta_{ \text{ren} } }{ h_{3} - h_{1} } 
\right\}
\nonumber \\
&\times&
\sin \frac{ ( h_{3} - h_{1} ) x }{2} \sin \frac{ ( h_{2} - h_{1} ) x }{2} 
\sin \frac{ ( h_{3} - h_{2} ) x }{2}. 
\end{eqnarray}

\end{document}